\documentclass[aps,prl,reprint,groupedaddress,twocolumn,superscriptaddress]{revtex4-1}

\usepackage{graphicx}% Include figure files
\usepackage{overpic}

\usepackage{dcolumn}% Align table columns on decimal point
\usepackage{bm}% bold math
\usepackage{natbib}
\usepackage{hyperref}
\hypersetup{
    colorlinks=true,
    linkcolor=blue,
    citecolor=blue,
    urlcolor=blue
}

\begin{document}

\preprint{APS/123-QED}
\bibliographystyle{unsrt}
\title{Topological Flat bands and Fractional Chern Insulators in Rashba Systems with Tunable Superlattice Potentials}
\author{Bokai Liang}
\affiliation{Department of Physics, University of Science and Technology of China, Hefei, Anhui 230026,China }
\author{Wei Qin}
\email{qinwei5@ustc.edu.cn} 
\affiliation{Department of Physics, University of Science and Technology of China, Hefei, Anhui 230026,China }
\affiliation{International Center for Quantum Design of Functional Materials (ICQD), Hefei National Research Center for Physical Sciences at Microscale, University of Science and Technology of China, Hefei 230026, China}
\author{Zhenyu Zhang}
\affiliation{International Center for Quantum Design of Functional Materials (ICQD), Hefei National Research Center for Physical Sciences at Microscale, University of Science and Technology of China, Hefei 230026, China}
\affiliation{Hefei National Laboratory, University of Science and Technology of China, Hefei 230088, China}

\date{\today}% It is always \today, today,

\begin{abstract}
Fractional Chern insulators are interaction-driven topological states that realize the lattice analogue of the fractional quantum Hall effect without external magnetic fields.
Here we propose a programmable platform for engineering topological flat bands by combining a Zeeman field with a tunable electrostatic superlattice potential in a Rashba spin-orbit-coupled thin film. The Zeeman field generates a nearly flat segment of the Rashba band, which is isolated by the superlattice potential into a flat band with Chern number $\mathcal{C}=1$. Using many-body exact diagonalization, we show that this band supports fractional Chern insulators at filling factors $n = 1/3$ and $2/3$, both of which remain robust over broad parameter ranges. We further identify realistic material candidates and the corresponding device conditions that enable experimental realization. These results establish a versatile and experimentally accessible platform for engineering topological flat bands and exploring correlated topological phases.
\end{abstract}

\maketitle

\textit{Introduction---}Fractional Chern insulators (FCIs) represent the lattice analogy of the fractional quantum Hall effect, where strong electronic correlations stablize exotic quantum many-body phases with fractionalized quasiparticle excitations \cite{laughlin1983anomalous,moore1991nonabelions,stormer1999fractional,stormer1999nobel,jain1990theory,regnault2011fractional}. The emergence of FCIs requires the coexistence of nontrivial band topology and strong electron-electron interactions \cite{girvin1986magneto,tang2011high,parameswaran2012fractional,bergholtz2013topological,roy2014band}. Two-dimensional (2D) van der Waals (vdW) heterostructures have recently emerged as promising platforms for realizing FCIs through the formation of topological moir\'e minibands \cite{bistritzer2011moire, wu2019topological,song2019all,tarnopolsky2019origin,andrei2021marvels,mak2022semiconductor}. Compelling experimental evidence of FCIs have been reported in moir\'e graphene systems \cite{spanton2018observation,xie2021fractional},
twisted transition metal dichalcogenides \cite{park2023observation,cai2023signatures,xu2023observation,redekop2024direct,ji2024local}, and rhombohedral graphene multilayers \cite{lu2024fractional,chen2024tunable,lu2025extended,choi2025superconductivity}. In these systems, FCIs can arise in the absence of external magnetic fields, provided that Coulomb interactions drive spontaneous spin-valley ferromagnetism \cite{li2021spontaneous,reddy2023fractional,crepel2023anomalous,yu2024fractional,dong2024theory,dong2024anomalous,zhou2024fractional}.  

The stability of FCIs is crucial for exploring their exotic properties and is governed by both intrinsic and extrinsic factors. Intrinsically, competing phases driven by charge, spin, and valley fluctuations may destabilize FCIs \cite{li2021spontaneous,reddy2023fractional,crepel2023anomalous, yu2024fractional}. A non-degenerate, isolated flat band with quantum geometry closely resembling that of the Landau levels is widely recognized as essential for stabilizing FCIs \cite{liu2022recent}. Extrinsically, moir\'e systems are inevitably subject to disorder, strain, and lattice relaxations \cite{kennes2021moire,song2021direct,he2021moire,lau2022reproducibility,nakatsuji2022moire}, all of which can compromise FCI stability. For example, the recently observed FCIs in rhombohedral graphene/hBN moir\'e superlattices are supplanted by an extended quantum anomalous Hall phase at lower temperatures \cite{lu2025extended}, likely due to disorder-induced localization \cite{huang2024impurity,das2024thermal}. 
These considerations highlight the need for ultra-clean systems hosting well-isolated, single-flavor flat bands to achieve robust FCIs.

Electrostatically engineered superlattices have recently emerged as a promising platform for generating flat bands \cite{forsythe2018band, barcons2022engineering,suri2023superlattice,ghorashi2023topological,wang2024dispersion,sun2024signature,tan2024designing,yang2025engineering,PhysRevB.111.045148,ghyq-sz16,Wang:2026aa}. The underlying principle is that imposing a superlattice potential on a system with nontrivial band topology can produce topological flat bands.
In this Letter, we build on this principle and show that Rashba systems, proximity-coupled to a magnetic insulator and subject to an electrostatically superlattice potential, provide a versatile and experimentally accessible platform for realizing robust topological flat bands and FCIs.
Our model calculations show that the interplay between Zeeman coupling and the superlattice potential gives rise to an isolated, nondegenerate flat band with Chern number $\mathcal{C}=1$, persisting across a broad parameter space. Furthermore, many-body exact diagonalization (ED) calculations demonstrate that Coulomb interactions stabilize FCIs at filling factors $n = 1/3$ and $2/3$, characterized by excitation gaps that remain sizable over a wide range of parameters. 
Finally, based on a scaling analysis, we identify realistic material candidates and delineate the corresponding experimental conditions for realizing these predictions. 
%\textcolor{green}{Unlike moiré platforms, where flat Chern bands rely on twist-induced interference or interaction-driven symmetry breaking, our proposal achieves an isolated topological flat band through explicit symmetry control and electrostatic band truncation, providing a robust and tunable platform for realizing FCI.}

\begin{figure}
    \centering
    \includegraphics[width=\columnwidth]{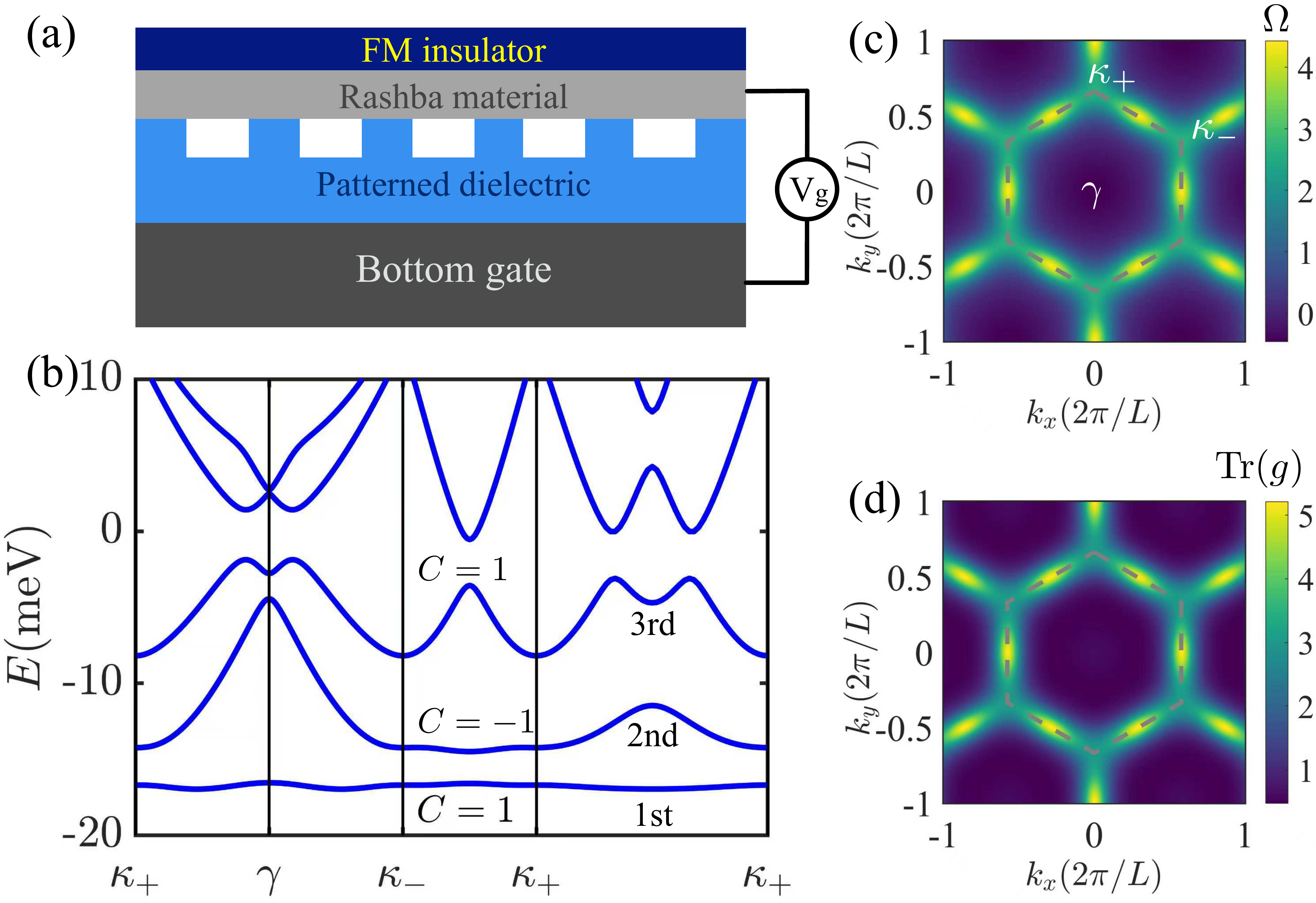}
    \caption{(a) Schematic of the proposed setup, consisting of a Rashba material sandwiched between a ferromagnetic (FM) insulator and a patterned dielectric layer. (b) Band structure for a superlattice potential with period $L=15$ nm. (c) Berry curvature $\Omega$ of the 1st band shown in (b), with the dashed hexagon denoting the BZ. (d) Trace of the Fubini-Study metric $\text{Tr}(g)$ of the 1st band. The units for both $\Omega$ and $\text{Tr}(g)$ are $2\pi/A_{\text{BZ}}$, with $A_{\text{BZ}}$ denoting the area of BZ. These results are obtained for $\tilde{\lambda}=0.4$,  $\tilde{V}_z=0.12$, and $\tilde{U}_0=0.03$, corresponding to $\lambda = 0.76$ eV$\cdot$\AA, $V_z = 22.9 $ meV, and $U_0=5.7$ meV, respectively. }
    \label{fig:figure1}	
\end{figure}
    
\textit{Model---}The proposed setup is schematically illustrated in Fig.~\ref{fig:figure1}(a), where a 2D semiconductor with Rashba spin-orbit coupling (SOC) is sandwiched between a ferromagnetic insulator and a periodically patterned dielectric material. The low-energy physics of the system is described by 
    \begin{equation}
    \mathcal{H}_0(\bm{k})=\frac{\hbar^2 \bm{k}^2}{2m^*}+\lambda (k_y\sigma_x-k_x\sigma_y)+\frac{V_z}{2} \sigma_z,
    \label{H0}
    \end{equation}
    where $m^*$ is the electron effective mass, $\lambda$ is the Rashba SOC strength, $V_z$ is the proximity-induced Zeeman coupling, and $\sigma_{x,y,z}$ are the Pauli matrices acting on spin. The presence of $V_z$ breaks time-reversal symmetry and lifts the Kramers degeneracy.

By patterning the dielectric material into a superlattice, an externally applied gate voltage can generate a periodic electrostatic potential in the Rashba thin film \cite{forsythe2018band,wang2024dispersion,ghorashi2023topological,tan2024designing,suri2023superlattice}. 
In this study, we consider a triangular superlattice as a representative example, while discussions on a square superlattice are provided in the Supplementary Material (SM) \cite{SM2024}. The lowest-harmonic components of a triangular superlattice potential are given by
    \begin{equation}
    U(\bm{r})=2 U_0 \sum_{n=0}^2 \cos \left(\bm{G}_n \cdot \bm{r}\right),
    \label{Vout}
    \end{equation}
where $U_0$ denotes the potential depth, $\bm{G}_n=(4\pi/\sqrt{3}L) (\cos n\phi, \sin n\phi)$ with $\phi = 2\pi/3$ are reciprocal lattice vectors, and $L$ is the superlattice period. In the following calculations, we adopt $L = 15$ nm and $m^*=0.2 m_e$, a typical value for the electron effective mass in semiconductors \cite{spitzer1957determination}. The total Hamiltonian, $\mathcal{H} = \mathcal{H}_0+U(\bm{r})$, can be expressed in a dimensionless form as 
    \begin{equation}
    \tilde{\mathcal{H}}= \bm{\tilde{k}}^2 +\tilde{\lambda} (\tilde{k}_y\sigma_x - \tilde{k}_x\sigma_y)  + \frac{\tilde{V}_z}{2}\sigma_z  +\tilde{U}(\bm{r}),
    \label{eq:dimensionlessH}
    \end{equation}
where $\tilde{\mathcal{H}} = \mathcal{H}/E_0$, $\tilde{k}_{x,y}=k_{x,y}/k_0$, $\tilde{\lambda} = \lambda k_0/E_0$, $\tilde{V}_z = V_z/E_0$, $\tilde{U}(\bm{r}) = U(\bm{r}) /E_0$, $k_0 =1$ nm$^{-1}$, and $E_0=\hbar^2k_0^2/2m^* \approx 190.5$ meV.
    Fig.~\ref{fig:figure1}(b) shows a representative band structure, where an isolated flat band emerges at the bottom of the spectrum. This flat band is topologically nontrivial, carrying a Chern number $\mathcal{C} =1$. As shown in Figs.~\ref{fig:figure1}(c) and (d), the corresponding Berry curvature $\Omega_{\bm{k}}$ and trace of the Fubini-Study metric $\text{Tr}(g_{\bm{k}})$ distribute predominantly along the Brillouin zone (BZ) boundary. The trace condition violation $T=A_{\text{BZ}}/2\pi \int d^2\bm{k} [\text{Tr}(g_{\bm{k}})-\Omega_{\bm{k}}]$ is estimated to be $\sim$0.54, a relatively small value and comparable with those reported in graphene systems \cite{dong2024theory,dong2024anomalous,zhou2024fractional}.

\begin{figure}
    \includegraphics[width=\columnwidth]{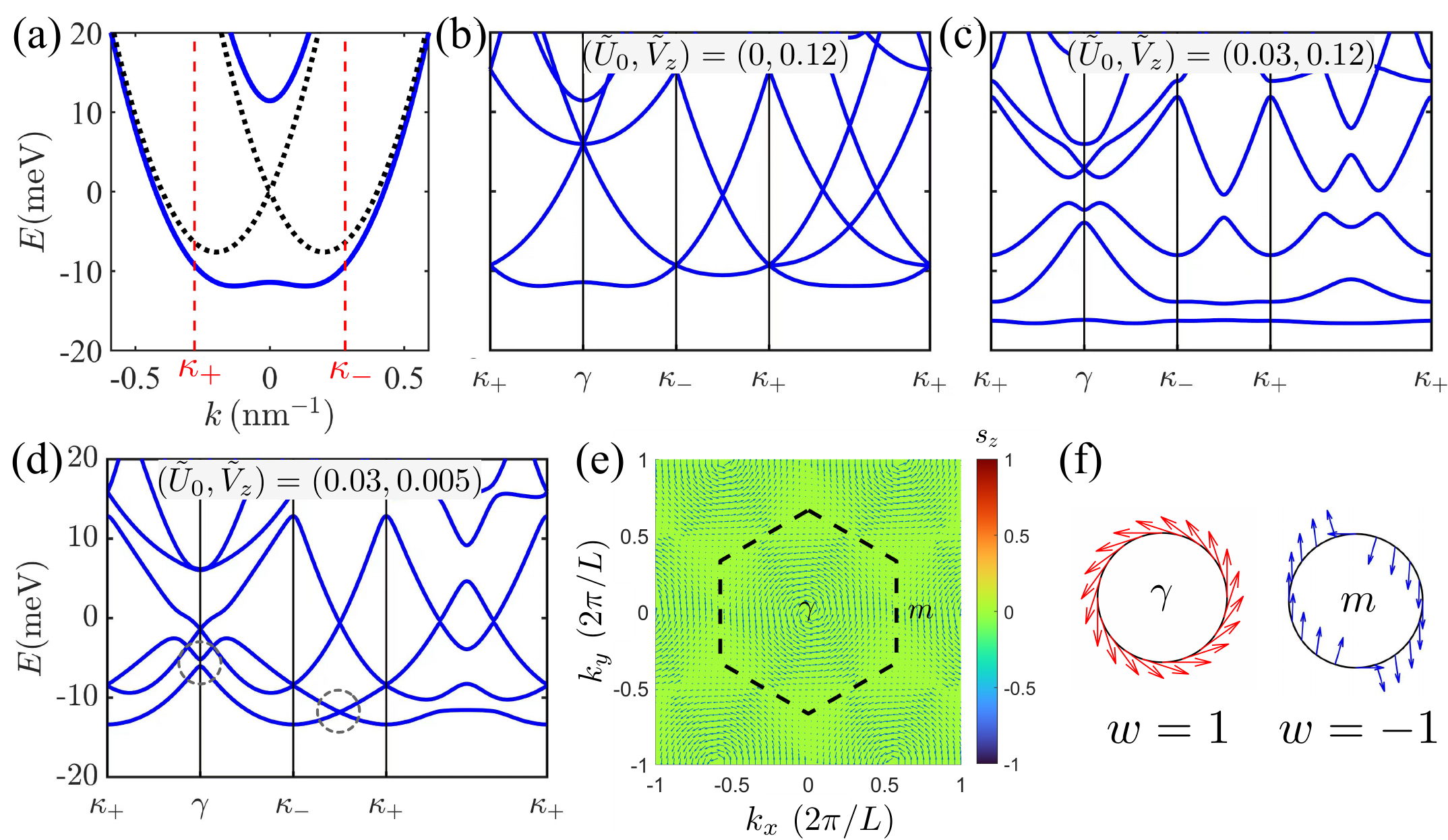}    
    \caption{(a) Rashba bands without (dotted) and with (solid) a Zeeman field, $\tilde{V}_z = 0.12$. The vertical dahsed lines mark the wavevectors at BZ corners. (b,c) Folded Rashba bands within the first BZ for (b) $\tilde{U}_0=0$ and (c) $\tilde{U}_0=0.03$. (d) Band structure for $\tilde{V}_z = 0.005$, where the dahsed circles highlight gapped Dirac cones. (e) Spin texture of the lowest-energy band shown in (d). The dashed hexagon denotes the first BZ. (f) Spin winding around closed loops encircling the (left) $\gamma$ and (right) $m$ points. The arrows denote the in-plane spin directions, and $w$ is the corresponding winding number. All results are obtained for $\tilde{\lambda} = 0.4$. }
    \label{fig:figure4}
\end{figure}

\textit{Flat Chern band---}
To elucidate the origin of the isolated flat Chern band, we examine the Rashba band folding induced by superlattice potential. As illustrated in Fig.~\ref{fig:figure4}(a), the Zeeman field lifts the degeneracy of the Rashba band crossing and generates a nearly flat segment around the band bottom of the lower Rashba band when $\tilde{V}_z$ becomes comprable to $\tilde{\lambda}^2$ \cite{SM2024}. A suitably designed periodic superlattice then selects this nearly flat segment by folding the Rashba bands into the first BZ, as shown in Fig.~\ref{fig:figure4}(b). Bragg scattering induced by superlattice potential further opens gaps along high-symmetry lines and at high-symmetry points in the BZ, thereby separating the nearly flat segment from the remaining dispersive bands and yielding an isolated band with an extremely small bandwidth, as shown in Fig.~\ref{fig:figure4}(c). Detailed derivations and demonstrations of this flat-band formation mechanism are provided in the SM \cite{SM2024}.

The nontrivial topology of the flat band originates from the gapped Dirac cones of the folded Rashba bands. In the absence of Zeeman field, band crossings at $\gamma$ and $m$ points are protected by time-reversal symmetry and are captured by effective Dirac Hamiltonians \cite{SM2024}. As shown in Fig.~\ref{fig:figure4}(d), a small $\tilde{V}_z$ acts as a mass term that gaps these Dirac cones, generating Berry curvature concentrated around the $\gamma$ and $m$ points \cite{SM2024}. Figures~\ref{fig:figure4}(e) and (f) show the spin texture and the corresponding spin winding around the $\gamma$ and $m$ points, yielding winding numbers $w_{\gamma}=1$ and $w_{m}=-1$, respectively. Since there are three symmetry-related $m$ points within the BZ, the Chern number of 1st band is
$\mathcal{C} = -\frac{1}{2}\text{sign}(\tilde{V}_z)(w_{\gamma}+3w_{m}) = 1$ for $\tilde{V}_z>0$ \cite{SM2024}.
As $\tilde{V}_z$ is further increased, the gapped Chern band evolves continuously into the isolated flat band shown in Fig.~\ref{fig:figure4}(c) without closing the band gap. Therefore, the flat band inherits the same nontrivial topology, yielding $\mathcal{C}=1$, in agreement with the numerical calculation.

\begin{figure}
\centering
\includegraphics[width=\columnwidth]{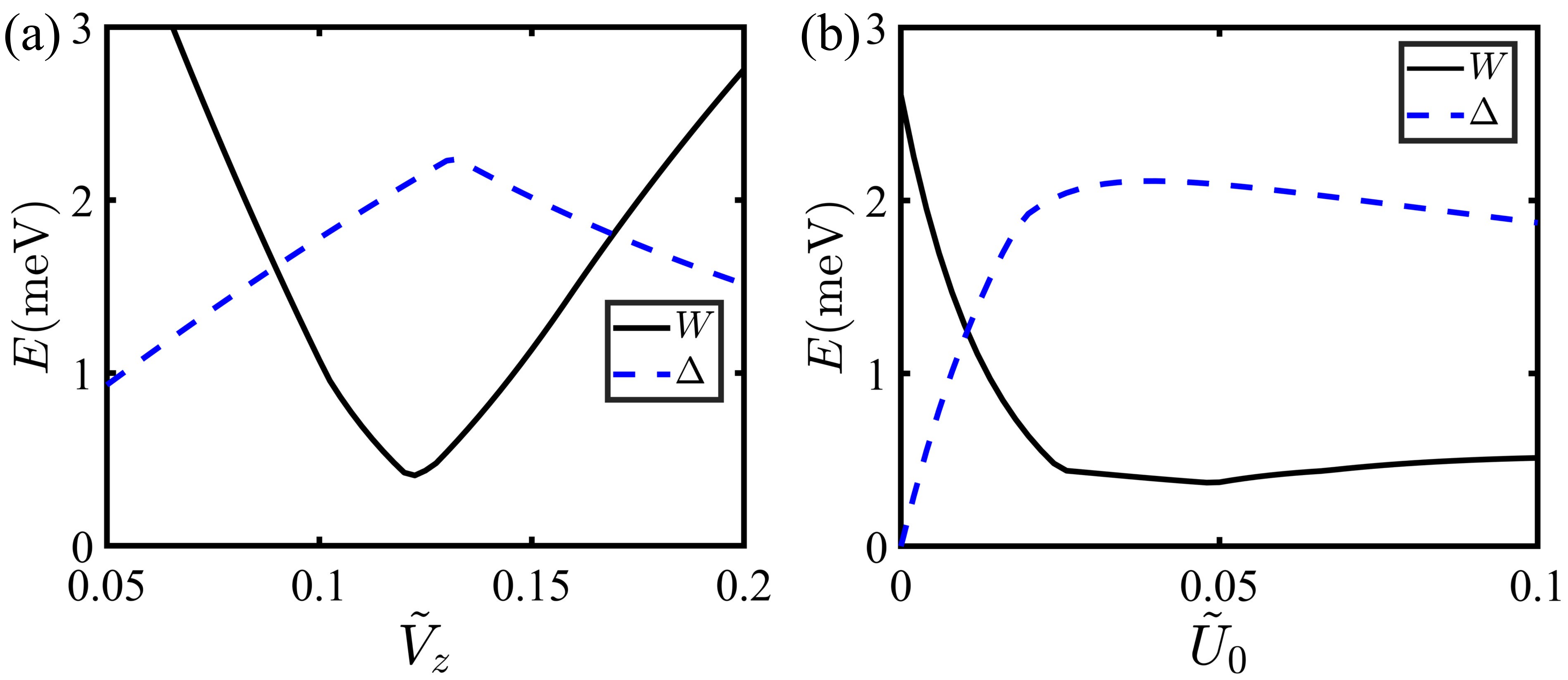}
\caption{(a,b) Bandwidth $W$ of the 1st band and direct band gap $\Delta$ as functions of (a) $\tilde{V}_z$ with $\tilde{U}_0=0.03$ and (b) $\tilde{U}_0$ with $\tilde{V}_z=0.12$. These results are obtained for $\tilde{\lambda}=0.4$. }
\label{fig:figure2}
\end{figure}

\textit{Phase diagram---}We next explore two key characters of the 1st band shown in Fig.~\ref{fig:figure1}(c), namely, the bandwidth $W$ and the direct band gap $\Delta$ separating it from neighboring bands. A small $W$ enhances interaction effects, while a large $\Delta$ suppresses interband mixing, both of which are crucial for stabilizing FCIs \cite{liu2022recent}. 
As shown in Fig.~\ref{fig:figure2}(a), $W$ and $\Delta$ exhibit a V-shaped and dome-like dependences upon increasing $\tilde{V}_z$ with $\tilde{U}_0=0.03$. Notably, $W$ reaches a minimum of $\sim$0.4 meV at $\tilde{V}_z =$ 0.12, nearly coinciding with a maximum $\Delta$ of $\sim$2.2 meV. This concurrence of minimal bandwidth and maximal band gap defines the optimal condition for realizing an isolated flat band. 
As shown in 
Fig.~\ref{fig:figure2}(b), for $\tilde{V_z}=0.12$, $W$ decreases rapidly with increasing $\tilde{U}_0$, approaching saturation once $\tilde{U}_0$ exceeds a threshold value of $\sim$0.02, whereas $\Delta$ rises quickly before saturating near $\sim$2 meV.  

\begin{figure}
    \includegraphics[width=\columnwidth]{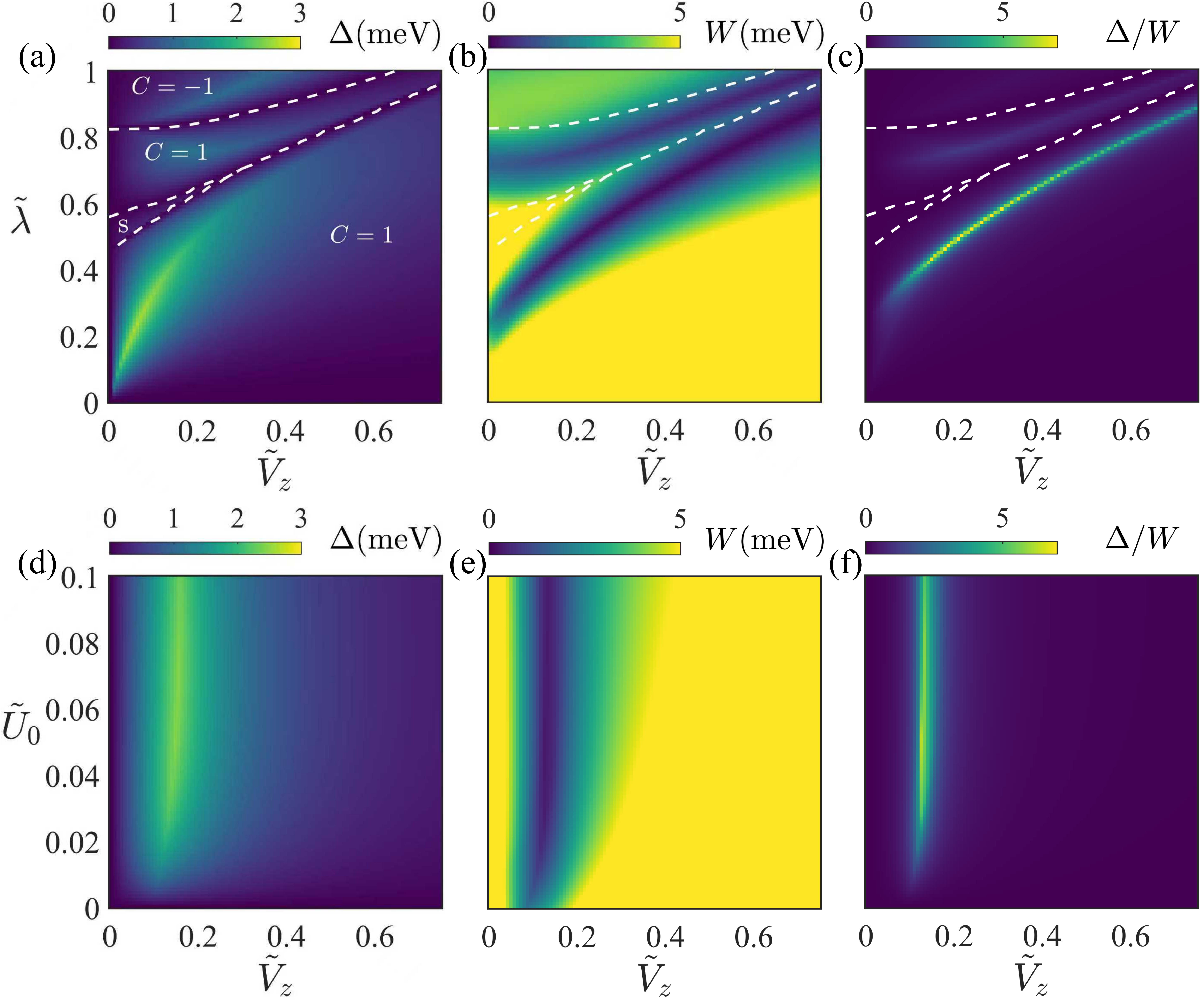}    
    \caption{(a)-(c) Phase diagrams of (a) direct band gap $\Delta$, (b) bandwidth $W$, and (c) their ratio $\Delta/W$ as functions of $\tilde{\lambda}$ and $\tilde{V}_z$. White dashed lines indicate gap-closing boundaries, separating the parameter sapce into several regions with corresponding Chern number $\mathcal{C}$ marked. The region labeled $s$ in (a) denotes $\mathcal{C}=-2$. Results in (a)-(c) are obtained with $\tilde{U}_0=0.03$. (d)-(f) Phase diagrams of (d) $\Delta$, (e) $W$, and (f) $\Delta/W$ as functions of \ $\tilde{U}_0$ and $\tilde{V}_z$, calculated with $\tilde{\lambda} = 0.4$.}
    \label{fig:figure3}
\end{figure}

Figure~\ref{fig:figure3} presents a systematic study of the 1st band by varying $\tilde{V}_z$, $\tilde{\lambda}$, and $\tilde{U}_0$. In Fig.~\ref{fig:figure3}(a), the phase diagram of $\Delta$ consists of distinct regions separated by gap-closing boundaries, where the $\mathcal{C}=1$ phase extends over a broad parameter space. Fig.~\ref{fig:figure3}(b) depicts the phase diagram of $W$, where the dark region indicates that the 1st band is ultra-flat.
The quality of the isolated flat band is characterized by the ratio $\Delta/W$, as shown in Fig.~\ref{fig:figure3}(c), where a bright arc emerges, defining the optimal condition in ($\tilde{V}_z$, $\tilde{\lambda}$) plane. The nearly vanishing bandwidth along this arc implies that Coulomb interactions dominate over electron kinetic energy and may stabilize FCIs. To examine the robustness of the optimal condition, we calculate phase diagrams by fixing $\tilde{\lambda}$ while varying 
$\tilde{V}_z$ and $\tilde{U}_0$. As shown in Fig.~\ref{fig:figure3}(d)-(f), the optimal regime remains robust once $\tilde{U}_0$ exceeds a threshold of $\sim$0.02, consistent with the result shown in Fig.~\ref{fig:figure2}(b). 
Additional phase diagrams for fixed $\tilde{V}_z$ while varying $\tilde{\lambda}$ and $\tilde{U}_0$ are provided in the SM \cite{SM2024}. These results demonstrate that the formation of the isolated flat band is relatively insensitive to $\tilde{U}_0$ beyond the threshold, implying that the electrostatic potential can be applied with substantial flexibility in experiments.

\begin{figure}[t]
    \centering
    \includegraphics[width=\columnwidth]{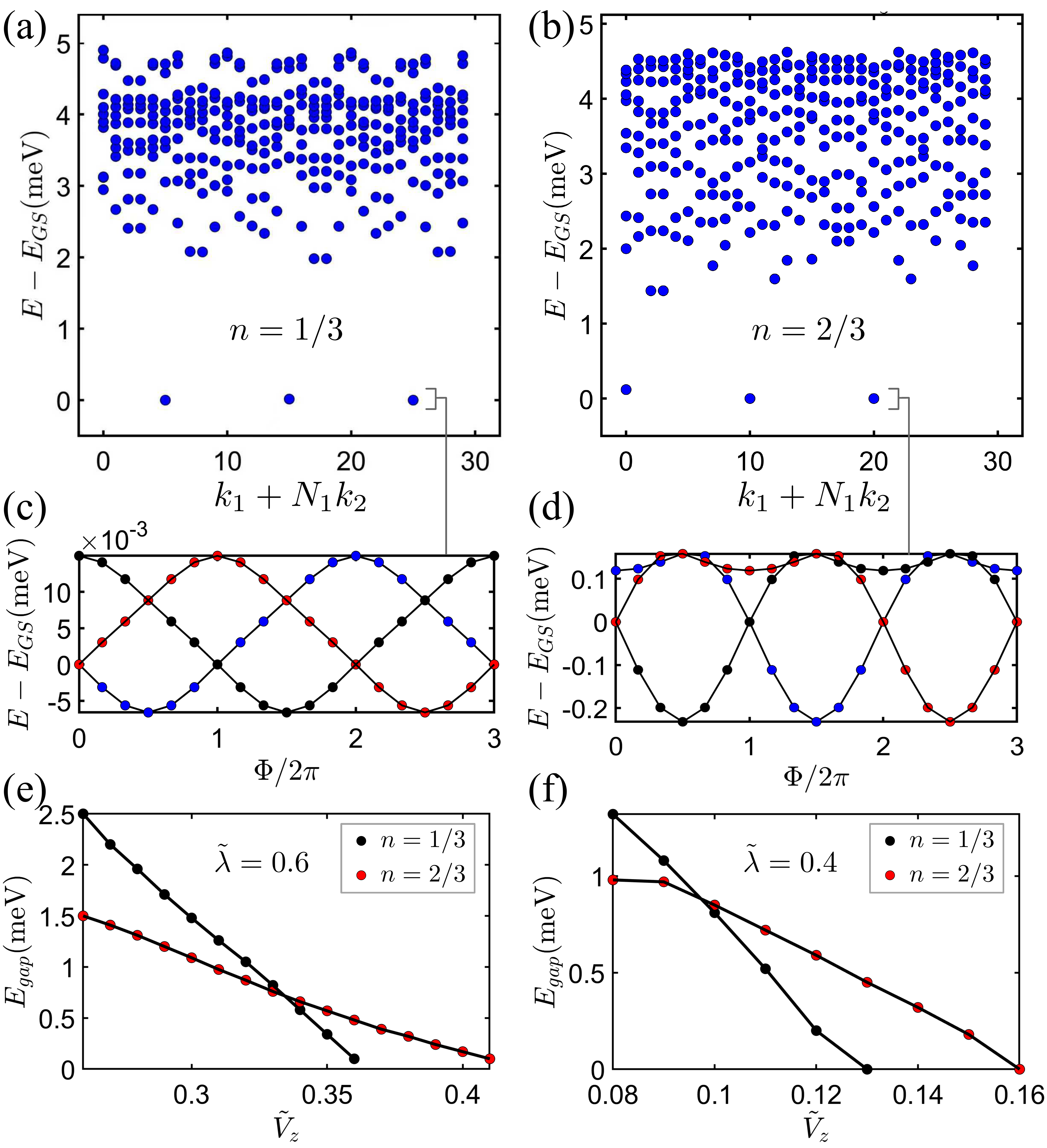}
    \caption{(a,b) Many-body spectra for filling factors (a) $n=1/3$ and (b) $n=2/3$. (c,d) Spectral flow of the ground states under magnetic flux insertion, with the three nearly degenerate states distinguished by colors. Results in (a)-(d) are obtained for $\tilde{\lambda}=0.6$, $\tilde{V}_z=0.28$, and $\tilde{U}_0=0.03$. (e,f) Excitation gaps $E_{gap} = E_{4}-E_3$ as functions of $\tilde{V}_z$ for (e) $\tilde{\lambda}=0.6$ and (f) $\tilde{\lambda}=0.4$, where $E_{i}$ denotes the $i$-th lowest energy state. All these results are obtained on a $5\times 6$ cluster with dielectric constant $\epsilon=5$.}
    \label{fig:figure5}
\end{figure}

\textit{Evidences of FCI---}Having established the emergence of topological flat band in this system, we now investigate possible many-body ground states, focusing on FCIs stabilized by electron-electron interactions. Specifically, we consider the 1st band with $\mathcal{C}=1$ and an unscreened Coulomb potential $V_{\bm{q}} = 2 \pi e^2/\epsilon q$, where $\epsilon$ denotes the environmental dielectric constant. Projecting the Coulomb interaction onto the 1st band yields the Hamiltonian \cite{SM2024} 
\begin{equation}
    H=\sum_{\bm{k}} \varepsilon_{\bm{k}} c_{\bm{k}}^{\dagger} c_{\bm{k}}+\frac{1}{2} \sum_{\bm{k}^{\prime}  \bm{p}^{\prime} \bm{p}\bm{k} } V_{ \bm{k}^{\prime}  \bm{p}^{\prime} \bm{p}\bm{k} } c_{\bm{k}^{\prime} }^{\dagger} c_{\bm{p}^{\prime} }^{\dagger} c_{\bm{p} } c_{\bm{k}},
\label{Hmanybody}
\end{equation}
where $\varepsilon_{\bm{k}}$ denotes the band dispersion and $c_{\bm{k}}^{\dagger}$ ($c_{\bm{k}}$) is the electron creation (annihilation) operator. 
The many-body ED calculations are performed at filling factors $n$, which is defined as the number of electrons per unit cell.
The Hilbert space is decomposed into subspaces labeled by crystal momentum $\bm{k}=k_1\bm{T}_1+k_2\bm{T}_2$, where $\bm{T}_{1,2}$ are the reciprocal basis vectors \cite{SM2024}. As shown in Figs.~\ref{fig:figure5}(a) and (b), the many-body spectra exhibit a threefold nearly degenerate ground-state manifold at both 1/3 and 2/3 fillings, consistent with the topological ground-state degeneracy of a fractional quantum Hall state on the torus. The corresponding excitation gaps are $\sim$2 meV and $\sim$1.5 meV for $n=1/3$ and $2/3$, respectively. This charateristic ground-state degeneracy and finite excitation gaps are also observed for 15-, 24-, and 27-unit-cell clusters \cite{SM2024}, providing further evidence for the stability of FCIs.

The spectral flow of the degenerate ground states is obtained by inserting magnetic flux through the torus, which shifts the crystal momentum as $\bm{k}' = \bm{k}+(\Phi/2\pi)\bm{T}_{i}$, where $\Phi=\phi/\phi_0$ and $\phi$ ($\phi_0$) denotes the inserted flux (flux quantum). 
As shown in Figs.~\ref{fig:figure5}(c) and (d), the ground states cyclically permute under insertion of three flux quanta, consistent with the characteristic behavior of fractional quantum Hall states. 
As shown in Fig.~\ref{fig:figure5}(e) and (f), the excitation gaps for both $n=1/3$ and $2/3$ remain rubust over a wide range of $\tilde{V}_z$. The gaps also persist upon varying $\tilde{U}_0$, as shown in the SM \cite{SM2024}. 
These results demonstrate the robustness of FCIs against variations in model parameters.
We further assess the effect of band mixing by allowing one additional electron to occupy the 2nd band. As detailed in the SM \cite{SM2024}, the ground-state degeneracy persists with only slightly reduced excitation gaps, confirming that the FCIs remain stable against interaction-induced band mixing.

\textit{Scaling analysis and material candidates---}The preceding calculations are performed at fixed superlattice period $L_0=15$ nm. Varying $L$ modifies the reciprocal lattice vectors and hence the BZ. This effect can be captured by rescaling the wavevector with a factor $\eta = L/L_0$, yielding a unified BZ independent of $L$. The rescaling leads to a renormalized Hamiltonian  
  \begin{equation}
    \tilde{\mathcal{H}}= \eta^{-2} [\bm{\tilde{k}}^2 +\eta \tilde{\lambda}(\tilde{k}_y\sigma_x - \tilde{k}_x\sigma_y)  + \frac{\eta^2\tilde{V}_z}{2}\sigma_z  +\eta^2\tilde{U}(\bm{r}) ],
    \label{eq:renormalizedH}
    \end{equation}
where $\bm{\tilde{k}}$ is defined identically to that in Eq.~(\ref{eq:dimensionlessH}). Comparing with the case of $L_0$, varying $L$ renormalizes the model parameters and rescales the overall energy. Since the terms inside the square bracket of Eq.~(\ref{eq:renormalizedH}) retain the same structure as those on the right-hand side of Eq.~(\ref{eq:dimensionlessH}), the resulting band structure is formally identical but governed by the renormalized parameters. This analysis indicates that the optimal conditions for realizing an isolated topological flat band satisfy
\begin{equation}
\eta \tilde{\lambda} = \alpha_1, ~~~~\eta^2\tilde{V}_z = \alpha_2,
\label{eq:opcondition}
\end{equation}
where $\alpha_{1,2}$ are correlated parameters that correspond to the vertical and horizontal coordinates along the bright arc shown in Fig.~\ref{fig:figure3}(c). For a given material, $\lambda$ and $V_z$ are typically fixed. Equation~(\ref{eq:opcondition}) thus highlights a key advantage that the effective parameters $\tilde{\lambda}$ and $\tilde{V}_z$ can be tuned by varying the superlattice period, offering a practical knob for achieving the optimal flat-band condition. 

In addition to band width, the band gap $\Delta$ separating the flat band from other bands is crucial for stabilizing FCIs. 
Equation~(\ref{eq:renormalizedH}) implies that $\Delta$ scales with the $E_0/\eta^2$. Substituting Eq.~(\ref{eq:opcondition}) and the definitions of $\tilde{\lambda}$ and $\tilde{V}_z$ into this energy yields $2m^*\lambda^2 /\hbar^2 \alpha_1^2 $ and  $V_z/\alpha_2$, respectively. These results suggest that materials with large $m^*$, $\lambda$, and $V_z$ are favorable for designing flat band with large $\Delta$.
Based on these relations, Table~\ref{table1} summarizes candidate Rashba materials and the corresponding engineered conditions for achieving topological flat bands in our proposed setup. 
The optimal condition corresponds to the arc in Fig.~\ref{fig:figure3}(c), providing flexibility in the choice of engineering parameters, as listed in Table~\ref{table1} and further discussed in the SM \cite{SM2024}.

\begin{table*}[t]
\caption{\label{table1}
Candidate Rashba materials for realizing topological flat bands. The SOC strength $\lambda$ and electron effective mass $m^*$ are taken from Ref.~\cite{bordoloi2024promises}, while the superlattice period $L$, poential depth $U_0$, Zeeman field $V_z$, and band gap $\Delta$ are estimated using Eq.~(\ref{eq:opcondition}) by varying $(\alpha_{1},\alpha_{2})$ from $(0.35,0.08)$ to $(0.65,0.40)$ along the optimal flat-band condition (arc) shown in Fig.~\ref{fig:figure3}(c).
}
\begin{ruledtabular}
\begin{tabular}{c c c |  c c c c}
    Materials & $\lambda$ (eV$\cdot$\AA) & $m^*/m_e$ & $L$ (nm) & $U_0$ (meV)& $V_z$ (meV) & $\Delta$ (meV) \\
\hline
    %h-TaN & 4.23 & 0.063 & 7.5 -- 13.9 & 72.6 -- 21.1  &193 -- 277 & 24 -- 3.6 \\
    %h-NbN & 2.90 & 0.094 & 7.3 -- 13.6 & 51.1 -- 15.7 &136 -- 194 & 17 -- 2.5 \\
    AlBi  & 2.80 & 0.043 & 16.7 -- 31.0 & 21.6 -- 6.2 &57.5 -- 82.3 & 7.2 -- 1.1 \\
    PbSi  & 2.70 & 0.019 & 39.4 -- 73.1 &8.7 -- 2.5 &23.5 -- 33.7 & 2.9 -- 0.4 \\
    BiSb  & 2.30 & 0.037 & 23.2 -- 43.1 & 12.9 -- 3.7 &34.0 -- 48.6 & 4.2 -- 0.6 \\
    PbBi  & 1.60 & 0.119 & 10.5 -- 19.5 & 19.6 -- 5.7 &52.2 -- 74.8 & 6.5 -- 1.0 \\
    GaTe  & 1.00 & 0.229 & 8.8 -- 16.3 & 14.7 -- 4.2 &39.2 -- 56.1 & 4.9 -- 0.7 \\
    BiTeI & 1.97 & 0.157 & 6.5 -- 12.0 & 39.2 -- 11.4  &104 -- 150 & 13 -- 1.9 \\
    SbTeI & 1.39 & 0.134 & 10.7 -- 19.9 &16.8 -- 4.9 &44.4 -- 63.6 & 5.5 -- 0.8 \\
    %TiS$_2$Se & 1.08 & 0.522 & 3.6 -- 6.6 & 39.1 -- 11.3 &104 -- 150 & 13 -- 1.93 \\
    %WSeTe & 0.92 & 0.936 & 2.3 -- 4.3 &51.1 -- 14.8 &136 -- 194 & 17 -- 2.5 \\
    SnSTe & 0.76 & 0.186 & 14.3 -- 26.5 & 6.8 -- 2.0 &18.1 -- 26.0 & 2.3 -- 0.3 \\
    %ZrS$_2$Se & 0.72 & 0.563 & 5.0 -- 9.2 &18.6 -- 5.4 & 49.6 -- 71 & 6.2 -- 0.9 \\
\end{tabular}
\end{ruledtabular}
\label{Table1}
\end{table*}

\textit{Discussion---}We have generalized the design of topological flat bands to 2D Rashba materials subjected to electrostatic superlattice potentials. A previous study has shown that such superlattice potentials can induce $\text{Z}_{2}$ band topology in Rashba systems \cite{yang2025engineering}, but the formation of isolated flat bands is precluded by time-reversal symmetry. In this work, the Zeeman field plays a dual role by generating a nearly flat segment in the lower Rashba band and endowing it with Berry curvature through gapping the Dirac cones of the folded Rashba bands. The superlattice potential then isolates this flat segment, yielding a flat band with $\mathcal{C}=1$. We further demonstrate that this flat band supports robust FCIs at filling factors $n=1/3$ and $2/3$. 

 To assess the robustness of our proposal, we examine the effects of several perturbations, including potential fluctuations, magnetic inhomogeneity, and higher-order Fourier components of the superlattice potential. As detailed in the SM \cite{SM2024}, mild perturbations have only minor quantitative effects on the isolation and flatness of the 1st band. Although the FCI excitation gaps are moderately reduced, the characteristic threefold ground-state degeneracy at both 1/3 and 2/3 fillings remains robust. Furthermore, incorporating metallic-gate screening into the Coulomb potential suppresses the excitation gaps while leaving the threefold ground-state degeneracy intact \cite{SM2024}. These results demonstrate that the proposed flat Chern band and the associated FCIs are robust under realistic experimental conditions.

%By introducing a Zeeman field that lifts Kramers degeneracy, our scheme produces an isolated flat band with $\mathcal{C}=1$, arising from the subtle interplay between the superlattice potential and Zeeman coupling.

%\textcolor{blue}{The Zeeman field plays a dual role by generating a nearly flat segment in the lower Rashba band and endowing it with nontrivial topology through gapping the Dirac points of the folded Rashba bands, while the superlattice potential trancates this segment through band folding and Bragg hybridization to form an isolated flat Chern band.}

Experimental feasibility of our proposal can be further assessed from several perspectives. First, high-quality 2D Rashba materials can be epitaxially grown or mechanically exfoliated with ultrahigh electron mobility using established experimental techniques \cite{joyce1974growth,thompson1985secondary,sberveglieri1995recent},
making them well suited for exploring correlation-driven topological phases.
Second, recent advances in electrostatic superlattice engineering have enabled patterned dielectric layers (e.g., $\text{SiO}_2$) with potential periods of tens of nanometers \cite{forsythe2018band,sun2024signature, wang2024dispersion, barcons2022engineering} and modulation amplitudes $U_0$ ranging from a few meV \cite{forsythe2018band,Wang:2026aa} to over 10 meV \cite{wang2024dispersion}. 
Both the characteristic periods and amplitudes are comparable to those required for the candidate materials listed in Table~\ref{table1}. 
 Third, the proposed sandwich structure closely resembles heterostructures designed for realizing topological superconductors \cite{PhysRevLett.105.077001,PhysRevLett.104.040502,potter2011majorana,nakosai2012topological,klinovaja2012transition,manchon2015new,volpez2019second, bihlmayer2022rashba},
where proximity-induced Zeeman splittings ranging from a few to several tens of meV have been experimentally demonstrated
\cite{tokuyasu1988proximity, manna2014two,zhou2018proximity,huang2020emergent,tang2020magnetic,liu2021magnetic,hu2024tunable}.
 Although achieving Zeeman fields exceeding 100 meV solely via magnetic proximity remains challenging, magnetic doping provides a complementary route \cite{Chen:2010aa,Checkelsky:2012aa,Chang:2013aa,Qin:2014aa,Qin:2016aa}. For example, V-doped BiTeI exhibits a Zeeman splitting of $\sim$90 meV \cite{Ishizaka:2011aa,Crepaldi:2012aa,Landolt:2012aa,Klimovskikh:2017aa}, while Cr-doped dichalcogenide halides generate exchange fields of several tens of meV \cite{Hou:2022aa}. 
  Furthermore, combining magnetic proximity with magnetic doping may provide an effective strategy for achieving even larger Zeeman fields \cite{Li:2015aa,Liu:2015aa,Grutter:2021aa}.

Compared with moir\'e systems \cite{spanton2018observation,xie2021fractional,park2023observation,cai2023signatures,xu2023observation,redekop2024direct,ji2024local,lu2024fractional,chen2024tunable,lu2025extended,choi2025superconductivity}, where the band structure is fixed once the sample is fabricated, 
the superlattice potential in our proposal can be tuned \textit{in situ} by externally applied electrostatic fields \cite{suri2023superlattice,ghorashi2023topological,wang2024dispersion,tan2024designing}.
The geometry and period of the superlattice as well as the electronic degrees of freedom such as spin, valley, and sublattice can in principle be deliberately designed or dynamically adjusted. These capabilities open a multidimensional parameter space for stabilizing FCIs and may enable the realization of the exotic non-Abelian states.

%In summary, the Rashba-based superlattice systems provide a programmable and tunable platform for realizing isolated topological flat bands that can harbor FCIs and potentially other correlated phases.

\textit{Acknowledgement---}We thank Prof. Chong Wang and Prof. Mikhail M. Otrokov for helpful discussions. This work is supported by National Natural Science Foundation of China (Grants No.12474134 and Grant No. 12488101) and the Innovation Program for Quantum Science and Technology (Grant No.2021ZD0302800).

\bibliographystyle{apsrev4-1-title}
\bibliography{ref}

@article{Grutter:2021aa,
	Author = {Grutter, A. J. and He, Q. L.},
	Doi = {10.1103/PhysRevMaterials.5.090301},
	Issue = {9},
	Journal = {Phys. Rev. Mater.},
	Month = {Sep},
	Numpages = {21},
	Pages = {090301},
	Publisher = {American Physical Society},
	Title = {Magnetic proximity effects in topological insulator heterostructures: Implementation and characterization},
	Url = {https://link.aps.org/doi/10.1103/PhysRevMaterials.5.090301},
	Volume = {5},
	Year = {2021}}

@article{Liu:2015aa,
	Author = {Liu, Wenqing and He, Liang and Xu, Yongbing and Murata, Koichi and Onbasli, Mehmet C. and Lang, Murong and Maltby, Nick J. and Li, Shunpu and Wang, Xuefeng and Ross, Caroline A. and Bencok, Peter and van der Laan, Gerrit and Zhang, Rong and Wang, Kang. L.},
	Booktitle = {Nano Letters},
	Da = {2015/01/14},
	Date = {2015/01/14},
	Doi = {10.1021/nl504480g},
	Isbn = {1530-6984},
	Journal = {Nano Letters},
	Journal1 = {Nano Lett.},
	M3 = {doi: 10.1021/nl504480g},
	Month = {01},
	Number = {1},
	Pages = {764--769},
	Publisher = {American Chemical Society},
	Title = {Enhancing Magnetic Ordering in Cr-Doped Bi2Se3 Using High-TC Ferrimagnetic Insulator},
	Ty = {JOUR},
	Url = {https://doi.org/10.1021/nl504480g},
	Volume = {15},
	Year = {2015},
	Year1 = {2015}}

@article{Li:2015aa,
	Author = {Li, Mingda and Chang, Cui-Zu and Kirby, Brian. J. and Jamer, Michelle E. and Cui, Wenping and Wu, Lijun and Wei, Peng and Zhu, Yimei and Heiman, Don and Li, Ju and Moodera, Jagadeesh S.},
	Doi = {10.1103/PhysRevLett.115.087201},
	Issue = {8},
	Journal = {Phys. Rev. Lett.},
	Month = {Aug},
	Numpages = {5},
	Pages = {087201},
	Publisher = {American Physical Society},
	Title = {Proximity-Driven Enhanced Magnetic Order at Ferromagnetic-Insulator--Magnetic-Topological-Insulator Interface},
	Url = {https://link.aps.org/doi/10.1103/PhysRevLett.115.087201},
	Volume = {115},
	Year = {2015}}

@article{Wang:2026aa,
	Author = {Wang, Daisy Q. and Krix, Zeb and Tkachenko, Olga A. and Tkachenko, Vitaly A. and Chen, Chong and Farrer, Ian and Ritchie, David A. and Sushkov, Oleg P. and Hamilton, Alexander R. and Klochan, Oleh},
	Da = {2026/05/11},
	Doi = {10.1038/s41567-026-03291-7},
	Id = {Wang2026},
	Isbn = {1745-2481},
	Journal = {Nature Physics},
	Title = {Correlated insulator in the kagome flat band of a two-dimensional electrostatic crystal},
	Ty = {JOUR},
	Url = {https://doi.org/10.1038/s41567-026-03291-7},
	Year = {2026}}

@article{Hou:2022aa,
	Author = {Hou, Yusheng and Xue, Feng and Qiu, Liang and Wang, Zhe and Wu, Ruqian},
	Da = {2022/05/26},
	Doi = {10.1038/s41524-022-00802-x},
	Id = {Hou2022},
	Isbn = {2057-3960},
	Journal = {npj Computational Materials},
	Number = {1},
	Pages = {120},
	Title = {Multifunctional two-dimensional van der Waals Janus magnet Cr-based dichalcogenide halides},
	Ty = {JOUR},
	Url = {https://doi.org/10.1038/s41524-022-00802-x},
	Volume = {8},
	Year = {2022}}

@article{Klimovskikh:2017aa,
	Author = {Klimovskikh, I. I. and Shikin, A. M. and Otrokov, M. M. and Ernst, A. and Rusinov, I. P. and Tereshchenko, O. E. and Golyashov, V. A. and S{\'a}nchez-Barriga, J. and Varykhalov, A. Yu. and Rader, O. and Kokh, K. A. and Chulkov, E. V.},
	Da = {2017/06/13},
	Doi = {10.1038/s41598-017-03507-0},
	Id = {Klimovskikh2017},
	Isbn = {2045-2322},
	Journal = {Scientific Reports},
	Number = {1},
	Pages = {3353},
	Title = {Giant Magnetic Band Gap in the Rashba-Split Surface State of Vanadium-Doped BiTeI: A Combined Photoemission and Ab Initio Study},
	Ty = {JOUR},
	Url = {https://doi.org/10.1038/s41598-017-03507-0},
	Volume = {7},
	Year = {2017}}

@article{Crepaldi:2012aa,
	Author = {Crepaldi, A. and Moreschini, L. and Aut\`es, G. and Tournier-Colletta, C. and Moser, S. and Virk, N. and Berger, H. and Bugnon, Ph. and Chang, Y. J. and Kern, K. and Bostwick, A. and Rotenberg, E. and Yazyev, O. V. and Grioni, M.},
	Doi = {10.1103/PhysRevLett.109.096803},
	Issue = {9},
	Journal = {Phys. Rev. Lett.},
	Month = {Aug},
	Numpages = {5},
	Pages = {096803},
	Publisher = {American Physical Society},
	Title = {Giant Ambipolar Rashba Effect in the Semiconductor BiTeI},
	Url = {https://link.aps.org/doi/10.1103/PhysRevLett.109.096803},
	Volume = {109},
	Year = {2012}}

@article{Landolt:2012aa,
	Author = {Landolt, Gabriel and Eremeev, Sergey V. and Koroteev, Yury M. and Slomski, Bartosz and Muff, Stefan and Neupert, Titus and Kobayashi, Masaki and Strocov, Vladimir N. and Schmitt, Thorsten and Aliev, Ziya S. and Babanly, Mahammad B. and Amiraslanov, Imamaddin R. and Chulkov, Evgueni V. and Osterwalder, J\"urg and Dil, J. Hugo},
	Doi = {10.1103/PhysRevLett.109.116403},
	Issue = {11},
	Journal = {Phys. Rev. Lett.},
	Month = {Sep},
	Numpages = {5},
	Pages = {116403},
	Publisher = {American Physical Society},
	Title = {Disentanglement of Surface and Bulk Rashba Spin Splittings in Noncentrosymmetric BiTeI},
	Url = {https://link.aps.org/doi/10.1103/PhysRevLett.109.116403},
	Volume = {109},
	Year = {2012}}

@article{Ishizaka:2011aa,
	Author = {Ishizaka, K. and Bahramy, M. S. and Murakawa, H. and Sakano, M. and Shimojima, T. and Sonobe, T. and Koizumi, K. and Shin, S. and Miyahara, H. and Kimura, A. and Miyamoto, K. and Okuda, T. and Namatame, H. and Taniguchi, M. and Arita, R. and Nagaosa, N. and Kobayashi, K. and Murakami, Y. and Kumai, R. and Kaneko, Y. and Onose, Y. and Tokura, Y.},
	Da = {2011/07/01},
	Doi = {10.1038/nmat3051},
	Id = {Ishizaka2011},
	Isbn = {1476-4660},
	Journal = {Nature Materials},
	Number = {7},
	Pages = {521--526},
	Title = {Giant Rashba-type spin splitting in bulk BiTeI},
	Ty = {JOUR},
	Url = {https://doi.org/10.1038/nmat3051},
	Volume = {10},
	Year = {2011}}

@article{Qin:2016aa,
	Author = {Qin, Wei and Xiao, Di and Chang, Kai and Shen, Shun-Qing and Zhang, Zhenyu},
	Da = {2016/12/19},
	Doi = {10.1038/srep39188},
	Id = {Qin2016},
	Isbn = {2045-2322},
	Journal = {Scientific Reports},
	Number = {1},
	Pages = {39188},
	Title = {Disorder-induced topological phase transitions in two-dimensional spin-orbit coupled superconductors},
	Ty = {JOUR},
	Url = {https://doi.org/10.1038/srep39188},
	Volume = {6},
	Year = {2016}}

@article{Qin:2014aa,
	Author = {Qin, Wei and Zhang, Zhenyu},
	Doi = {10.1103/PhysRevLett.113.266806},
	Issue = {26},
	Journal = {Phys. Rev. Lett.},
	Month = {Dec},
	Numpages = {5},
	Pages = {266806},
	Publisher = {American Physical Society},
	Title = {Persistent Ferromagnetism and Topological Phase Transition at the Interface of a Superconductor and a Topological Insulator},
	Url = {https://link.aps.org/doi/10.1103/PhysRevLett.113.266806},
	Volume = {113},
	Year = {2014}}

@article{Checkelsky:2012aa,
	Author = {Checkelsky, Joseph G. and Ye, Jianting and Onose, Yoshinori and Iwasa, Yoshihiro and Tokura, Yoshinori},
	Da = {2012/10/01},
	Doi = {10.1038/nphys2388},
	Id = {Checkelsky2012},
	Isbn = {1745-2481},
	Journal = {Nature Physics},
	Number = {10},
	Pages = {729--733},
	Title = {Dirac-fermion-mediated ferromagnetism in a topological insulator},
	Ty = {JOUR},
	Url = {https://doi.org/10.1038/nphys2388},
	Volume = {8},
	Year = {2012}}

@article{Chen:2010aa,
	Author = {Y. L. Chen and J.-H. Chu and J. G. Analytis and Z. K. Liu and K. Igarashi and H.-H. Kuo and X. L. Qi and S. K. Mo and R. G. Moore and D. H. Lu and M. Hashimoto and T. Sasagawa and S. C. Zhang and I. R. Fisher and Z. Hussain and Z. X. Shen},
	Doi = {10.1126/science.1189924},	
	Journal = {Science},
	Number = {5992},
	Pages = {659-662},
	Title = {Massive Dirac Fermion on the Surface of a Magnetically Doped Topological Insulator},
	Url = {https://www.science.org/doi/abs/10.1126/science.1189924},
	Volume = {329},
	Year = {2010}}

@article{Chang:2013aa,
	Author = {Cui-Zu Chang and Jinsong Zhang and Xiao Feng and Jie Shen and Zuocheng Zhang and Minghua Guo and Kang Li and Yunbo Ou and Pang Wei and Li-Li Wang and Zhong-Qing Ji and Yang Feng and Shuaihua Ji and Xi Chen and Jinfeng Jia and Xi Dai and Zhong Fang and Shou-Cheng Zhang and Ke He and Yayu Wang and Li Lu and Xu-Cun Ma and Qi-Kun Xue},
	Doi = {10.1126/science.1234414},
	Journal = {Science},
	Number = {6129},
	Pages = {167-170},
	Title = {Experimental Observation of the Quantum Anomalous Hall Effect in a Magnetic Topological Insulator},
	Url = {https://www.science.org/doi/abs/10.1126/science.1234414},
	Volume = {340},
	Year = {2013}}

@misc{SM2024,
	Title = {See Supplemental Material at xxxx},
	Year = {2020}}

@article{yang2025engineering,
	Author = {Yang, Kaijie and Liu, Yunzhe and Schindler, Frank and Liu, Chao-Xing},
	Doi = {10.1103/PhysRevB.111.L241104},
	Issue = {24},
	Journal = {Phys. Rev. B},
	Month = {Jun},
	Numpages = {8},
	Pages = {L241104},
	Publisher = {American Physical Society},
	Title = {Engineering miniband topology via band folding in moir\'e superlattice materials},
	Url = {https://link.aps.org/doi/10.1103/PhysRevB.111.L241104},
	Volume = {111},
	Year = {2025}}

@article{laughlin1983anomalous,
	Author = {Laughlin, R. B.},
	Doi = {10.1103/PhysRevLett.50.1395},
	Issue = {18},
	Journal = {Phys. Rev. Lett.},
	Month = {May},
	Numpages = {0},
	Pages = {1395--1398},
	Publisher = {American Physical Society},
	Title = {Anomalous Quantum Hall Effect: An Incompressible Quantum Fluid with Fractionally Charged Excitations},
	Url = {https://link.aps.org/doi/10.1103/PhysRevLett.50.1395},
	Volume = {50},
	Year = {1983}}

@article{girvin1986magneto,
	Author = {Girvin, S. M. and MacDonald, A. H. and Platzman, P. M.},
	Doi = {10.1103/PhysRevB.33.2481},
	Issue = {4},
	Journal = {Phys. Rev. B},
	Month = {Feb},
	Numpages = {0},
	Pages = {2481--2494},
	Publisher = {American Physical Society},
	Title = {Magneto-roton theory of collective excitations in the fractional quantum Hall effect},
	Url = {https://link.aps.org/doi/10.1103/PhysRevB.33.2481},
	Volume = {33},
	Year = {1986}}

@inbook{liu2022recent,
	Author = {Liu, Zhao and Bergholtz, Emil J.},
	Booktitle = {Encyclopedia of Condensed Matter Physics},
	Doi = {10.1016/b978-0-323-90800-9.00136-0},
	Isbn = {9780323914086},
	Pages = {515--538},
	Publisher = {Elsevier},
	Title = {Recent developments in fractional Chern insulators},
	Url = {http://dx.doi.org/10.1016/B978-0-323-90800-9.00136-0},
	Year = {2024}}

@article{stormer1999fractional,
	Author = {Stormer, Horst L. and Tsui, Daniel C. and Gossard, Arthur C.},
	Doi = {10.1103/RevModPhys.71.S298},
	Issue = {2},
	Journal = {Rev. Mod. Phys.},
	Month = {Mar},
	Numpages = {0},
	Pages = {S298--S305},
	Publisher = {American Physical Society},
	Title = {The fractional quantum Hall effect},
	Url = {https://link.aps.org/doi/10.1103/RevModPhys.71.S298},
	Volume = {71},
	Year = {1999}}

@article{stormer1999nobel,
	Author = {Stormer, Horst L.},
	Doi = {10.1103/RevModPhys.71.875},
	Issue = {4},
	Journal = {Rev. Mod. Phys.},
	Month = {Jul},
	Numpages = {0},
	Pages = {875--889},
	Publisher = {American Physical Society},
	Title = {Nobel Lecture: The fractional quantum Hall effect},
	Url = {https://link.aps.org/doi/10.1103/RevModPhys.71.875},
	Volume = {71},
	Year = {1999}}

@article{jain1990theory,
	Author = {Jain, J. K.},
	Doi = {10.1103/PhysRevB.41.7653},
	Issue = {11},
	Journal = {Phys. Rev. B},
	Month = {Apr},
	Numpages = {0},
	Pages = {7653--7665},
	Publisher = {American Physical Society},
	Title = {Theory of the fractional quantum Hall effect},
	Url = {https://link.aps.org/doi/10.1103/PhysRevB.41.7653},
	Volume = {41},
	Year = {1990}}

@article{moore1991nonabelions,
	Author = {Gregory Moore and Nicholas Read},
	Doi = {https://doi.org/10.1016/0550-3213(91)90407-O},
	Issn = {0550-3213},
	Journal = {Nuclear Physics B},
	Number = {2},
	Pages = {362-396},
	Title = {Nonabelions in the fractional quantum hall effect},
	Url = {https://www.sciencedirect.com/science/article/pii/055032139190407O},
	Volume = {360},
	Year = {1991}}

@article{regnault2011fractional,
	Author = {Regnault, N. and Bernevig, B. Andrei},
	Doi = {10.1103/PhysRevX.1.021014},
	Issue = {2},
	Journal = {Phys. Rev. X},
	Month = {Dec},
	Numpages = {14},
	Pages = {021014},
	Publisher = {American Physical Society},
	Title = {Fractional Chern Insulator},
	Url = {https://link.aps.org/doi/10.1103/PhysRevX.1.021014},
	Volume = {1},
	Year = {2011}}

@article{parameswaran2012fractional,
	Author = {Parameswaran, S. A. and Roy, R. and Sondhi, S. L.},
	Doi = {10.1103/PhysRevB.85.241308},
	Issue = {24},
	Journal = {Phys. Rev. B},
	Month = {Jun},
	Numpages = {4},
	Pages = {241308},
	Publisher = {American Physical Society},
	Title = {Fractional Chern insulators and the ${W}_{\ensuremath{\infty}}$ algebra},
	Url = {https://link.aps.org/doi/10.1103/PhysRevB.85.241308},
	Volume = {85},
	Year = {2012}}

@article{bergholtz2013topological,
	Author = {Bergholtz, Emil J and Liu, Zhao},
	Journal = {Int. J. Mod. Phys. B},
	Number = {24},
	Pages = {1330017},
	Publisher = {World Scientific},
	Title = {Topological flat band models and fractional Chern insulators},
	Url = {https://doi.org/10.1142/S021797921330017X},
	Volume = {27},
	Year = {2013}}

@article{roy2014band,
	Author = {Roy, Rahul},
	Doi = {10.1103/PhysRevB.90.165139},
	Issue = {16},
	Journal = {Phys. Rev. B},
	Month = {Oct},
	Numpages = {7},
	Pages = {165139},
	Publisher = {American Physical Society},
	Title = {Band geometry of fractional topological insulators},
	Url = {https://link.aps.org/doi/10.1103/PhysRevB.90.165139},
	Volume = {90},
	Year = {2014}}

@article{park2023observation,
	Author = {Park, Heonjoon and Cai, Jiaqi and Anderson, Eric and Zhang, Yinong and Zhu, Jiayi and Liu, Xiaoyu and Wang, Chong and Holtzmann, William and Hu, Chaowei and Liu, Zhaoyu and others},
	Journal = {Nature},
	Number = {7981},
	Pages = {74--79},
	Publisher = {Nature Publishing Group UK London},
	Title = {Observation of fractionally quantized anomalous Hall effect},
	Url = {https://doi.org/10.1038/s41586-023-06536-0},
	Volume = {622},
	Year = {2023}}

@article{cai2023signatures,
	Author = {Cai, Jiaqi and Anderson, Eric and Wang, Chong and Zhang, Xiaowei and Liu, Xiaoyu and Holtzmann, William and Zhang, Yinong and Fan, Fengren and Taniguchi, Takashi and Watanabe, Kenji and others},
	Journal = {Nature},
	Number = {7981},
	Pages = {63--68},
	Publisher = {Nature Publishing Group UK London},
	Title = {Signatures of fractional quantum anomalous Hall states in twisted MoTe2},
	Url = {https://doi.org/10.1038/s41586-023-06289-w},
	Volume = {622},
	Year = {2023}}

@article{xu2023observation,
	Author = {Xu, Fan and Sun, Zheng and Jia, Tongtong and Liu, Chang and Xu, Cheng and Li, Chushan and Gu, Yu and Watanabe, Kenji and Taniguchi, Takashi and Tong, Bingbing and Jia, Jinfeng and Shi, Zhiwen and Jiang, Shengwei and Zhang, Yang and Liu, Xiaoxue and Li, Tingxin},
	Doi = {10.1103/PhysRevX.13.031037},
	Issue = {3},
	Journal = {Phys. Rev. X},
	Month = {Sep},
	Numpages = {12},
	Pages = {031037},
	Publisher = {American Physical Society},
	Title = {Observation of Integer and Fractional Quantum Anomalous Hall Effects in Twisted Bilayer ${\mathrm{MoTe}}_{2}$},
	Url = {https://link.aps.org/doi/10.1103/PhysRevX.13.031037},
	Volume = {13},
	Year = {2023}}

@article{lu2024fractional,
	Author = {Lu, Zhengguang and Han, Tonghang and Yao, Yuxuan and Reddy, Aidan P and Yang, Jixiang and Seo, Junseok and Watanabe, Kenji and Taniguchi, Takashi and Fu, Liang and Ju, Long},
	Journal = {Nature},
	Number = {8000},
	Pages = {759--764},
	Publisher = {Nature Publishing Group UK London},
	Title = {Fractional quantum anomalous Hall effect in multilayer graphene},
	Url = {https://doi.org/10.1038/s41586-023-07010-7},
	Volume = {626},
	Year = {2024}}

@article{dong2024theory,
	Author = {Dong, Zhihuan and Patri, Adarsh S and Senthil, Todadri},
	Journal = {Physical Review Letters},
	Number = {20},
	Pages = {206502},
	Publisher = {APS},
	Title = {Theory of quantum anomalous Hall phases in pentalayer rhombohedral graphene moir{\'e} structures},
	Url = {https://link.aps.org/doi/10.1103/PhysRevLett.133.206502},
	Volume = {133},
	Year = {2024}}

@article{dong2024anomalous,
	Author = {Dong, Junkai and Wang, Taige and Wang, Tianle and Soejima, Tomohiro and Zaletel, Michael P. and Vishwanath, Ashvin and Parker, Daniel E.},
	Doi = {10.1103/PhysRevLett.133.206503},
	Issue = {20},
	Journal = {Phys. Rev. Lett.},
	Month = {Nov},
	Numpages = {11},
	Pages = {206503},
	Publisher = {American Physical Society},
	Title = {Anomalous Hall Crystals in Rhombohedral Multilayer Graphene. I. Interaction-Driven Chern Bands and Fractional Quantum Hall States at Zero Magnetic Field},
	Url = {https://link.aps.org/doi/10.1103/PhysRevLett.133.206503},
	Volume = {133},
	Year = {2024}}

@article{reddy2023fractional,
	Author = {Reddy, Aidan P. and Alsallom, Faisal and Zhang, Yang and Devakul, Trithep and Fu, Liang},
	Doi = {10.1103/PhysRevB.108.085117},
	Issue = {8},
	Journal = {Phys. Rev. B},
	Month = {Aug},
	Numpages = {10},
	Pages = {085117},
	Publisher = {American Physical Society},
	Title = {Fractional quantum anomalous Hall states in twisted bilayer ${\mathrm{MoTe}}_{2}$ and ${\mathrm{WSe}}_{2}$},
	Url = {https://link.aps.org/doi/10.1103/PhysRevB.108.085117},
	Volume = {108},
	Year = {2023}}

@article{crepel2023anomalous,
	Author = {Cr\'epel, Valentin and Fu, Liang},
	Doi = {10.1103/PhysRevB.107.L201109},
	Issue = {20},
	Journal = {Phys. Rev. B},
	Month = {May},
	Numpages = {5},
	Pages = {L201109},
	Publisher = {American Physical Society},
	Title = {Anomalous Hall metal and fractional Chern insulator in twisted transition metal dichalcogenides},
	Url = {https://link.aps.org/doi/10.1103/PhysRevB.107.L201109},
	Volume = {107},
	Year = {2023}}

@article{yu2024fractional,
	Author = {Yu, Jiabin and Herzog-Arbeitman, Jonah and Wang, Minxuan and Vafek, Oskar and Bernevig, B. Andrei and Regnault, Nicolas},
	Doi = {10.1103/PhysRevB.109.045147},
	Issue = {4},
	Journal = {Phys. Rev. B},
	Month = {Jan},
	Numpages = {37},
	Pages = {045147},
	Publisher = {American Physical Society},
	Title = {Fractional Chern insulators versus nonmagnetic states in twisted bilayer ${\mathrm{MoTe}}_{2}$},
	Url = {https://link.aps.org/doi/10.1103/PhysRevB.109.045147},
	Volume = {109},
	Year = {2024}}

@article{li2021spontaneous,
	Author = {Li, Heqiu and Kumar, Umesh and Sun, Kai and Lin, Shi-Zeng},
	Doi = {10.1103/PhysRevResearch.3.L032070},
	Issue = {3},
	Journal = {Phys. Rev. Res.},
	Month = {Sep},
	Numpages = {6},
	Pages = {L032070},
	Publisher = {American Physical Society},
	Title = {Spontaneous fractional Chern insulators in transition metal dichalcogenide moir\'e superlattices},
	Url = {https://link.aps.org/doi/10.1103/PhysRevResearch.3.L032070},
	Volume = {3},
	Year = {2021}}

@article{ghorashi2023topological,
	Author = {Ghorashi, Sayed Ali Akbar and Dunbrack, Aaron and Abouelkomsan, Ahmed and Sun, Jiacheng and Du, Xu and Cano, Jennifer},
	Doi = {10.1103/PhysRevLett.130.196201},
	Issue = {19},
	Journal = {Phys. Rev. Lett.},
	Month = {May},
	Numpages = {8},
	Pages = {196201},
	Publisher = {American Physical Society},
	Title = {Topological and Stacked Flat Bands in Bilayer Graphene with a Superlattice Potential},
	Url = {https://link.aps.org/doi/10.1103/PhysRevLett.130.196201},
	Volume = {130},
	Year = {2023}}

@article{wang2024dispersion,
	Author = {Wang, Shuai and Zhan, Zhen and Fan, Xiaodong and Li, Yonggang and Pantale\'on, Pierre A. and Ye, Chaochao and He, Zhiping and Wei, Laiming and Li, Lin and Guinea, Francisco and Yuan, Shengjun and Zeng, Changgan},
	Doi = {10.1103/PhysRevLett.133.066302},
	Issue = {6},
	Journal = {Phys. Rev. Lett.},
	Month = {Aug},
	Numpages = {7},
	Pages = {066302},
	Publisher = {American Physical Society},
	Title = {Dispersion-Selective Band Engineering in an Artificial Kagome Superlattice},
	Url = {https://link.aps.org/doi/10.1103/PhysRevLett.133.066302},
	Volume = {133},
	Year = {2024}}

@article{suri2023superlattice,
	Author = {Suri, Nishchay and Wang, Chong and Hunt, Benjamin M. and Xiao, Di},
	Doi = {10.1103/PhysRevB.108.155409},
	Issue = {15},
	Journal = {Phys. Rev. B},
	Month = {Oct},
	Numpages = {7},
	Pages = {155409},
	Publisher = {American Physical Society},
	Title = {Superlattice engineering of topology in massive Dirac fermions},
	Url = {https://link.aps.org/doi/10.1103/PhysRevB.108.155409},
	Volume = {108},
	Year = {2023}}

@article{forsythe2018band,
	Author = {Forsythe, Carlos and Zhou, Xiaodong and Watanabe, Kenji and Taniguchi, Takashi and Pasupathy, Abhay and Moon, Pilkyung and Koshino, Mikito and Kim, Philip and Dean, Cory R},
	Journal = {Nat. Nano.},
	Number = {7},
	Pages = {566--571},
	Publisher = {Nature Publishing Group UK London},
	Title = {Band structure engineering of 2D materials using patterned dielectric superlattices},
	Url = {https://doi.org/10.1038/s41565-018-0138-7},
	Volume = {13},
	Year = {2018}}

@article{tan2024designing,
	Author = {Tan, Tixuan and Reddy, Aidan P. and Fu, Liang and Devakul, Trithep},
	Doi = {10.1103/PhysRevLett.133.206601},
	Issue = {20},
	Journal = {Phys. Rev. Lett.},
	Month = {Nov},
	Numpages = {8},
	Pages = {206601},
	Publisher = {American Physical Society},
	Title = {Designing Topology and Fractionalization in Narrow Gap Semiconductor Films via Electrostatic Engineering},
	Url = {https://link.aps.org/doi/10.1103/PhysRevLett.133.206601},
	Volume = {133},
	Year = {2024}}

@article{bihlmayer2022rashba,
	Author = {Bihlmayer, Gustav and No{\"e}l, Paul and Vyalikh, Denis V and Chulkov, Evgueni V and Manchon, Aur{\'e}lien},
	Journal = {Nat. Rev. Phys.},
	Number = {10},
	Pages = {642--659},
	Publisher = {Nature Publishing Group UK London},
	Title = {Rashba-like physics in condensed matter},
	Url = {https://doi.org/10.1038/s42254-022-00490-y},
	Volume = {4},
	Year = {2022}}

@article{manchon2015new,
	Author = {Manchon, Aurelien and Koo, Hyun Cheol and Nitta, Junsaku and Frolov, Sergey M and Duine, Rembert A},
	Journal = {Nature Mater.},
	Number = {9},
	Pages = {871--882},
	Publisher = {Nature Publishing Group UK London},
	Title = {New perspectives for Rashba spin--orbit coupling},
	Url = {https://doi.org/10.1038/nmat4360},
	Volume = {14},
	Year = {2015}}

@article{xiao2010berry,
	Author = {Xiao, Di and Chang, Ming-Che and Niu, Qian},
	Doi = {10.1103/RevModPhys.82.1959},
	Issue = {3},
	Journal = {Rev. Mod. Phys.},
	Month = {Jul},
	Numpages = {0},
	Pages = {1959--2007},
	Publisher = {American Physical Society},
	Title = {Berry phase effects on electronic properties},
	Url = {https://link.aps.org/doi/10.1103/RevModPhys.82.1959},
	Volume = {82},
	Year = {2010}}

@article{nakosai2012topological,
	Author = {Nakosai, Sho and Tanaka, Yukio and Nagaosa, Naoto},
	Doi = {10.1103/PhysRevLett.108.147003},
	Issue = {14},
	Journal = {Phys. Rev. Lett.},
	Month = {Apr},
	Numpages = {5},
	Pages = {147003},
	Publisher = {American Physical Society},
	Title = {Topological Superconductivity in Bilayer Rashba System},
	Url = {https://link.aps.org/doi/10.1103/PhysRevLett.108.147003},
	Volume = {108},
	Year = {2012}}

@article{volpez2019second,
	Author = {Volpez, Yanick and Loss, Daniel and Klinovaja, Jelena},
	Doi = {10.1103/PhysRevLett.122.126402},
	Issue = {12},
	Journal = {Phys. Rev. Lett.},
	Month = {Mar},
	Numpages = {6},
	Pages = {126402},
	Publisher = {American Physical Society},
	Title = {Second-Order Topological Superconductivity in $\ensuremath{\pi}$-Junction Rashba Layers},
	Url = {https://link.aps.org/doi/10.1103/PhysRevLett.122.126402},
	Volume = {122},
	Year = {2019}}

@article{klinovaja2012transition,
	Author = {Klinovaja, Jelena and Stano, Peter and Loss, Daniel},
	Doi = {10.1103/PhysRevLett.109.236801},
	Issue = {23},
	Journal = {Phys. Rev. Lett.},
	Month = {Dec},
	Numpages = {5},
	Pages = {236801},
	Publisher = {American Physical Society},
	Title = {Transition from Fractional to Majorana Fermions in Rashba Nanowires},
	Url = {https://link.aps.org/doi/10.1103/PhysRevLett.109.236801},
	Volume = {109},
	Year = {2012}}

@article{potter2011majorana,
	Author = {Potter, Andrew C. and Lee, Patrick A.},
	Doi = {10.1103/PhysRevB.83.094525},
	Issue = {9},
	Journal = {Phys. Rev. B},
	Month = {Mar},
	Numpages = {8},
	Pages = {094525},
	Publisher = {American Physical Society},
	Title = {Majorana end states in multiband microstructures with Rashba spin-orbit coupling},
	Url = {https://link.aps.org/doi/10.1103/PhysRevB.83.094525},
	Volume = {83},
	Year = {2011}}

@article{huang2020emergent,
	Author = {Huang, Bevin and McGuire, Michael A and May, Andrew F and Xiao, Di and Jarillo-Herrero, Pablo and Xu, Xiaodong},
	Journal = {Nature Mater.},
	Number = {12},
	Pages = {1276--1289},
	Publisher = {Nature Publishing Group UK London},
	Title = {Emergent phenomena and proximity effects in two-dimensional magnets and heterostructures},
	Url = {https://doi.org/10.1038/s41563-020-0791-8},
	Volume = {19},
	Year = {2020}}

@article{wu2019topological,
	Author = {Wu, Fengcheng and Lovorn, Timothy and Tutuc, Emanuel and Martin, Ivar and MacDonald, A. H.},
	Doi = {10.1103/PhysRevLett.122.086402},
	Issue = {8},
	Journal = {Phys. Rev. Lett.},
	Month = {Feb},
	Numpages = {5},
	Pages = {086402},
	Publisher = {American Physical Society},
	Title = {Topological Insulators in Twisted Transition Metal Dichalcogenide Homobilayers},
	Url = {https://link.aps.org/doi/10.1103/PhysRevLett.122.086402},
	Volume = {122},
	Year = {2019}}

@article{song2019all,
	Author = {Song, Zhida and Wang, Zhijun and Shi, Wujun and Li, Gang and Fang, Chen and Bernevig, B. Andrei},
	Doi = {10.1103/PhysRevLett.123.036401},
	Issue = {3},
	Journal = {Phys. Rev. Lett.},
	Month = {Jul},
	Numpages = {6},
	Pages = {036401},
	Publisher = {American Physical Society},
	Title = {All Magic Angles in Twisted Bilayer Graphene are Topological},
	Url = {https://link.aps.org/doi/10.1103/PhysRevLett.123.036401},
	Volume = {123},
	Year = {2019}}

@article{bistritzer2011moire,
	Author = {Rafi Bistritzer and Allan H. MacDonald},
	Doi = {10.1073/pnas.1108174108},
	Journal = {Proc. Natl. Acad. Sci. U.S.A.},
	Number = {30},
	Pages = {12233-12237},
	Title = {Moir{\'e} bands in twisted double-layer graphene},
	Url = {https://www.pnas.org/doi/abs/10.1073/pnas.1108174108},
	Volume = {108},
	Year = {2011}}

@article{mak2022semiconductor,
	Author = {Mak, Kin Fai and Shan, Jie},
	Journal = {Nat. Nanotechnol.},
	Number = {7},
	Pages = {686--695},
	Publisher = {Nature Publishing Group UK London},
	Title = {Semiconductor moir{\'e} materials},
	Url = {https://doi.org/10.1038/s41565-022-01165-6},
	Volume = {17},
	Year = {2022}}

@article{bordoloi2024promises,
	Author = {Bordoloi, Arjyama and Garcia-Castro, AC and Romestan, Zachary and Romero, Aldo H and Singh, Sobhit},
	Journal = {J. Appl. Phys.},
	Number = {22},
	Publisher = {AIP Publishing},
	Title = {Promises and technological prospects of two-dimensional Rashba materials},
	Url = {https://doi.org/10.1063/5.0212170},
	Volume = {135},
	Year = {2024}}

@article{wu2018hubbard,
	Author = {Wu, Fengcheng and Lovorn, Timothy and Tutuc, Emanuel and MacDonald, A. H.},
	Doi = {10.1103/PhysRevLett.121.026402},
	Issue = {2},
	Journal = {Phys. Rev. Lett.},
	Month = {Jul},
	Numpages = {5},
	Pages = {026402},
	Publisher = {American Physical Society},
	Title = {Hubbard Model Physics in Transition Metal Dichalcogenide Moir\'e Bands},
	Url = {https://link.aps.org/doi/10.1103/PhysRevLett.121.026402},
	Volume = {121},
	Year = {2018}}

@article{sun2024signature,
	Author = {Sun, Jiacheng and Akbar Ghorashi, Sayed Ali and Watanabe, Kenji and Taniguchi, Takashi and Camino, Fernando and Cano, Jennifer and Du, Xu},
	Journal = {Nano Let.},
	Number = {43},
	Pages = {13600--13606},
	Publisher = {ACS Publications},
	Title = {Signature of correlated insulator in electric field controlled superlattice},
	Url = {https://doi.org/10.1021/acs.nanolett.4c03238},
	Volume = {24},
	Year = {2024}}

@article{barcons2022engineering,
	Author = {Barcons Ruiz, David and Herzig Sheinfux, Hanan and Hoffmann, Rebecca and Torre, Iacopo and Agarwal, Hitesh and Kumar, Roshan Krishna and Vistoli, Lorenzo and Taniguchi, Takashi and Watanabe, Kenji and Bachtold, Adrian and others},
	Journal = {Nat. Commun.},
	Number = {1},
	Pages = {6926},
	Publisher = {Nature Publishing Group UK London},
	Title = {Engineering high quality graphene superlattices via ion milled ultra-thin etching masks},
	Url = {https://doi.org/10.1038/s41467-022-34734-3},
	Volume = {13},
	Year = {2022}}

@article{tokuyasu1988proximity,
	Author = {Tokuyasu, T. and Sauls, J. A. and Rainer, D.},
	Doi = {10.1103/PhysRevB.38.8823},
	Issue = {13},
	Journal = {Phys. Rev. B},
	Month = {Nov},
	Numpages = {0},
	Pages = {8823--8833},
	Publisher = {American Physical Society},
	Title = {Proximity effect of a ferromagnetic insulator in contact with a superconductor},
	Url = {https://link.aps.org/doi/10.1103/PhysRevB.38.8823},
	Volume = {38},
	Year = {1988}}

@article{manna2014two,
	Author = {P.K. Manna and S.M. Yusuf},
	Doi = {https://doi.org/10.1016/j.physrep.2013.10.002},
	Issn = {0370-1573},
	Journal = {Phys. Rep.},
	Number = {2},
	Pages = {61-99},
	Title = {Two interface effects: Exchange bias and magnetic proximity},
	Url = {https://www.sciencedirect.com/science/article/pii/S0370157313003773},
	Volume = {535},
	Year = {2014}}

@article{lau2022reproducibility,
	Author = {Lau, Chun Ning and Bockrath, Marc W and Mak, Kin Fai and Zhang, Fan},
	Journal = {Nature},
	Number = {7895},
	Pages = {41--50},
	Publisher = {Nature Publishing Group UK London},
	Title = {Reproducibility in the fabrication and physics of moir{\'e} materials},
	Url = {https://doi.org/10.1038/s41586-021-04173-z},
	Volume = {602},
	Year = {2022}}

@article{nakatsuji2022moire,
	Author = {Nakatsuji, Naoto and Koshino, Mikito},
	Doi = {10.1103/PhysRevB.105.245408},
	Issue = {24},
	Journal = {Phys. Rev. B},
	Month = {Jun},
	Numpages = {11},
	Pages = {245408},
	Publisher = {American Physical Society},
	Title = {Moir\'e disorder effect in twisted bilayer graphene},
	Url = {https://link.aps.org/doi/10.1103/PhysRevB.105.245408},
	Volume = {105},
	Year = {2022}}

@article{kennes2021moire,
	Author = {Kennes, Dante M and Claassen, Martin and Xian, Lede and Georges, Antoine and Millis, Andrew J and Hone, James and Dean, Cory R and Basov, DN and Pasupathy, Abhay N and Rubio, Angel},
	Journal = {Nat. Phys.},
	Number = {2},
	Pages = {155--163},
	Publisher = {Nature Publishing Group UK London},
	Title = {Moir{\'e} heterostructures as a condensed-matter quantum simulator},
	Url = {https://doi.org/10.1038/s41567-020-01154-3},
	Volume = {17},
	Year = {2021}}

@article{song2021direct,
	Author = {Tiancheng Song and Qi-Chao Sun and Eric Anderson and Chong Wang and Jimin Qian and Takashi Taniguchi and Kenji Watanabe and Michael A. McGuire and Rainer St{\"o}hr and Di Xiao and Ting Cao and J{\"o}rg Wrachtrup and Xiaodong Xu},
	Doi = {10.1126/science.abj7478},
	Journal = {Science},
	Number = {6571},
	Pages = {1140-1144},
	Title = {Direct visualization of magnetic domains and moir{\'e} magnetism in twisted 2D magnets},
	Url = {https://www.science.org/doi/abs/10.1126/science.abj7478},
	Volume = {374},
	Year = {2021}}

@article{he2021moire,
	Author = {He, Feng and Zhou, Yongjian and Ye, Zefang and Cho, Sang-Hyeok and Jeong, Jihoon and Meng, Xianghai and Wang, Yaguo},
	Doi = {10.1021/acsnano.0c10435},
	Journal = {ACS Nano},
	Number = {4},
	Pages = {5944-5958},
	Title = {Moir{\'e} Patterns in 2D Materials: A Review},
	Url = {https://doi.org/10.1021/acsnano.0c10435},
	Volume = {15},
	Year = {2021}}

@article{tarnopolsky2019origin,
	Author = {Tarnopolsky, Grigory and Kruchkov, Alex Jura and Vishwanath, Ashvin},
	Doi = {10.1103/PhysRevLett.122.106405},
	Issue = {10},
	Journal = {Phys. Rev. Lett.},
	Month = {Mar},
	Numpages = {6},
	Pages = {106405},
	Publisher = {American Physical Society},
	Title = {Origin of Magic Angles in Twisted Bilayer Graphene},
	Url = {https://link.aps.org/doi/10.1103/PhysRevLett.122.106405},
	Volume = {122},
	Year = {2019}}

@article{andrei2021marvels,
	Author = {Andrei, Eva Y and Efetov, Dmitri K and Jarillo-Herrero, Pablo and MacDonald, Allan H and Mak, Kin Fai and Senthil, T and Tutuc, Emanuel and Yazdani, Ali and Young, Andrea F},
	Journal = {Nat. Rev. Mater.},
	Number = {3},
	Pages = {201--206},
	Publisher = {Nature Publishing Group UK London},
	Title = {The marvels of moir{\'e} materials},
	Url = {https://doi.org/10.1038/s41578-021-00284-1},
	Volume = {6},
	Year = {2021}}

@article{tang2011high,
	Author = {Tang, Evelyn and Mei, Jia-Wei and Wen, Xiao-Gang},
	Doi = {10.1103/PhysRevLett.106.236802},
	Issue = {23},
	Journal = {Phys. Rev. Lett.},
	Month = {Jun},
	Numpages = {4},
	Pages = {236802},
	Publisher = {American Physical Society},
	Title = {High-Temperature Fractional Quantum Hall States},
	Url = {https://link.aps.org/doi/10.1103/PhysRevLett.106.236802},
	Volume = {106},
	Year = {2011}}

@article{spitzer1957determination,
	Author = {Spitzer, W. G. and Fan, H. Y.},
	Doi = {10.1103/PhysRev.106.882},
	Issue = {5},
	Journal = {Phys. Rev.},
	Month = {Jun},
	Numpages = {0},
	Pages = {882--890},
	Publisher = {American Physical Society},
	Title = {Determination of Optical Constants and Carrier Effective Mass of Semiconductors},
	Url = {https://link.aps.org/doi/10.1103/PhysRev.106.882},
	Volume = {106},
	Year = {1957}}

@article{xie2021fractional,
	Author = {Xie, Yonglong and Pierce, Andrew T and Park, Jeong Min and Parker, Daniel E and Khalaf, Eslam and Ledwith, Patrick and Cao, Yuan and Lee, Seung Hwan and Chen, Shaowen and Forrester, Patrick R and others},
	Journal = {Nature},
	Number = {7889},
	Pages = {439--443},
	Publisher = {Nature Publishing Group UK London},
	Title = {Fractional Chern insulators in magic-angle twisted bilayer graphene},
	Url = {https://doi.org/10.1038/s41586-021-04002-3},
	Volume = {600},
	Year = {2021}}

@article{spanton2018observation,
	Author = {Spanton, Eric M and Zibrov, Alexander A and Zhou, Haoxin and Taniguchi, Takashi and Watanabe, Kenji and Zaletel, Michael P and Young, Andrea F},
	Journal = {Science},
	Number = {6384},
	Pages = {62--66},
	Publisher = {American Association for the Advancement of Science},
	Title = {Observation of fractional Chern insulators in a van der Waals heterostructure},
	Url = {https://www.science.org/doi/abs/10.1126/science.aan8458},
	Volume = {360},
	Year = {2018}}

@article{redekop2024direct,
	Author = {Redekop, Evgeny and Zhang, Canxun and Park, Heonjoon and Cai, Jiaqi and Anderson, Eric and Sheekey, Owen and Arp, Trevor and Babikyan, Grigory and Salters, Samuel and Watanabe, Kenji and others},
	Journal = {Nature},
	Number = {8039},
	Pages = {584--589},
	Publisher = {Nature Publishing Group UK London},
	Title = {Direct magnetic imaging of fractional Chern insulators in twisted MoTe2},
	Url = {https://doi.org/10.1038/s41586-024-08153-x},
	Volume = {635},
	Year = {2024}}

@article{ji2024local,
	Author = {Ji, Zhurun and Park, Heonjoon and Barber, Mark E and Hu, Chaowei and Watanabe, Kenji and Taniguchi, Takashi and Chu, Jiun-Haw and Xu, Xiaodong and Shen, Zhi-Xun},
	Journal = {Nature},
	Number = {8039},
	Pages = {578--583},
	Publisher = {Nature Publishing Group UK London},
	Title = {Local probe of bulk and edge states in a fractional Chern insulator},
	Url = {https://doi.org/10.1038/s41586-024-08092-7},
	Volume = {635},
	Year = {2024}}

@article{lu2025extended,
	Author = {Lu, Zhengguang and Han, Tonghang and Yao, Yuxuan and Hadjri, Zach and Yang, Jixiang and Seo, Junseok and Shi, Lihan and Ye, Shenyong and Watanabe, Kenji and Taniguchi, Takashi and others},
	Journal = {Nature},
	Number = {8048},
	Pages = {1090--1095},
	Publisher = {Nature Publishing Group UK London},
	Title = {Extended quantum anomalous Hall states in graphene/hBN moir{\'e} superlattices},
	Url = {https://doi.org/10.1038/s41586-024-08470-1},
	Volume = {637},
	Year = {2025}}

@article{chen2024tunable,
	Author = {Chen, Yiwei and Huang, Yan and Li, Qingxin and Tong, Bingbing and Kuang, Guangli and Xi, Chuanying and Watanabe, Kenji and Taniguchi, Takashi and Liu, Guangtong and Zhu, Zheng and others},
	Journal = {Nat. Commun.},
	Number = {1},
	Pages = {6236},
	Publisher = {Nature Publishing Group UK London},
	Title = {Tunable even-and odd-denominator fractional quantum Hall states in trilayer graphene},
	Url = {https://doi.org/10.1038/s41467-024-50589-2},
	Volume = {15},
	Year = {2024}}

@article{choi2025superconductivity,
	Author = {Choi, Youngjoon and Choi, Ysun and Valentini, Marco and Patterson, Caitlin L and Holleis, Ludwig FW and Sheekey, Owen I and Stoyanov, Hari and Cheng, Xiang and Taniguchi, Takashi and Watanabe, Kenji and others},
	Journal = {Nature},
	Number = {8048},
	Pages = {342},
	Publisher = {Nature Publishing Group UK London},
	Title = {Superconductivity and quantized anomalous Hall effect in rhombohedral graphene},
	Url = {https://doi.org/10.1038/s41586-025-08621-y},
	Volume = {639},
	Year = {2025}}

@article{huang2024impurity,
	Author = {Huang, Ke and Sarma, Sankar Das and Li, Xiao},
	Journal = {arXiv preprint arXiv:2409.04349},
	Title = {Impurity-induced thermal crossover in fractional Chern insulators},
	Url = {https://arxiv.org/abs/2409.04349},
	Year = {2024}}

@article{das2024thermal,
	Author = {Das Sarma, Sankar and Xie, Ming},
	Doi = {10.1103/PhysRevB.110.155148},
	Issue = {15},
	Journal = {Phys. Rev. B},
	Month = {Oct},
	Numpages = {10},
	Pages = {155148},
	Publisher = {American Physical Society},
	Title = {Thermal crossover from a Chern insulator to a fractional Chern insulator in pentalayer graphene},
	Url = {https://link.aps.org/doi/10.1103/PhysRevB.110.155148},
	Volume = {110},
	Year = {2024}}

@article{joyce1974growth,
	Author = {Joyce, BA},
	Journal = {Rep. Prog. Phys.},
	Number = {3},
	Pages = {363},
	Publisher = {IOP Publishing},
	Title = {The growth and structure of semiconducting thin films},
	Url = {https://dx.doi.org/10.1088/0034-4885/37/3/002},
	Volume = {37},
	Year = {1974}}

@article{thompson1985secondary,
	Author = {Thompson, CV},
	Journal = {J. Appl. Phys.},
	Number = {2},
	Pages = {763--772},
	Publisher = {American Institute of Physics},
	Title = {Secondary grain growth in thin films of semiconductors: Theoretical aspects},
	Url = {https://doi.org/10.1063/1.336194},
	Volume = {58},
	Year = {1985}}

@article{sberveglieri1995recent,
	Author = {G Sberveglieri},
	Doi = {https://doi.org/10.1016/0925-4005(94)01278-P},
	Issn = {0925-4005},
	Journal = {Sensors and Actuators B: Chemical},
	Number = {2},
	Pages = {103-109},
	Title = {Recent developments in semiconducting thin-film gas sensors},
	Url = {https://www.sciencedirect.com/science/article/pii/092540059401278P},
	Volume = {23},
	Year = {1995}}

@article{PhysRevLett.105.077001,
	Author = {Lutchyn, Roman M. and Sau, Jay D. and Das Sarma, S.},
	Doi = {10.1103/PhysRevLett.105.077001},
	Issue = {7},
	Journal = {Phys. Rev. Lett.},
	Month = {Aug},
	Numpages = {4},
	Pages = {077001},
	Publisher = {American Physical Society},
	Title = {Majorana Fermions and a Topological Phase Transition in Semiconductor-Superconductor Heterostructures},
	Url = {https://link.aps.org/doi/10.1103/PhysRevLett.105.077001},
	Volume = {105},
	Year = {2010}}

@article{PhysRevLett.104.040502,
	Author = {Sau, Jay D. and Lutchyn, Roman M. and Tewari, Sumanta and Das Sarma, S.},
	Doi = {10.1103/PhysRevLett.104.040502},
	Issue = {4},
	Journal = {Phys. Rev. Lett.},
	Month = {Jan},
	Numpages = {4},
	Pages = {040502},
	Publisher = {American Physical Society},
	Title = {Generic New Platform for Topological Quantum Computation Using Semiconductor Heterostructures},
	Url = {https://link.aps.org/doi/10.1103/PhysRevLett.104.040502},
	Volume = {104},
	Year = {2010}}

@article{ghyq-sz16,
	Author = {Pantale\'on, Pierre A. and Zhan, Zhen and Morales, Siddhartha E. and Naumis, Gerardo G.},
	Doi = {10.1103/ghyq-sz16},
	Issue = {24},
	Journal = {Phys. Rev. B},
	Month = {Jun},
	Numpages = {14},
	Pages = {245303},
	Publisher = {American Physical Society},
	Title = {Designing flat bands and pseudo-Landau levels in GaAs with patterned gates},
	Url = {https://link.aps.org/doi/10.1103/ghyq-sz16},
	Volume = {111},
	Year = {2025}}

@article{PhysRevB.111.045148,
	Author = {Zhan, Zhen and Li, Yonggang and Pantale\'on, Pierre A.},
	Doi = {10.1103/PhysRevB.111.045148},
	Issue = {4},
	Journal = {Phys. Rev. B},
	Month = {Jan},
	Numpages = {14},
	Pages = {045148},
	Publisher = {American Physical Society},
	Title = {Designing band structures by patterned dielectric superlattices},
	Url = {https://link.aps.org/doi/10.1103/PhysRevB.111.045148},
	Volume = {111},
	Year = {2025}}

@article{zhou2024fractional,
	Author = {Zhou, Boran and Yang, Hui and Zhang, Ya-Hui},
	Doi = {10.1103/PhysRevLett.133.206504},
	Issue = {20},
	Journal = {Phys. Rev. Lett.},
	Month = {Nov},
	Numpages = {7},
	Pages = {206504},
	Publisher = {American Physical Society},
	Title = {Fractional Quantum Anomalous Hall Effect in Rhombohedral Multilayer Graphene in the Moir\'eless Limit},
	Url = {https://link.aps.org/doi/10.1103/PhysRevLett.133.206504},
	Volume = {133},
	Year = {2024}}

@article{hu2024tunable,
	Author = {Hu, Junxiong and Han, Yulei and Chi, Xiao and Omar, Ganesh Ji and Al Ezzi, Mohammed Mohammed Esmail and Gou, Jian and Yu, Xiaojiang and Andrivo, Rusydi and Watanabe, Kenji and Taniguchi, Takashi and others},
	Journal = {Advanced Materials},
	Number = {8},
	Pages = {2305763},
	Publisher = {Wiley Online Library},
	Title = {Tunable Spin-Polarized States in Graphene on a Ferrimagnetic Oxide Insulator},
	Url = {https://doi.org/10.1002/adma.202305763},
	Volume = {36},
	Year = {2024}}

@article{zhou2018proximity,
	Author = {Zhou, Baozeng and Ji, Shiwen and Tian, Zhen and Cheng, Weijia and Wang, Xiaocha and Mi, Wenbo},
	Journal = {Carbon},
	Pages = {25--31},
	Publisher = {Elsevier},
	Title = {Proximity effect induced spin filtering and gap opening in graphene by half-metallic monolayer Cr2C ferromagnet},
	Url = {https://doi.org/10.1016/j.carbon.2018.02.044},
	Volume = {132},
	Year = {2018}}

@article{tang2020magnetic,
	Author = {Tang, Chaolong and Zhang, Zhaowei and Lai, Shen and Tan, Qinghai and Gao, Wei-bo},
	Journal = {Advanced Materials},
	Number = {16},
	Pages = {1908498},
	Publisher = {Wiley Online Library},
	Title = {Magnetic proximity effect in graphene/CrBr3 van der Waals heterostructures},
	Url = {https://doi.org/10.1002/adma.201908498},
	Volume = {32},
	Year = {2020}}

@article{liu2021magnetic,
	Author = {Liu, Yuxin and Niu, Xuefan and Zhang, Rencong and Zhang, Qinghua and Teng, Jing and Li, Yongqing},
	Journal = {Chinese Physics Letters},
	Number = {5},
	Pages = {057303},
	Publisher = {IOP Publishing},
	Title = {Magnetic proximity effect in an antiferromagnetic insulator/topological insulator heterostructure with sharp interface},
	Url = {https://dx.doi.org/10.1088/0256-307X/38/5/057303},
	Volume = {38},
	Year = {2021}}

@article{reddy2023fractionalsup,
  title={Supplemental material of Fractional quantum anomalous Hall states in twisted bilayer MoTe 2 and WSe 2},
  author={Reddy, Aidan P and Alsallom, Faisal and Zhang, Yang and Devakul, Trithep and Fu, Liang},
  journal={Physical Review B},
  volume={108},
  number={8},
  pages={085117},
  year={2023},
  publisher={APS}
}

@article{reddy2024non,
  title={Supplemental material of Non-Abelian fractionalization in topological minibands},
  author={Reddy, Aidan P and Paul, Nisarga and Abouelkomsan, Ahmed and Fu, Liang},
  journal={Physical Review Letters},
  volume={133},
  number={16},
  pages={166503},
  year={2024},
  publisher={APS}
}

@article{laughlin1999nobel,
  title={Nobel lecture: Fractional quantization},
  author={Laughlin, Robert B},
  journal={Reviews of Modern Physics},
  volume={71},
  number={4},
  pages={863},
  year={1999},
  publisher={APS}
}

@article{simon2020topological,
  title={Topological Quantum: Lecture Notes and Proto-Book},
  author={Simon, Steven H},
  journal={Unpublished prototype.[online] Available at: http://www-thphys. physics. ox. ac. uk/people/SteveSimon},
  volume={26},
  pages={27--35},
  year={2020}
}

@article{lu2023synergistic,
  title={Synergistic correlated states and nontrivial topology in coupled graphene-insulator heterostructures},
  author={Lu, Xin and Zhang, Shihao and Wang, Yaning and Gao, Xiang and Yang, Kaining and Guo, Zhongqing and Gao, Yuchen and Ye, Yu and Han, Zheng and Liu, Jianpeng},
  journal={Nature Communications},
  volume={14},
  number={1},
  pages={5550},
  year={2023},
  publisher={Nature Publishing Group UK London}
}

@article{tseng2022gate,
  title={Gate-tunable proximity effects in graphene on layered magnetic insulators},
  author={Tseng, Chun-Chih and Song, Tiancheng and Jiang, Qianni and Lin, Zhong and Wang, Chong and Suh, Jaehyun and Watanabe, Kenji and Taniguchi, Takashi and McGuire, Michael A and Xiao, Di and others},
  journal={Nano Letters},
  volume={22},
  number={21},
  pages={8495--8501},
  year={2022},
  publisher={ACS Publications}
}

@article{yang2023unconventional,
  title={Unconventional correlated insulator in CrOCl-interfaced Bernal bilayer graphene},
  author={Yang, Kaining and Gao, Xiang and Wang, Yaning and Zhang, Tongyao and Gao, Yuchen and Lu, Xin and Zhang, Shihao and Liu, Jianpeng and Gu, Pingfan and Luo, Zhaoping and others},
  journal={Nature communications},
  volume={14},
  number={1},
  pages={2136},
  year={2023},
  publisher={Nature Publishing Group UK London}
}

@article{wang2022quantum,
  title={Quantum Hall phase in graphene engineered by interfacial charge coupling},
  author={Wang, Yaning and Gao, Xiang and Yang, Kaining and Gu, Pingfan and Lu, Xin and Zhang, Shihao and Gao, Yuchen and Ren, Naijie and Dong, Baojuan and Jiang, Yuhang and others},
  journal={Nature nanotechnology},
  volume={17},
  number={12},
  pages={1272--1279},
  year={2022},
  publisher={Nature Publishing Group UK London}
}

@article{wang2026correlated,
  title={Correlated insulator in the kagome flat band of a two-dimensional electrostatic crystal},
  author={Wang, Daisy Q and Krix, Zeb and Tkachenko, Olga A and Tkachenko, Vitaly A and Chen, Chong and Farrer, Ian and Ritchie, David A and Sushkov, Oleg P and Hamilton, Alexander R and Klochan, Oleh},
  journal={Nature Physics},
  pages={1--8},
  year={2026},
  publisher={Nature Publishing Group UK London}
}

\end{document}

\bibliographystyle{unsrt}
\title{Supplemental Material for "Topological Flat Minibands and Fractional Chern Insulators in Rashba Systems with Tunable Superlattice Potentials"}

\author{Bokai Liang}
\affiliation{Department of Physics, University of Science and Technology of China, Hefei, Anhui 230026,China }
\author{Wei Qin}
\email{qinwei5@ustc.edu.cn} 
\affiliation{Department of Physics, University of Science and Technology of China, Hefei, Anhui 230026,China }
\affiliation{International Center for Quantum Design of Functional Materials (ICQD), Hefei National Research Center for Physical Sciences at Microscale, University of Science and Technology of China, Hefei 230026, China}
\author{Zhenyu Zhang}
\affiliation{International Center for Quantum Design of Functional Materials (ICQD), Hefei National Research Center for Physical Sciences at Microscale, University of Science and Technology of China, Hefei 230026, China}
\affiliation{Hefei National Laboratory, University of Science and Technology of China, Hefei 230088, China}

\maketitle

\section{Evolution of the miniband structure}
\label{section1}
\begin{figure}
    \centering
    \includegraphics[width=0.99\columnwidth]{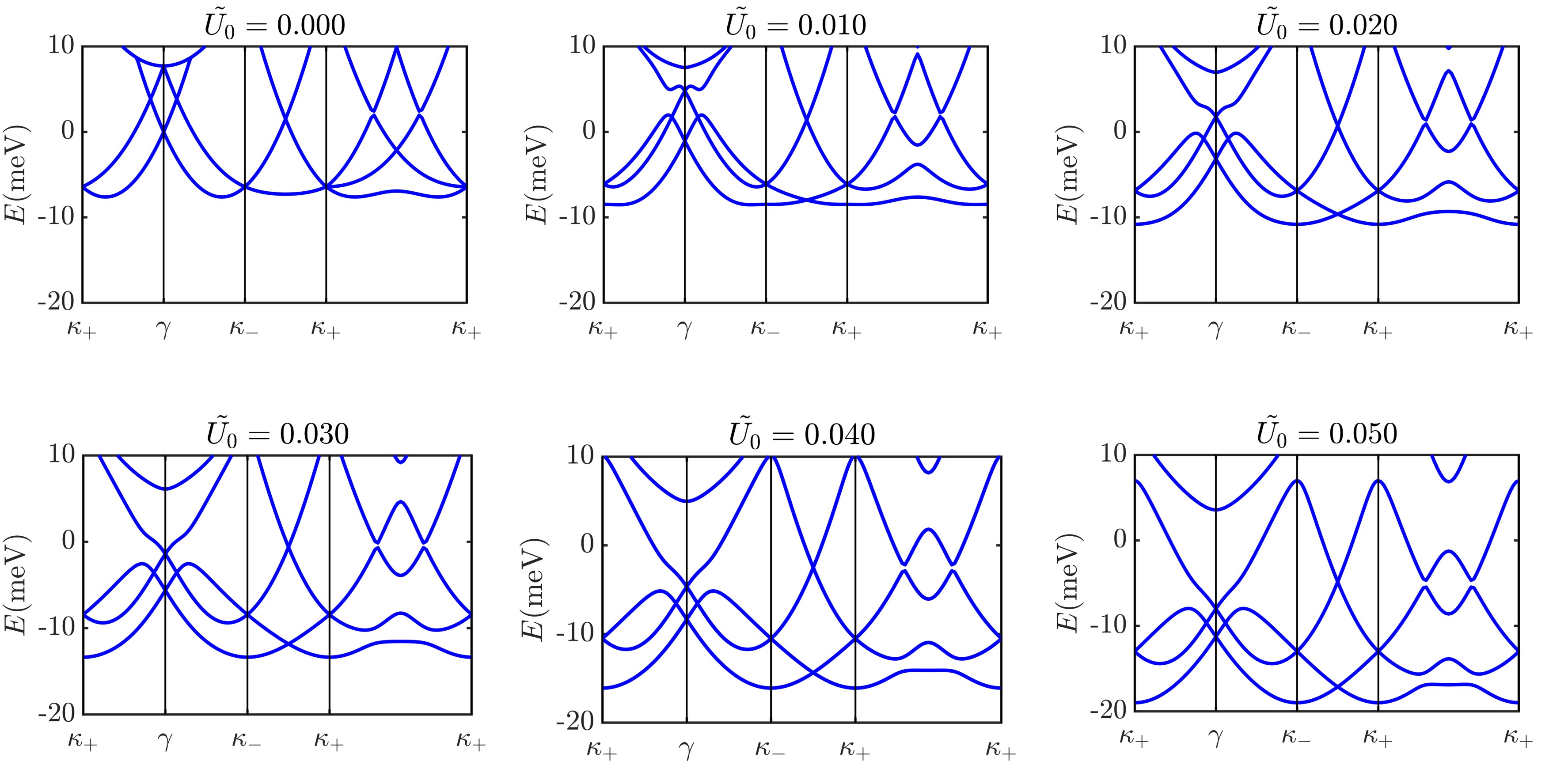}
    \caption{Evolution of the miniband structure with increasing superlattice potential $\tilde{U}_0$. These calculations are performed with Rashba SOC strength $\tilde{\lambda}=0.4$ and Zeeman coupling $\tilde{V}_z=0$.}
    \label{band1}	
\end{figure}

\begin{figure}
    \centering
    \includegraphics[width=0.666\columnwidth]{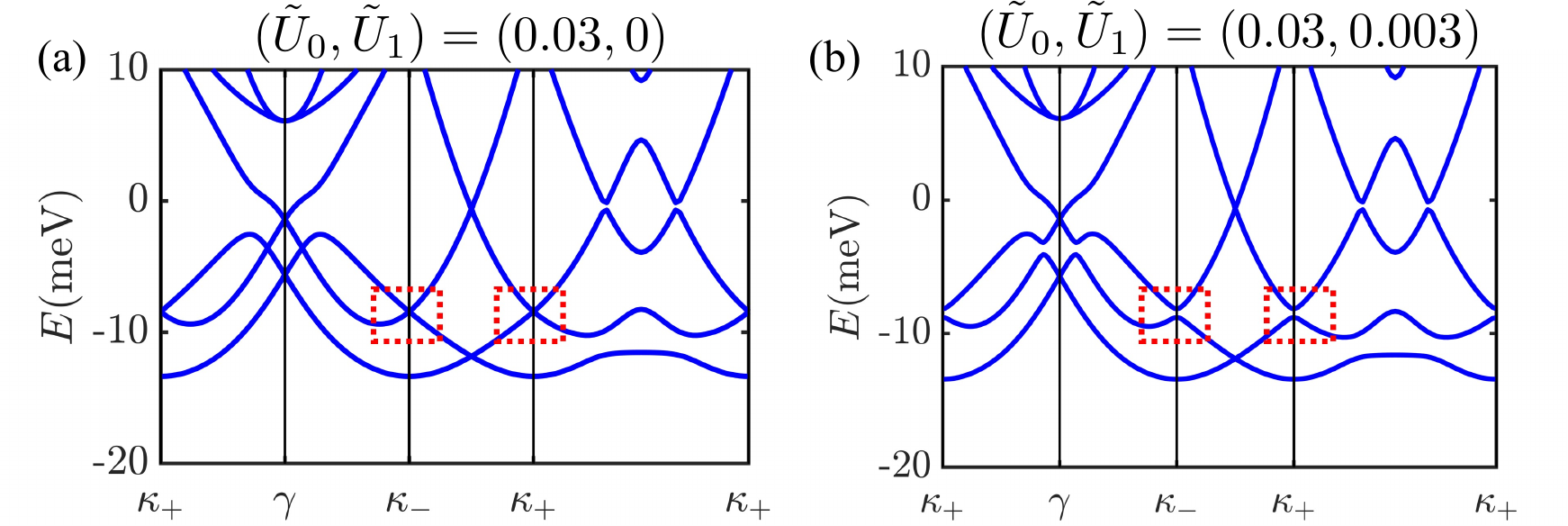}
    \caption{(a),(b) Miniband structures calculated (a) without and (b) with the additional superlattice potential $\tilde{U}_1$, which breaks in-plane inversion symmetry. Other parameters are chosen as $\tilde{U}_0 =0.03$, $\tilde{\lambda}=0.4$, and $\tilde{V}_z =0$  }
    \label{band4}	
\end{figure}

\begin{figure}
    \centering
    \includegraphics[width=0.99\columnwidth]{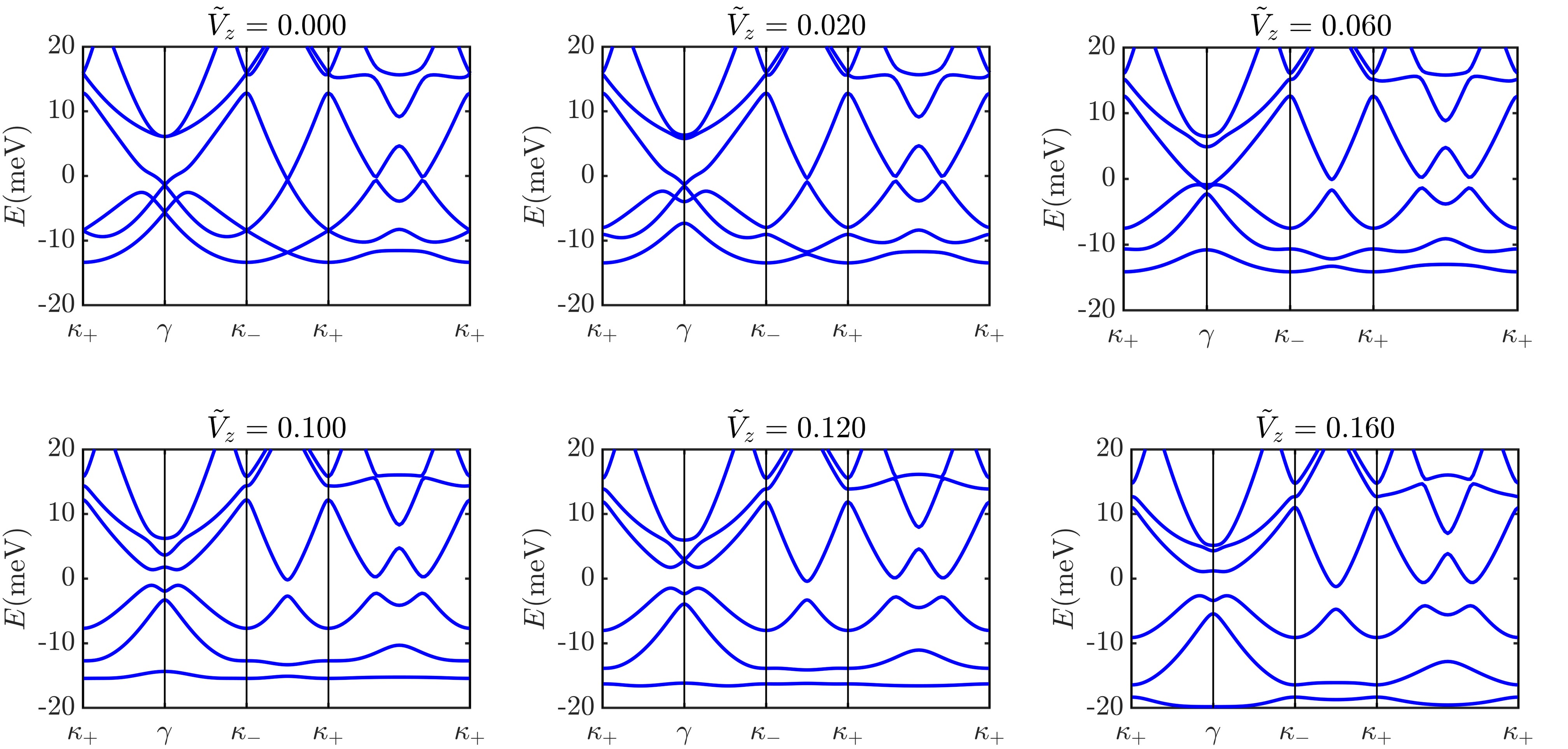}
    \caption{Evolution of the miniband structure with increasing Zeeman coupling $\tilde{U}_0$. These calculations are performed with Rashba SOC strength $\tilde{\lambda}=0.4$ and superlattice potential $\tilde{U}_0=0.03$.}
    \label{band3}	
\end{figure}

Here we provide additional details on the evolution of the miniband structure as model parameters are varied. Figure~\ref{band1} shows the dependence of the miniband on the superlattice potential ($\tilde{U}_0$) in the absence of Zeeman coupling ($\tilde{V}_z=0$), with Rashba spin-orbit coupling (SOC) strength fixed at $\tilde{\lambda}=0.4$. As discussed in the main text, a finite $\tilde{U}_0$ induces Bragg scatterings at reciprocal lattice vectors, lifting the band degeneracies caused by band folding along high-symmetry lines of the mini Brillouin zone (MBZ). Nevertheless, as shown in Fig.~\ref{band1}, certain degeneracies persist at high-symmetry points due to symmetry protection. Specifically, the degeneracies at time-reversal invariant points $\gamma$ and $m$ are protected by time-reversal symmetry ($\mathcal{T}$), whereas those at $\kappa_{\pm}$ are protected by the combined symmetry $\mathcal{T}\mathcal{I} $, where $\mathcal{I} $ denotes the in-plane inversion. This distinction can be verified by adding a perturbative superlattice potential to Eq.~(2) in the main text, yielding
\begin{equation}
U(\bm{r})=2 U_0 \sum_{n=0}^2 \cos \left(\bm{G}_n \cdot \bm{r}\right) + 2U_1 \sum_{n=0}^{2}\sin(\bm{G}_n \cdot \bm{r}),
\end{equation}
where the second term on the right hand side of the above equation explicitly breaks $\mathcal{I}$. As shown in Fig \ref{band4}, this perturbation lifts the band degeneracies at $\kappa_{\pm}$, while those at $\gamma$ and $m$ remain robust.  Finally, the inclusion of $\tilde{V}_z$ breaks both $\mathcal{T}$ and $\mathcal{T}\mathcal{I} $, thereby removing all symmetry-protected degeneracies at the high-symmetry points, as illustrated in Fig.~\ref{band3}.


%We see that gap is opened at the $\kappa_{+}$ and $\kappa_{-}$ point. While at $\gamma$ points and the $M$ point, the band remains degenerate, which could be attributed to the protection of the time reversal symmetry at $\gamma$ and $M$ points. 

%In Fig \ref{band2}, we set $\tilde{U}_0$ to be zero and change $\tilde{V}_z$. We see that $\tilde{V}_z$ opens gap at all points in BZ except the corner points $\kappa_+$ and $\kappa_-$. 

By focusing on the lowest-energy miniband (labeled as the 1st in Fig.~1(b) of the main text), we can gain qualitative understanding of the emergence of an isolated flat miniband. As illustrated in Fig \ref{band1}, the energies of the 1st miniband at the $\kappa_{\pm}$ points are primarily governed by the superlattice potential $\tilde{U}_0$, whereas those at the $\gamma$ and $m$ points are predominantly controlled by the Zeeman coupling $\tilde{V}_z$. For a given value of $\tilde{U}_0$, increasing $\tilde{V}_z$ lowers the band energies at $\gamma$ and $m$, thereby flattening the overall dispersion of the1st miniband. At an optimal value of $\tilde{V}_{z}$, the bandwidth of this miniband reaches a minimum, as illustrated in Fig \ref{band3} for $\tilde{V}_z = 0.12 $.

%\begin{figure}[t]
   % \centering
    %\includegraphics[width=0.8\columnwidth]{band2.jpg}
    %\caption{The effect of Zeenman splitting on the band structure. The Rashba coefficient $\tilde{\lambda}$ is fixed to be 0.4, the strength of the superlattice potential $\tilde{U}_0$ is fixed to be zero.}
    %\label{band2}	
%\end{figure}

%In the main text, we point out that the degeneracy at $\kappa_+$ and $\kappa_- $ points (as plotted in Fig \ref{band4}(a)) are protected by the time reversal symmetry and the Inversion symmetry. In Fig \ref{band3}, we have see that as we turn on the $\tilde{V}_z$ and breaks the time reversal symmetry, the degeneracy is destroyed. In Fig \ref{band4}(b), we add a potential term $V'(\bm{r})=2D \sum_{n=0}^{2}\sin(\bm{G}_n \cdot \bm{r})$, where $D=0.6meV$. This term breaks the inversion symmetry and we see that the degeneracy is also destroyed.

\section{The original of Flat miniband}
\label{section2}

\begin{figure}[t]
    \centering
    \includegraphics[width=0.75\columnwidth]{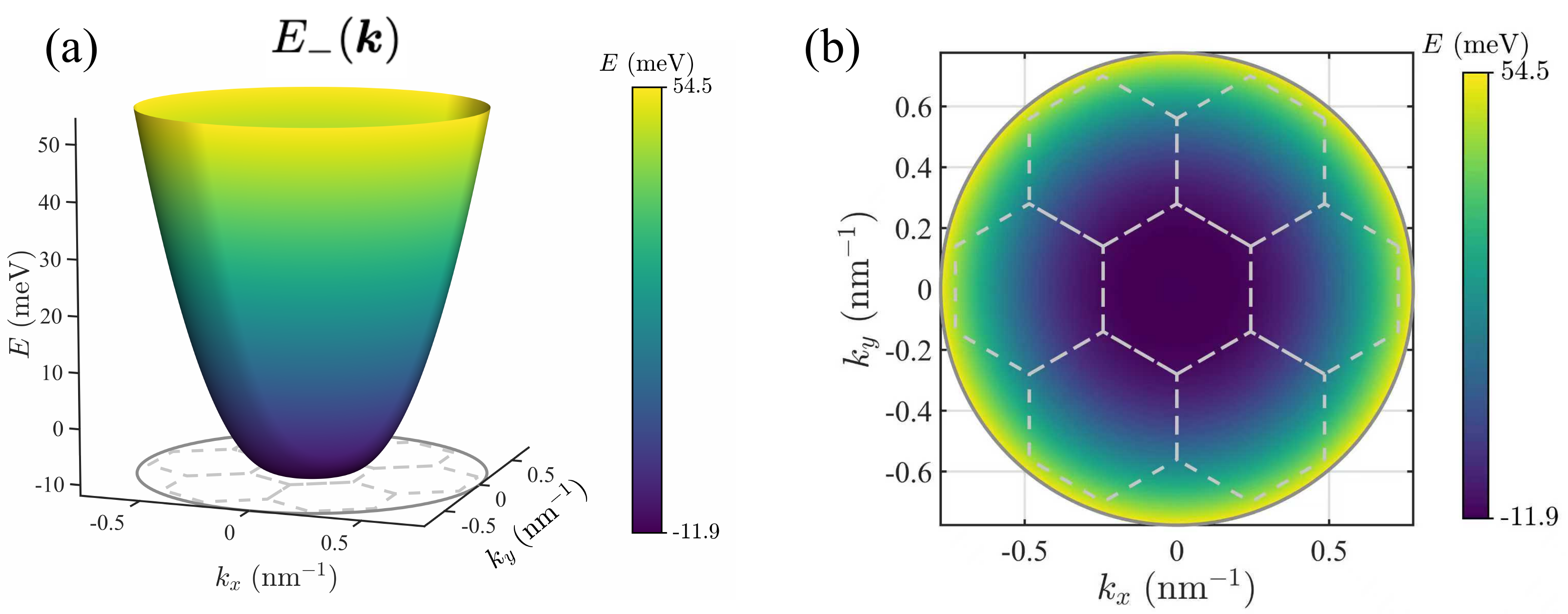}
    \caption{ (a) Three-dimensional view and (b) corresponding color map of the lower Rashba band $E_{-}(\bm{k})$, where a nearly flat segment develops near the band bottom. The dashed hexagons indicate the Brillouin zone (BZ) defined by the periodic superlattice potential, which truncates this flat segment within the BZ to form a flat band.
    These results are obtained by choosing Rashba SOC strength $\tilde{\lambda}=0.4$ and Zeeman coupling $\tilde{V}_z=0.12$.}	
    \label{flatsegment}
\end{figure}

\begin{figure}[t]
    \centering
    \includegraphics[width=\columnwidth]{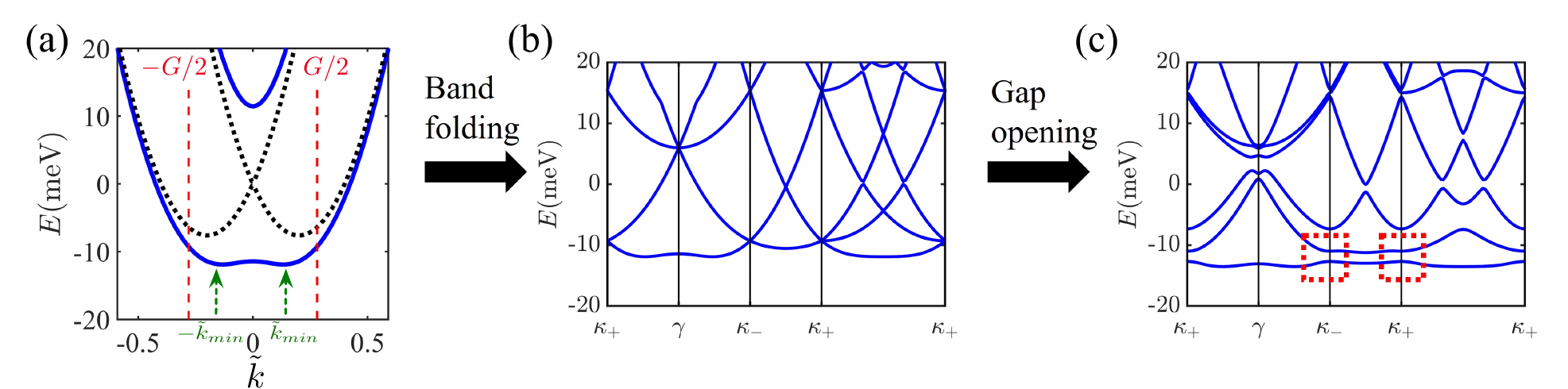}
    \caption{\textbf{Formation of an isolated flat band}. (a) Rashba band in the presence of Zeeman coupling $V_z = 22.86$ meV ($\tilde{V}_z=0.12$). Here $G$ denotes the reciprocal lattice vector for a superlattice potential with period $L=15$ nm, and $k_{min}$ marks the wavevector where the Rashba band reaches its minimum energy. (b) Folded band structure for $\tilde{U}_0=0$. (c) Band gap openings along the MBZ boundary (red rectangulars) induced by a finite $\tilde{U}_0=0.015$. These results are obtained by choosing Rashba SOC strength $\tilde{\lambda}=0.4$.}	
    \label{flatband}
\end{figure}

In this section, we analyze the origin of the flat band in the proposed system and derive the optimal condition for realizing an isolated flat band.
In the absence of superlattice potential, the Rashba Hamiltonian of Eq.~(1) in the main text can be written as
\begin{equation}
\tilde{\mathcal{H}}_0= \bm{\tilde{k}}^2 +\tilde{\lambda} (\tilde{k}_y\sigma_x - \tilde{k}_x\sigma_y)  + \frac{\tilde{V}_z}{2}\sigma_z,
\end{equation}
where $\tilde{\mathcal{H}}_0 = \mathcal{H}_0/E_0$, $E_0$ and other tilde notations are defined in the main text. The energy spectra of $\mathcal{H}_0(\bm{k})$ are derived as
\begin{equation}
E_{\pm}(\bm{k}) = E_0 \left(\tilde{k}^2 \pm \sqrt{ (\tilde{V}_z/2)^2+\tilde{\lambda}^2 \tilde{k}^2}\right),
\label{eq:Edispersion}
\end{equation} 
where $\pm$ denotes the upper and lower Rashba bands. 
By expanding $E_{-}(\bm{k})$ around $\bm{k}=0$, we have 
\begin{equation}
    E_{-}(\bm{k}) \approx E_0 \left[-\frac{\tilde{V}_z}{2} + \left(1-\frac{\tilde{\lambda}^2}{\tilde{V}_z} \right)\tilde{k}^2 + \frac{\tilde{\lambda}^4}{\tilde{V}_z^3} \tilde{k}^4 +\cdots \right].
\end{equation}
This result indicates that $E_{-}(\bm{k})$ develops a nearly flat segment around the band bottom when $\tilde{V}_z \sim \tilde{\lambda}^2$, as illustrated in Fig~\ref{flatsegment}(a). In particular, when $\tilde{V}_z = \tilde{\lambda}^2$, the spectrum becomes quartic, $E_{-}(\bm{k}) \propto \tilde{k}^4 $,
reflecting a strong suppression of the band dispersion. 
Based on this observation, a flat band can be realized by truncating this nearly flat segment through the introduction of a periodic superlattice potential, as illustrated in Fig~\ref{flatsegment}(b), thereby forming an isolated flat band within the first Brillouin zone (1st BZ).
Figure~\ref{flatband} illustates the three steps leading to the formation of a flat band via the above mechanism. First, as shown in Fig~\ref{flatband}(a), the Zeeman field lifts the band degeneracy of the Rashba band structure, producing a nearly flat segment around the band bottom. 
Second, the periodic superlattice potential folds the Rashba bands into the 1st BZ, as illustrated in Fig~\ref{flatband}(b).
Third, Bragg scattering induced by the superlattice potential lifts the band degeneracies along high-symmetry lines and points in the BZ, isolating the flat segment from the remaining dispersive bands and thereby forming an isolated flat band.

%From this perspective, the flat band can be qualitatively understood as originating from the truncation of the nearly flat segement of the lower Rashba band by the superlattice potential. 

%Based on this understanding, the presence of isolated flat band can be arroximated as the presence of flat segement around the lower Rashba band bottom.

\begin{figure}[t]
    \centering
    \includegraphics[width=0.8\columnwidth]{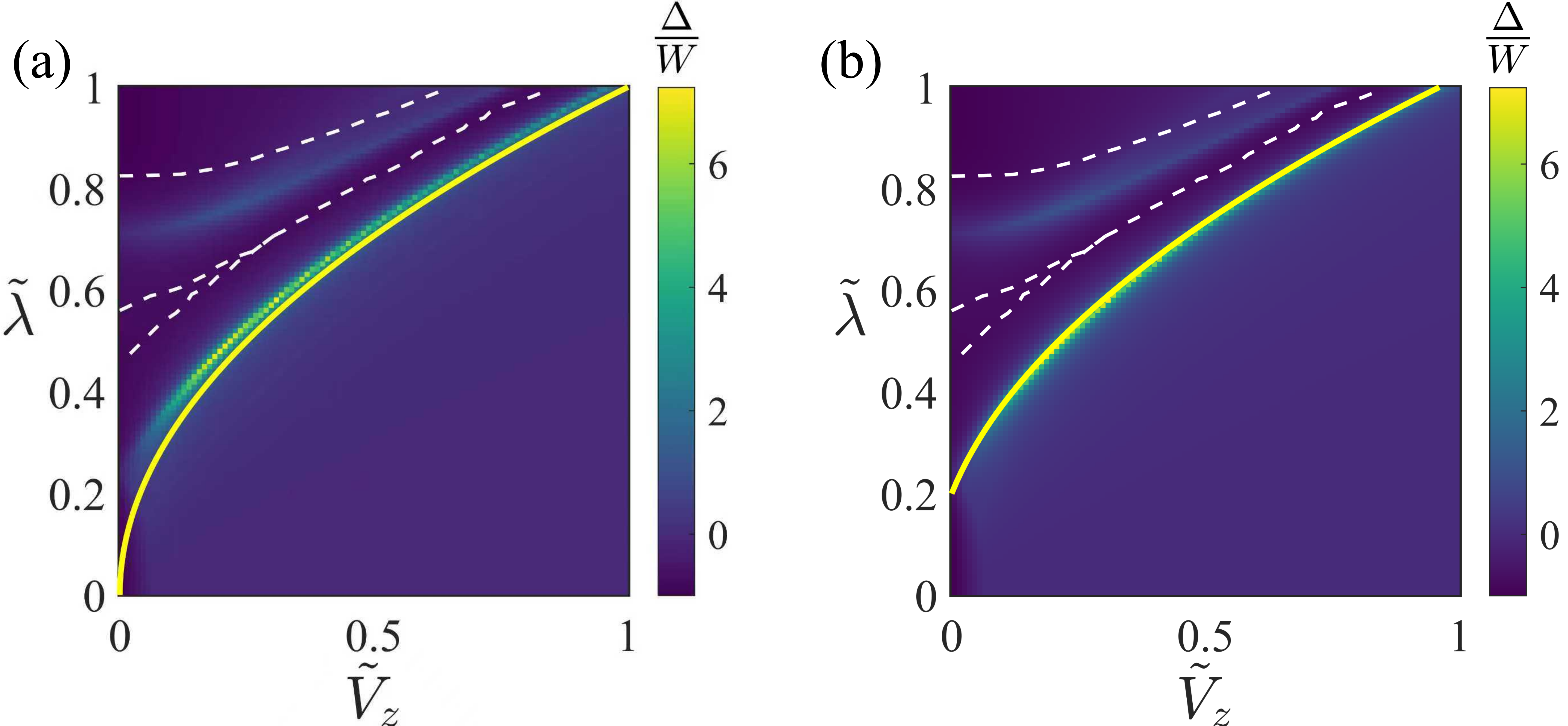}
    \caption{Numerical results of the ratio between direct band gap $\Delta$ and band width $W$ (reproduction of Fig.~4(c) in the main text). In panel (a), the yellow solid curve denotes the relation $\tilde{V}_z = \tilde{\lambda}^2$, corresponding to the condition for realizing quartic band dispersion $E_{-}(\bm{k}) \propto \tilde{k}^4$. In panel (b), the yellow solid curve indicates a slightly modified relation, $\tilde{V}_z = \tilde{\lambda}^2 - 0.04$. }
    \label{new ratio}	
\end{figure}

We investigate the relationship between the emergence of a nearly flat segment near the bottom of the lower Rashba band and the formation of an isolated flat band induced by the superlattice potential, by varying the model parameters.
Figure~\ref{new ratio} shows the numerical results for the ratio between the direct band gap $\Delta$ and the band width $W$ of the lowest-energy band as a function of $\tilde{V}_z$ and $\tilde{\lambda}$. As shown in Fig.~\ref{new ratio}(a), the optimal condition for the presence of isolated flat band is defined as the maxima value of $\Delta/W$, which follows similar a trend similar to the quartic dispersion condition $\tilde{V}_z = \tilde{\lambda}^2$. In Fig.~\ref{new ratio}(b), a slightly modified relation with a reduced Zeeman field,  $\tilde{V}_z = \tilde{\lambda}^2-0.04$, shows an excellent agreement with the numerically determined optimal flat-band condition. 
These results indicate that the flat band primarily originates from the nearly flat segement of the lower Rashba band.

To further elucidate this flat band generating mechanism, we perform an analytical derivation of the band energies.
The flatness of the lowest-energy band shown in Fig.~\ref{flatband}(c) is primarilty determined by the energy differences between band energyies at high-symmetry points and at the band minima. In the following, we evaluate the energies at these points by treating the superlattice potential $\tilde{U}_0$ as perturbation. The corresponding wavefunctions for $E_{\pm}(\bm{k})$ shown in Eq.~(\ref{eq:Edispersion}) are given by
\begin{equation}
\psi_{+}(\bm{k}) = \begin{pmatrix} \cos \frac{\theta_k}{2} \\ -i e^{i\phi_{\bm{k}}} \sin \frac{\theta_k}{2} \end{pmatrix}, \qquad
\psi_{-}(\bm{k}) = \begin{pmatrix} \sin \frac{\theta_k}{2} \\ i e^{i\phi_{\bm{k}}} \cos \frac{\theta_k}{2} \end{pmatrix},
\end{equation}
where $ (\cos \phi_{\bm{k}},\sin \phi_{\bm{k}}) = (\tilde{k}_x/\tilde{k},\tilde{k}_y/\tilde{k})$, and $\theta_k$ is defined as 
\begin{equation}
\cos \theta_k = \frac{\tilde{V}_z/2}{\sqrt{\left(\tilde{V}_z/2\right)^2 + (\tilde{\lambda} \tilde{k})^2}},  ~~~
 \sin \theta_k = \frac{ \tilde{\lambda} \tilde{k}}{\sqrt{\left(\tilde{V}_z/2\right)^2 + (\tilde{\lambda} \tilde{k})^2}}.
\end{equation}
The lower Rashba band $E_{-}(\bm{k})$ exhibits minima at wavevectors $\tilde{k}_{\text{min}} = \pm \frac{1}{2}\sqrt{\tilde{\lambda}^2 - \tilde{V}_z^2/\tilde{\lambda}^2}$ for $\tilde{V}_z \leq \tilde{\lambda}^2$, as indicated in Fig.~\ref{flatband}(a). The coreesponding band bottom energy is therefore given by 
\begin{equation}
E_{\text{min}} = - \frac{E_0}{4}\left(\frac{\tilde{V}_z^2}{\tilde{\lambda}^2} + \tilde{\lambda}^2\right).
\end{equation}
For $\tilde{V}_z > \tilde{\lambda}^2$, $\tilde{k}_{\text{min}} =0$ and $E_{\text{min}}= -E_0\tilde{V}_z/2$.

\begin{figure}[t]
    \centering
    \includegraphics[width=0.9\columnwidth]{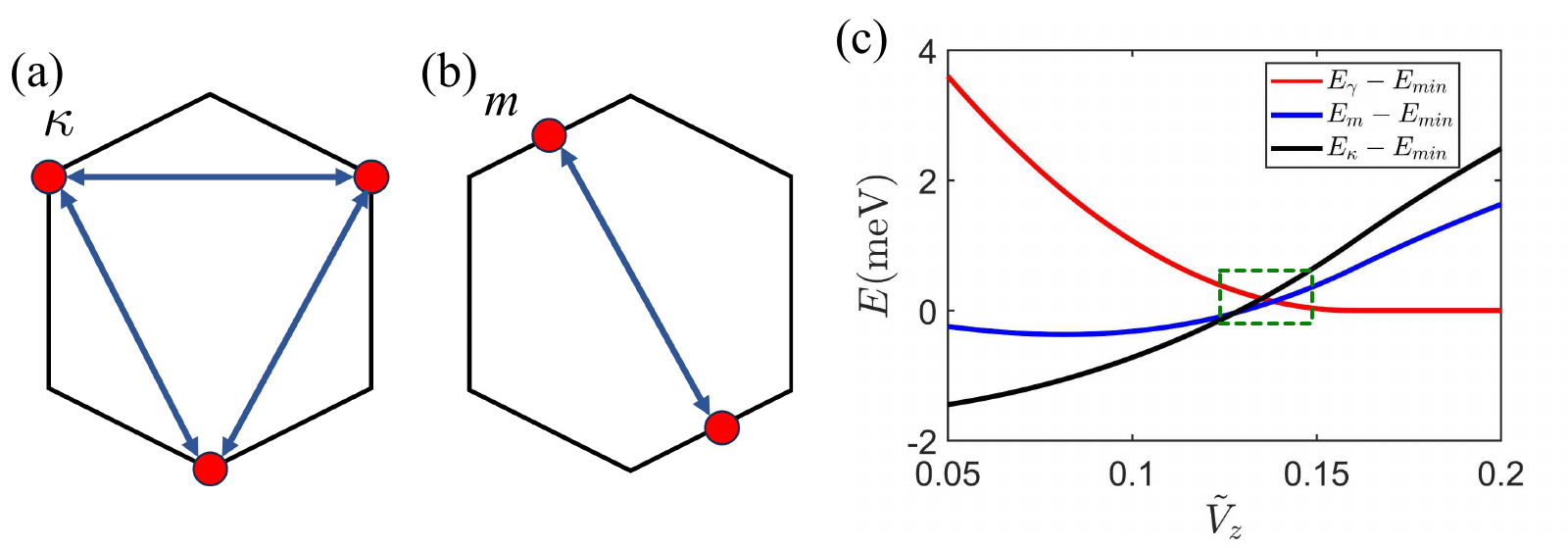}
    \caption{(a) Three equvalent $\kappa$ points connected by reciprocal lattice vectors. (b) Two equvalent $m$ points connected by a reciprocal lattice vector. (c) Eneergy differences between high-symmetry points and band mimimum. The green dashed rectangle highlights the parameter regime where the band is nearly flat. }
    \label{perturbation_theory_picture}	
\end{figure}

We now evaluate the band energies at high-symmetry points using perturbation theory, keeping terms up to first order of $\tilde{U}_0$.
At the $\gamma$ point, there are no degenerate states connected by a reciprocal lattice vector $\bm{G}$, so the first-order energy correction vanishes. The band energy at $\gamma$ point is therefore given by
\begin{equation}
E_{\gamma} = -\frac{1}{2}E_0 \tilde{V}_z.
\end{equation}
For $\tilde{V}_z \geq \tilde{\lambda}^2$, we have $E_{\gamma} = E_{\text{min}}$, and thus the energy difference $E_{\text{min}} - E_{\gamma} =0 $.
At the $\kappa$ point, there are three degenerate states labeled by $\alpha=1,2,3$ that are connected by reciprocal lattice vectors, as illustrated in Fig.~\ref{perturbation_theory_picture}(a). The matrix element of the superllatice potential between two states $\alpha$ and $\beta$ is derived as
\begin{equation}
\langle \psi_{-\alpha} | U_0 \sigma_0 | \psi_{-\beta} \rangle = U_0 \left[ \sin^2\left(\frac{\theta_k}{2}\right) + \cos^2\left(\frac{\theta_k}{2}\right) e^{i(\phi_\alpha - \phi_\beta)} \right].
\end{equation}
By diagonalizing the $3\times 3$ perturbation matrix, the band energy at the $\kappa$ point is given by
\begin{equation}
E_{\kappa} = E_0 \left( \tilde{k}_{\kappa}^2 - \sqrt{(\tilde{V}_z^2/2)^2 + \tilde{\lambda}^2 \tilde{k}_{\kappa}^2} \right) - U_0,
\end{equation}
where $k_{\kappa} = 4\pi/3L$.
At the $m$ point, two degenerate states are connected by the reciprocal lattice vectors, as illustrated in Fig.~\ref{perturbation_theory_picture}(b). Using first-order perturbation theory, the band energy at $m$ point is given by
\begin{equation}
E_{m} = E_0 \left( \tilde{k}_{m}^2 - \sqrt{(\tilde{V}_z^2/2)^2 + \tilde{\lambda}^2 \tilde{k}_{m}^2} \right) - U_0 \cos \theta_{k_m},
\end{equation}
where $k_{m} = 2\pi/\sqrt{3}L$.

\begin{figure}
    \centering
    \includegraphics[width=0.7\columnwidth]{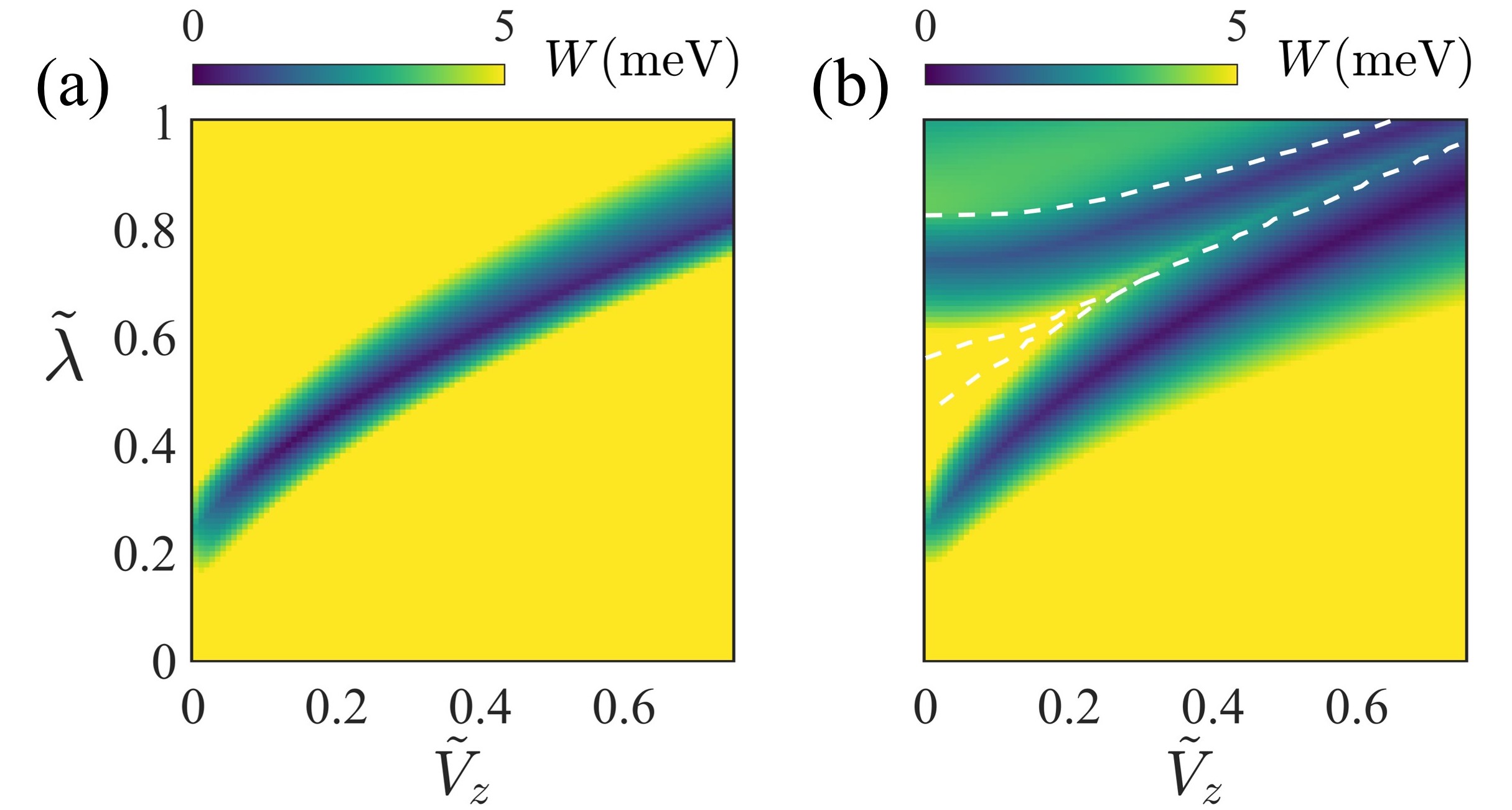}
    \caption{(a) Bandwidth ($W$) obtained from perturbation theory, where $W$ is defined as the maximaum of $|E_{\gamma}-E_{\text{min}}|$, $|E_{\kappa}-E_{\text{min}}|$, and $|E_{m}-E_{\text{min}}|$. (b) Bandwidth obtained by numerically solving the Hamiltonian in Eq.~(3) of the main text. These results are calculated for $\tilde{U}_0=0.015$.}
    \label{perturbation result}	
\end{figure}

Figure~\ref{perturbation_theory_picture}(c) shows the  band energies at high-symmetry points relative to the band minimum. Notably, there exists a parameter regime, hightlighted by the rectangle, in which all the three energy difference beome very small, implying that the band is nearly flat. 
Figure~\ref{perturbation result} compares the bandwidth obtained analytically from perturbation theory with that calculated numerically by solving Eq.~(3) in the main text.
In the vicinity of the optimal flat-band condition, the two results shows excellent agreement, suggesting that the superlattice potential plays the crucial role in isolating the flat segment of the lower Rashba band.

\section{Origin of the band topology}
\begin{figure}
    \centering
    \includegraphics[width=0.99\columnwidth]{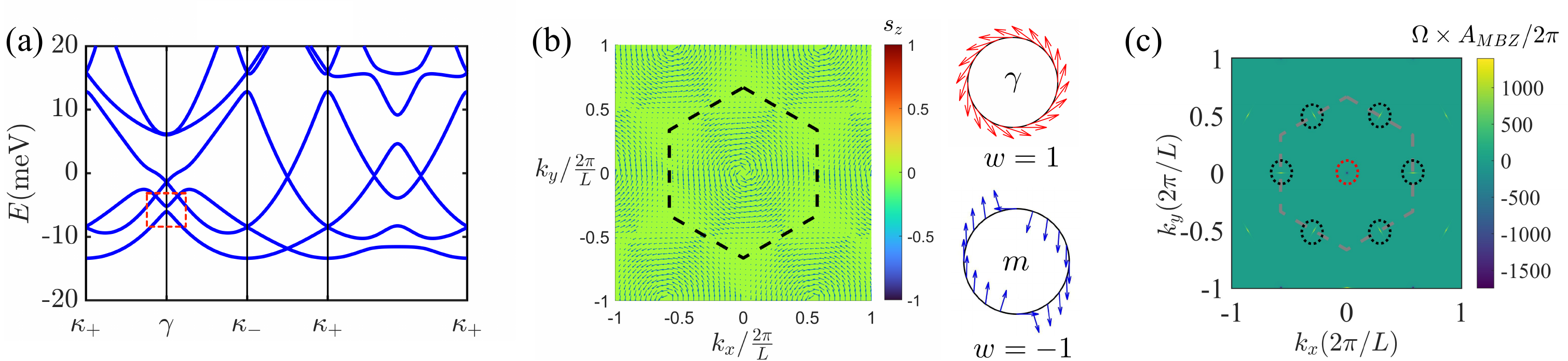}
    \caption{The origin of the nontrivial topology of the lowest-energy band. (a) A small Zeeman field opens band gaps at the $\gamma$ and $m$ points. (b) Spin texture of the lowest-energy band shown in (a). Right pannels: Spin windings around closed loops encircling (top) $\gamma$ and (bottom) $m$ points. The arrows denote the in-plane spin directions, and $w$ represents thecorresponding winding number. (c) Berry curvature distributon in the 1st BZ. All calculations are performed for $\tilde{V}_z$=0.005.}
    \label{bandtopoloy}	
\end{figure}

In this section, we investigate the origin of the nontrivial band topology of the lowest-enegy band. As shown in Fig.~\ref{band1}, in the absence of Zeeman field ($\tilde{V}_z=0$), the band degeracies at $\gamma$ and $m$ are protected by time-reversal symmetry and therefore cannot be lifted by the superlattice potential.
The low-energy physics around these time-reversal momenta are described by the Dirac-type Hamiltonian 
\begin{equation}
h(\bm{k}) = \nu_y k_y \sigma_x+\nu_x k_x \sigma_y,
\end{equation}
where velocities $v_{x,y}$ take different values at the $\gamma$ and $m$ points, determinging the linear dispersions and the associated of spin-momentum lockings.
The Dirac-type Hamitlonian represents a spin-$1/2$ monopole in the parameter space $(k_x,k_y,\tilde{V}_z)$ and carries a topological charge $|Q|=1$. Consequently, an infinitesimal Zeeman field acts as a mass term that breaks time-reversal symmetry and gaps the Dirac crossings at both the $\gamma$ and $m$ points, as illustrated in Fig.~\ref{bandtopoloy}(a). The resulting gapped band structure acquires nontrivial topology determined by the topological charges associated with these Dirac points. 

To identify the sign of these charges, we analyze the spin texture of the lowest-energy band. Figure~\ref{bandtopoloy}(b) shows the spin configuration in momentum space, while the right panels display the spin winding around the $\gamma$ and $m$ points. The corresponding winding numbers are $w_{\gamma}=1$ and $w_{m}=-1$, yielding topological charges $Q_{\gamma}=1$ and $Q_{m}=-1$, respectively. Since there are three symmetry-related $m$ points in the first Brillouin zone, the total Chern number of the lowest-energy band is given by $C = -\text{sgn}(\tilde{V}_z)(Q_{\gamma}+3Q_{m})/2 = 1$ for $\tilde{V}_z>0$.

Figure~\ref{bandtopoloy}(c) shows the calculated Berry curvature distribution. For small values of $\tilde{V}_z$, the Berry ccurvature is strongly concentrated around the $\gamma$ and $m$ points, reflecting the dominant contribution of the gapped Dirac cones to the band topology. Moreover, the sign distribution of the Berry curvature is consistent with the topological charges inferred from the spin-winding analysis.
As the Zeemann field is further increased, the lowest energy band involes into a flat band, as illustrated in Fig.~\ref{band3} without any gap closing. 
Since the Chern number is invariant under such an adiabatic deformation, the band topology remains unchanged throughout the evolution. We therefore conclude that the nontrivial topology of the lowest-energy flat band arises from the topological charges associated with the gapped Dirac points at the time-reversal-invariant momenta of the 1st BZ.

\section{Quantum geometry the miniband}
\begin{figure}[t]
    \centering
    \includegraphics[width=0.99\columnwidth]{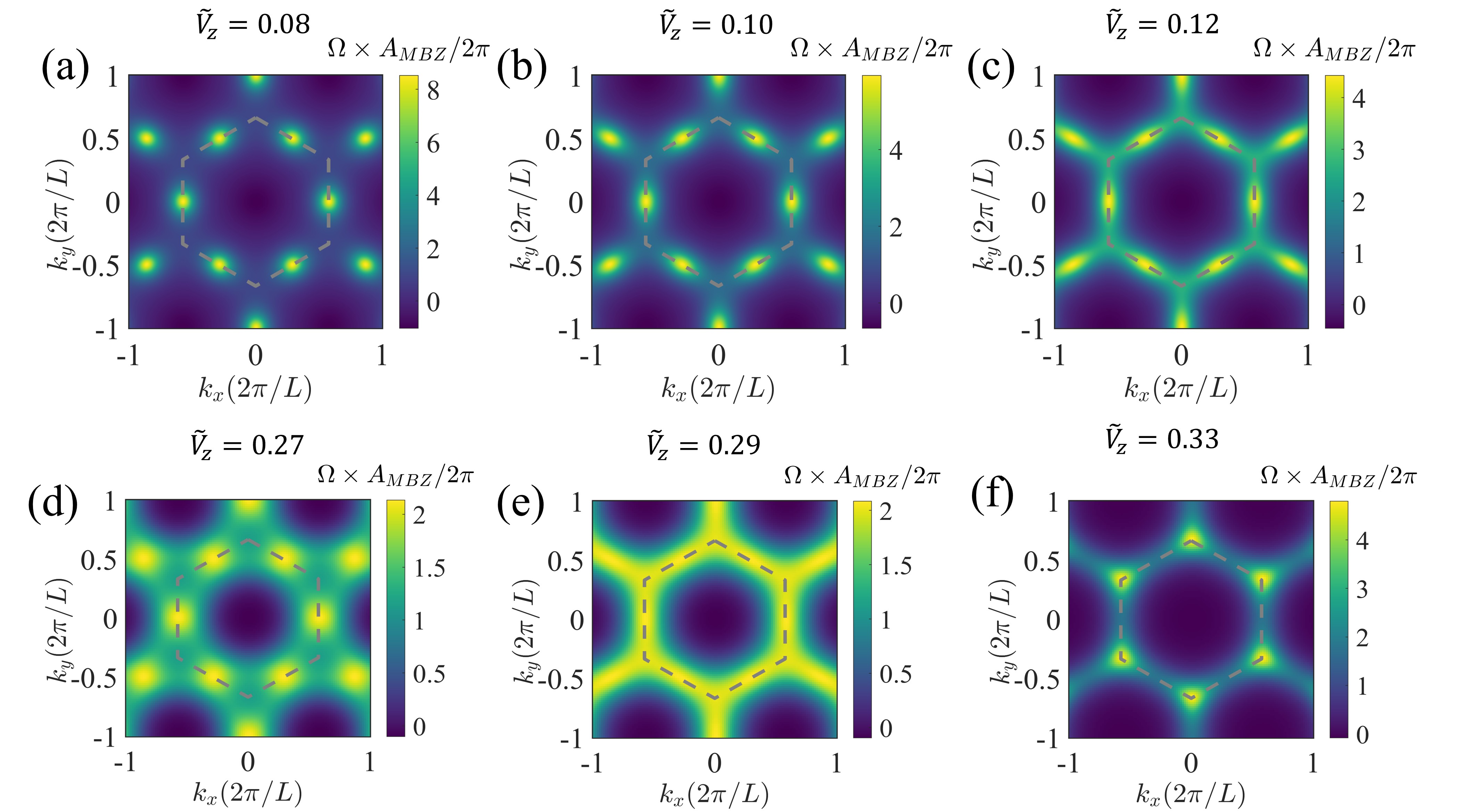}
    \caption{Evolution of the Berry curvature in the $\mathcal{C}=1$ region. In(a)-(c), $\tilde{\lambda}$ is fixed to be 0.4. In(d)-(f), $\tilde{\lambda}$ is fixed to be 0.6.$\tilde{U}_0$ in (a)-(f) are all fixed to be 0.03.}
    \label{topology1}	
\end{figure}

\begin{figure}[t]
    \centering
    \includegraphics[width=0.99\columnwidth]{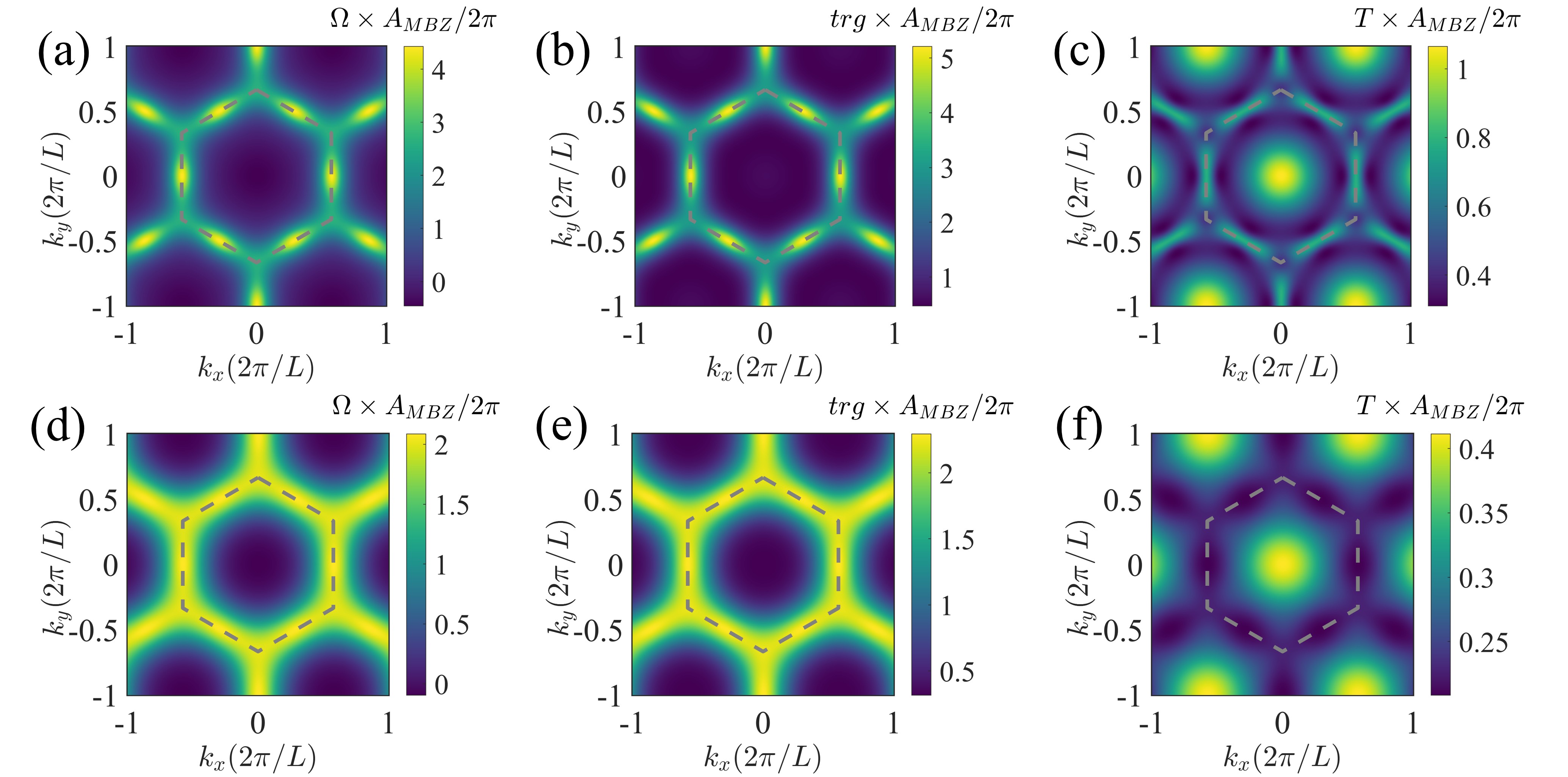}
    \caption{Additional results for quantum geometry.$g$ is the Fubini-Study matrix and $T(k)=trg-\Omega$. In(a)-(c), $\tilde{\lambda}=0.4$, $\tilde{V}_z=0.12$. In (d)-(f), $\tilde{\lambda}=0.6$, $\tilde{V}_z=0.29$.}
    \label{topology2}	
\end{figure}

In the main text, we show that the 1st miniband is topologically nontrivial with Chern number $\mathcal{C}=1$. Here we systematically investigate the evolution of Berry curvature $\Omega$ as a function of Zeeman coupling. As shown in Fig.~\ref{topology1}, with increasing $\tilde{V}_z$ , $\Omega$ initially concentrates near the $m$ points of the MBZ, then becomes nearly nearly uniformly distributed along the MBZ boundary at intermediate values of $\tilde{V}_z$, and finally shifts toward the $\kappa_{\pm}$ points of the MBZ. This behavior can be understood from the evolution of band structures with increasing $\tilde{V}_z$ as (see Fig.~\ref{band3}) by invoking that $\Omega_{\bm{k}} \propto 1/\Delta^{2}_{\bm{k}}$ \cite{xiao2010berry}, where $\Delta_{\bm{k}}$ denotes the band gap at wavevector $\bm{k}$. Specifically, at relative small value of $\tilde{V}_z$, the band gap opening is mainly around the $m$ point, resulting in $m$-centered $\Omega_{\bm{k}}$. At intermediate value of $\tilde{V}_z$, where the 1st miniband become ultra flat, the band gap opens along the MBZ boundary, leading to uniformly distributed $\Omega_{\bm{k}}$ along the boundary. When further increasing the $\tilde{V}_z$, the minima band gap moves to the $\kappa_{\pm}$ points, resulting in the corner centered Berry curvature. Figure~\ref{topology2} compare the Berry curvature and the Fubini-Study metric Tr($g_{\bm{k}}$). 
The distribution of the trace of Tr($g_{\bm{k}}$) exhibits similar distribution to that of $\Omega_{\bm{k}}$, therefore leading to small violation of the trace condition $T=A_{\text{MBZ}}/2\pi \int d^2\bm{k} T(\bm{k})$, where $ T(\bm{k}) = \text{Tr}(g_{\bm{k}})-\Omega_{\bm{k}} $ are shown in Figs.~\ref{topology2}(c) and (f).

\begin{figure}[t]
    \centering
    \includegraphics[width=0.75\columnwidth]{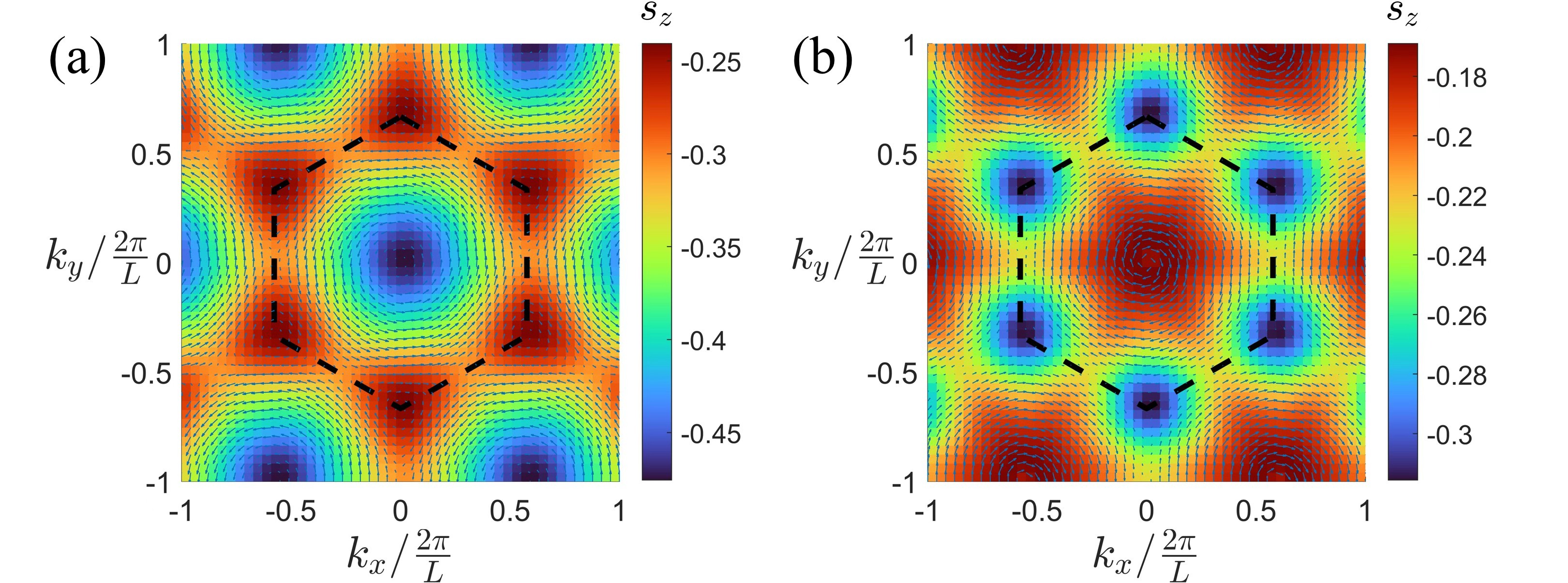}
    \caption{(a),(b) Spin texture of the (a) 1st and (b) 2nd minibands, where the colormap denotes the $z$-direction spin component and the quivers indicate the strength and direction of in-plane spin components. These results are calculated 
by choosing $\tilde{\lambda}=0.4$, $\tilde{V}_z=0.12$, and $\tilde{U}_0=0.03$. }
    \label{additional band}	
\end{figure}

We also calculate the spin textures of the 1st and 2nd minibands. As shown in In Figs.~\ref{additional band}(a) and (b), the in-plane spin components exhibit the typical Rashba character around the MBZ center and vanish at the high-symmetry points of the MBZ due to symmetry constraints. The $z$-component arises from the Zeeman coupling, which polarizes spin along the out-of-plane direction. Overall, the chiral spin texture of the Rashba system \cite{manchon2015new} is modulated by the presence of the superlattice potential.

\section{Phase diagram}
\begin{figure}[t]
    \centering
    \includegraphics[width=\columnwidth]{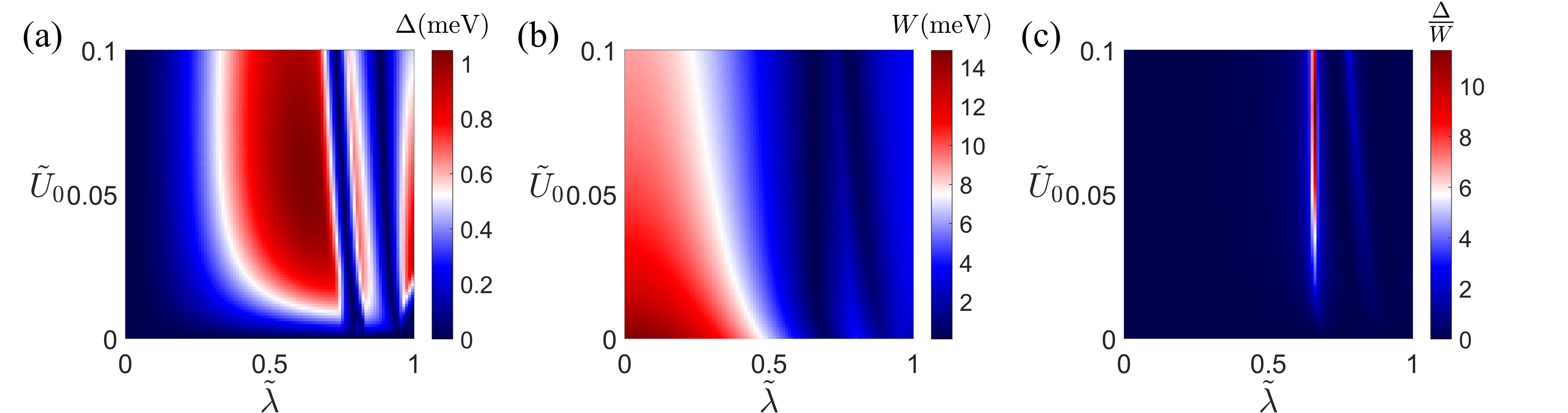}
    \caption{(a)-(c) Phase diagrams of (a) direct band gap $\Delta$, (b) bandwidth $W$, and (c) their ratio $\Delta/W$ as functions of superlattice potential $\tilde{U}_0$ and Rashba SOC strength $\tilde{\lambda}$. These results are obtained by choosing  Zeeman coupling $\tilde{V}_z=0.4$. }
    \label{phase diagram2}	
\end{figure}

\begin{figure}[t]
    \centering
    \includegraphics[width=\columnwidth]{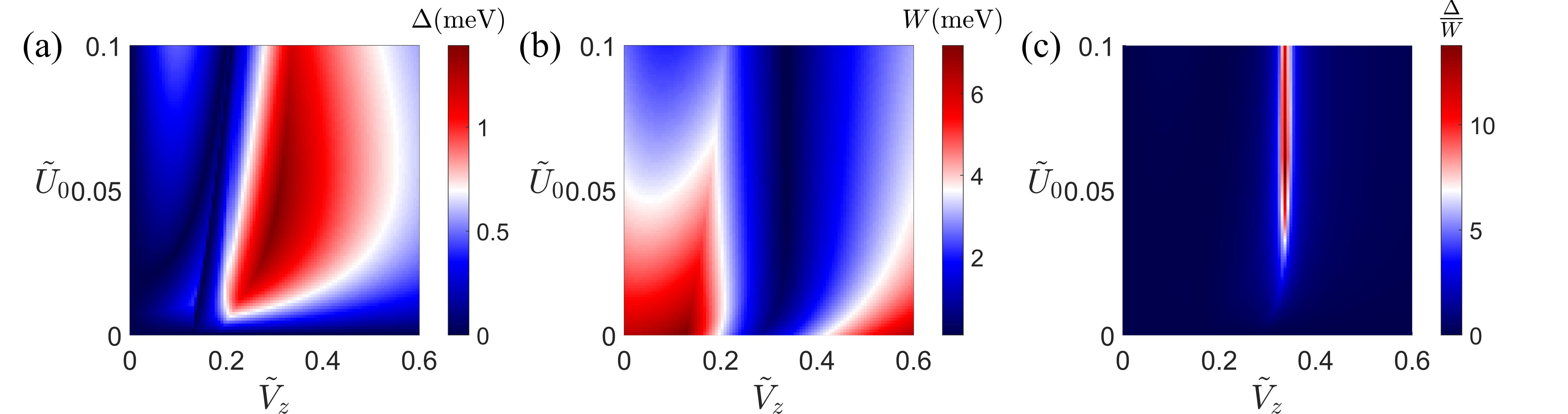}
    \caption{(a)-(c) Phase diagrams of (a) direct band gap $\Delta$, (b) bandwidth $W$, and (c) their ratio $\Delta/W$ as functions of superlattice potential $\tilde{U}_0$ and Zeeman coupling $\tilde{V}_z$. These results are obtained by choosing Rashba SOC strength $\tilde{\lambda}=0.6$.}
    \label{phase diagram3}	
\end{figure}

In this section, we present additional calculations of the direct band gap $\Delta$, bandwidth $W$, and their ratio $\Delta/W$ for different choices of model parameters. Figure~\ref{phase diagram2} shows the results for a fixed Zeeman coupling $\tilde{V}_z=0.4$. Figure~\ref{phase diagram3} displays phase diagrams analogy to Figs.~4(d)-(f) of the main text, but with a different Rashba SOC strength $\tilde{\lambda} = 0.6$. In both cases, the optimal condition for realizing an isolated flat miniband indicated by the bright regions in Figs.~\ref{phase diagram2}(c) and \ref{phase diagram3}(c). Notably, these bright regions form nearly vertical lines, indicating that the optimal condition is insensitive to variations in the superlattice potential $\tilde{U}_0$ once it exceeds a threshold value.     
 

\section{Band-projected interaction and exact diagonalization}
As explained in Eq.~(4) of the main text, the many-body calculations are performed using the 1st-miniband-projected Hamiltonian
\begin{equation}
 H=\sum_{\bm{k}} \varepsilon_{\bm{k}} c_{\bm{k}}^{\dagger} c_{\bm{k}}+\frac{1}{2} \sum_{\bm{k}^{\prime} \bm{p}^{\prime}  \bm{p}\bm{k}} V_{\bm{k}^{\prime} \bm{p}^{\prime}  \bm{p}\bm{k} } c_{\bm{k}^{\prime} }^{\dagger} c_{\bm{p}^{\prime} }^{\dagger} c_{\bm{p} } c_{\bm{k}}
\label{many body hamiltonian}
\end{equation}
 where $c_{\bm{k}}^{\dagger}$ ($c_{\bm{k}}$ ) are creation (annihilation) operators for electrons in the 1st miniband with wavevector $\bm{k}$, $\varepsilon_{\bm{k}}$ is the corresponding dispersion, and $V_{\boldsymbol{k}^{\prime} \boldsymbol{p}^{\prime} \boldsymbol{k} \boldsymbol{p}} = \left\langle\boldsymbol{k}^{\prime} ; \boldsymbol{p}^{\prime} \right| \hat{V}\left|\boldsymbol{p} ; \boldsymbol{k}  \right\rangle$, with $|\boldsymbol{k} \rangle$ representing the Bloch states. In this work, we adopt the bare Coulomb potential, $V(\bm{r})=e^2/\epsilon r $, to describe the electron-electron interaction. The exact diagonalization (ED) procedure proceeds in three steps.
 
 The first step is to calculate the band dispersion and the wave functions, which are obtained by diagonalizing the single-particle Hamiltonian $\tilde{H}_0$ (Eq.~(3) in the main text), yielding
\begin{equation}
\left| \boldsymbol{k} \right\rangle = \sum_{\boldsymbol{g}\sigma}z_{\boldsymbol{k}\boldsymbol{g}\sigma}\left|\boldsymbol{k}+\boldsymbol{g}, \sigma\right\rangle,
\end{equation}
where $\left|\boldsymbol{k}+\boldsymbol{g}, \sigma\right\rangle$ denotes the wave function in the plane-wave basis, $\boldsymbol{g}$ is the reciprocal vectors corresponding to the superlattice potential, and $\sigma$ is the spin index,. 

The second step is to impose periodic boundary condition to the superlattice. To make the problem numerically tractable, the total number of accessible momenta is restricted to a finite cluster defined by lattice vectors $\boldsymbol{L}_{1,2}$. The corresponding crystal momenta take the form $\boldsymbol{k}=k_1 \boldsymbol{T}_{1}+\boldsymbol{k}_2 \boldsymbol{T}_{2}$, where $\boldsymbol{T}_{i}=\frac{2\pi \epsilon_{ij}\boldsymbol{L}_{j} \times \hat{z}}{\left|\boldsymbol{L}_{1}\times \boldsymbol{L}_{2}\right|}$ are the basis vectors associated with the periodic boundary condition. For a given periodic condition, the Fock space is fixed and the Hamiltonian matrix elements are uniquely determined. In this work, we consider clusters of sizes $4\times6$, $5\times6$, and $3\times9$. Detailed configurations of these clusters are provided in Ref.~\cite{reddy2024non}.

The third step is to give the explicit expression of the interaction Hamiltonian. Following the definitions provided in Ref.~\cite{reddy2023fractionalsup}, the momentum $\bm{q}$ is decomposed into its "mesh" part $[\bm{q}]$ inside the MBZ and its reciprocal lattice (RL) part $\boldsymbol{g}_{\boldsymbol{q}} \equiv \boldsymbol{q}-\left[\boldsymbol{q}\right]$. Then the interaction matrix element then reads
\begin{equation}
    \begin{aligned}
        \left\langle\boldsymbol{k}_1 ; \boldsymbol{k}_2 \right| \hat{V}\left| \boldsymbol{k}_4, \boldsymbol{k}_3\right\rangle & =\frac{1}{A} \sum_{\boldsymbol{q}} V(\boldsymbol{q})\left\langle\boldsymbol{k}_1 \right| e^{i \boldsymbol{q} \cdot \boldsymbol{r}_1}\left|\boldsymbol{k}_3 \right\rangle\left\langle\boldsymbol{k}_2 \right| e^{-i \boldsymbol{q} \cdot \boldsymbol{r}_2}\left|\boldsymbol{k}_4 \right\rangle \\
        & =\frac{1}{A} \sum_{\boldsymbol{q}} V(\boldsymbol{q}) \delta_{\boldsymbol{k}_1,\left[\boldsymbol{k}_3+\boldsymbol{q}\right]} \delta_{\boldsymbol{k}_2,\left[\boldsymbol{k}_4-\boldsymbol{q}\right]}  \times F\left(\boldsymbol{k}_1, \boldsymbol{k}_3, \boldsymbol{g}_{\boldsymbol{k}_3+\boldsymbol{q}}\right) F\left(\boldsymbol{k}_2, \boldsymbol{k}_4,\boldsymbol{g}_{\boldsymbol{k}_4-\boldsymbol{q}}\right),
    \end{aligned}
\end{equation}
where the form factor 
\begin{equation}
    F\left(\boldsymbol{k}_1, \boldsymbol{k}_2, \boldsymbol{g}\right)=\sum_{\boldsymbol{g}^{\prime} \sigma} z_{\boldsymbol{k}_1\left(\boldsymbol{g}^{\prime}+\boldsymbol{g}\right) \sigma}^* z_{\boldsymbol{k}_2 \boldsymbol{g}^{\prime} \sigma} .
\end{equation}
In the numerical calculations, reciprocal lattice vectors are truncated up to $N=25$ reciprocal shells, the $\boldsymbol{q}=0$ component of the Coulomb interaction is omitted as what has been done in Ref.~\cite{reddy2023fractionalsup}.

The dimension of the Fock space is $d_F = C^{N_p}_{N}$, where $N$ is the total number of momentum meshes in the MBZ and $N_p$ is the number of filling electrons. This space further decomposes into momentum-conserving subspaces, each labeled by total crystal momentum and characterized by $k_1+N_1k_2$, with $N_1$ the total number of meshes along $\bm{T}_1$ for a given $k_2$. Here and $k_{1,2}$ are integers labeling the indices of meshes along $\bm{T}_{1,2}$ directions. 

As a hallmark of fractional states at filling $n=p/q$, one expects $q$-fold degenerate ground states that cyclically permute under insertion of 
$q$ flux quanta \cite{reddy2023fractional,laughlin1999nobel,simon2020topological}. A flux $\phi$ induces a momentum shift $\boldsymbol{p} \to \boldsymbol{p}+\frac{\Phi}{2\pi}\boldsymbol{T}_{i}$, where $\Phi = \phi/\phi_0$ and $\phi_0 = hc/e$ is the flux quantum.

\begin{figure}
    \centering
    \includegraphics[width=0.7\columnwidth]{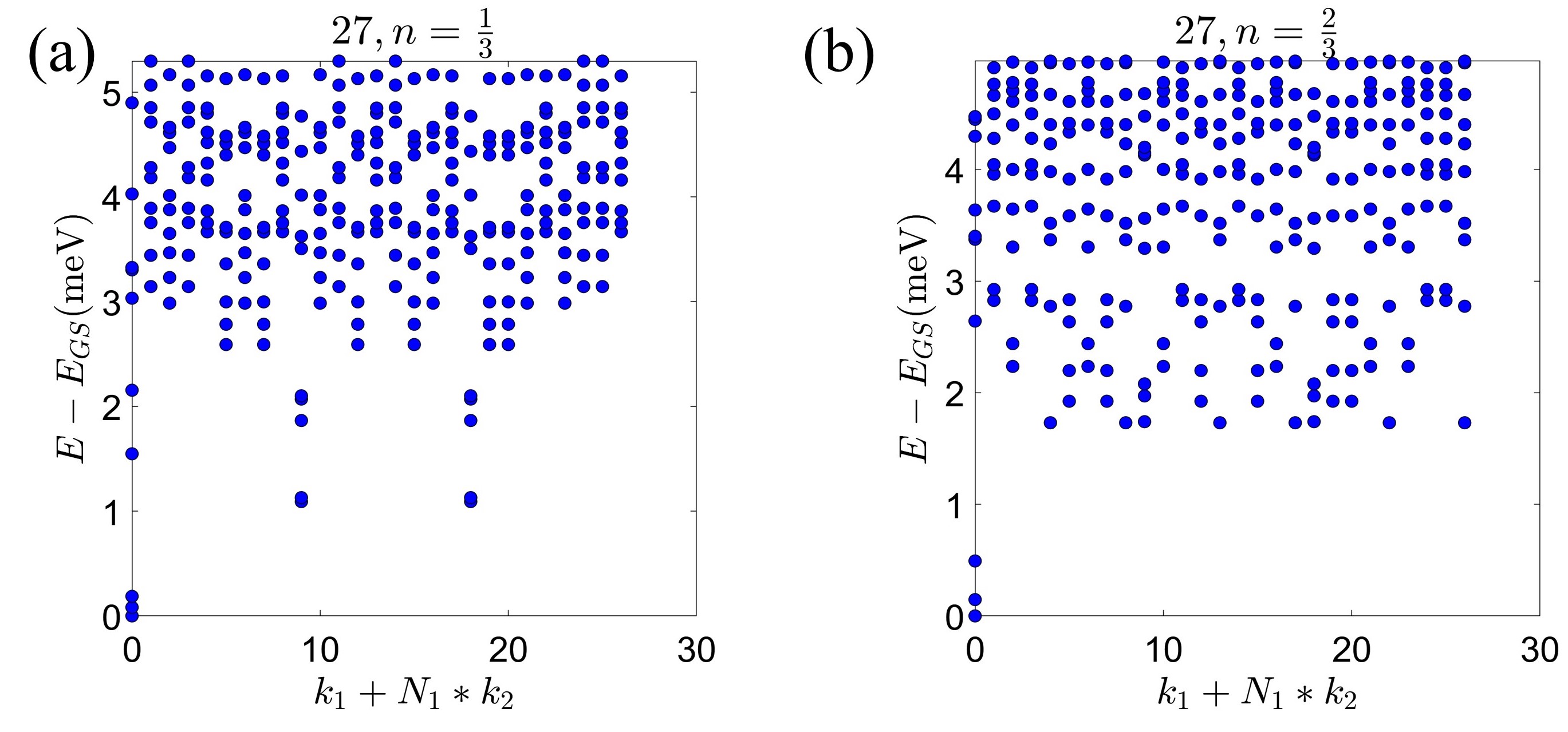}
    \caption{Many-body spectra for a 27-unit-cell cluster at filling factors (a) $n=1/3$ and (b) $n=2/3$, obtained with $\tilde{\lambda}=0.6$, $\tilde{V}_z=0.27$, and $\tilde{U}_0=0.03$.}
    \label{many body 27}	
\end{figure}

\section{Robustness of the FCI states}
In addition to the FCI states, it has been suggested that the charge-density-wave (CDW) states in a $5\times6$ cluster may also exhibit threefold degeneracy \cite{reddy2023fractionalsup}. To rule out this possibility, we perform ED calculations on a 27-unit-cell cluster. As shown in Fig.~\ref{many body 27}, the many-body spectrum display threefold degenerate ground states at zero total momentum for both filling factors $n=1/3$ and $n=2/3$, suggesting that the observed degenerate ground states are the FCI rather than CDW states. 

To provide further evidence for the robustness of FCI states, here we provide the ED results for different clusters. In Fig~\ref{many body collection} and Table \ref{tab:many_body_gap}, we show the ED results of $3 \times 5$, $4 \times 6$, $27$, $5 \times 6$, for all these clusters at filling $1/3$ and $2/3$, three robust degenerate ground states are displayed. 

\begin{figure}
    \centering
    \includegraphics[width=0.7\columnwidth]{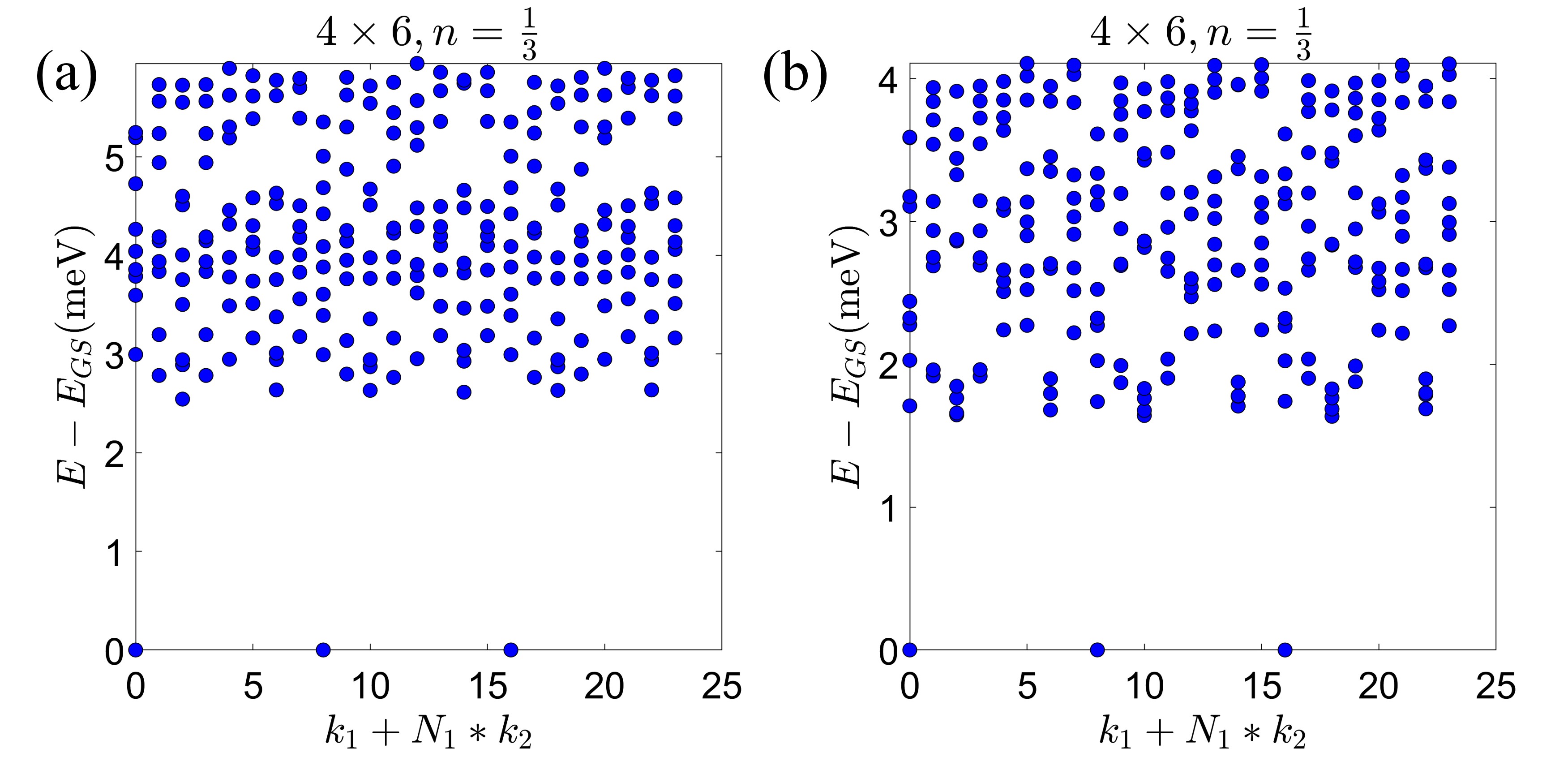}
    \caption{Many-body spectra for a 24-unit-cell cluster at filling factor $n=1/3$,calculated (a) without and (b) with band mixing by allowing one additional electron to occupy the 2nd miniband. Other parameters are $\tilde{\lambda}=0.6$, $\tilde{V}_z=0.28$, and $\tilde{U}_0=0.03$.}
    \label{many body band mixing}	
\end{figure}

To examine the effect of band mixing, we perform ED calculations by enlarging the Hilbert space to allow one additional electron to occupy the 2nd miniband. As shown in Fig.~\ref{many body band mixing}, this extension does not affect the ground-state degeneracy, while indeed slightly reduce the excitation gap from 2.5 meV to 1.65 meV. These results indicate that band mixing does not significantly suppress the stability of the FCI states.

\begin{figure}
    \centering
    \includegraphics[width=0.99\columnwidth]{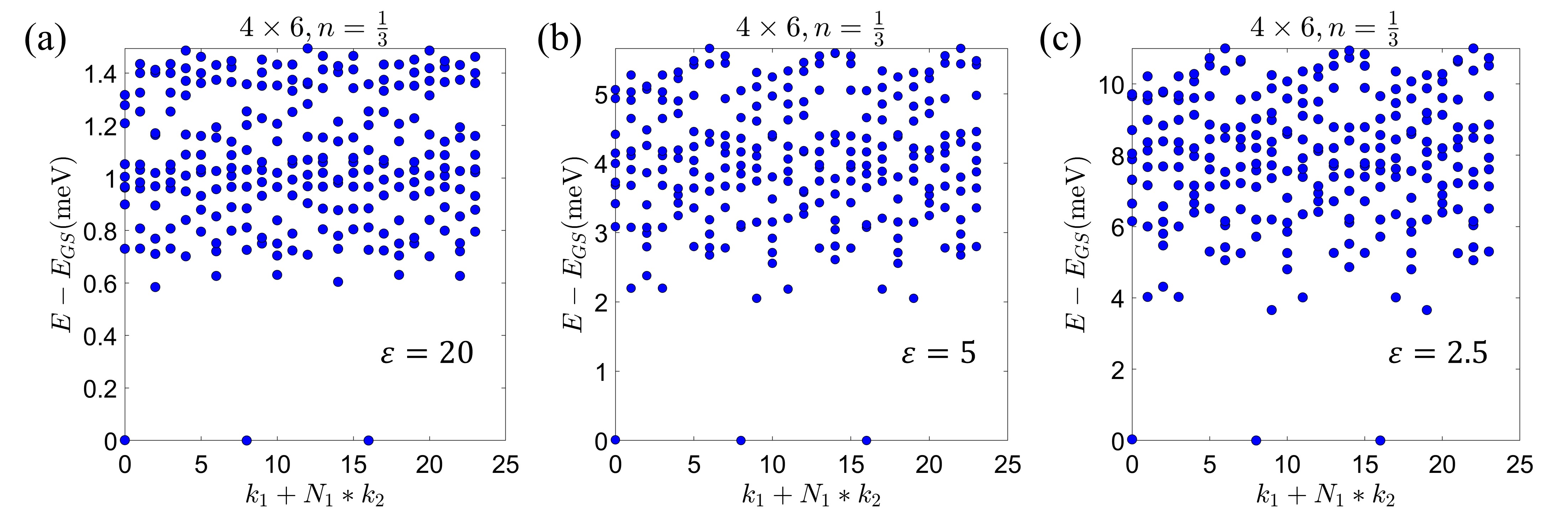}
    \caption{Many-body spectra for a 24-unit-cell cluster at filling factor $n=1/3$ for three values of dielectric constant $\epsilon$, obtained with $\tilde{\lambda}=0.6$, $\tilde{V}_z=0.28$, and $\tilde{U}_0=0.03$.}
    \label{many body dielectric}	
\end{figure}

\begin{figure}
    \centering
    \includegraphics[width=0.99\columnwidth]{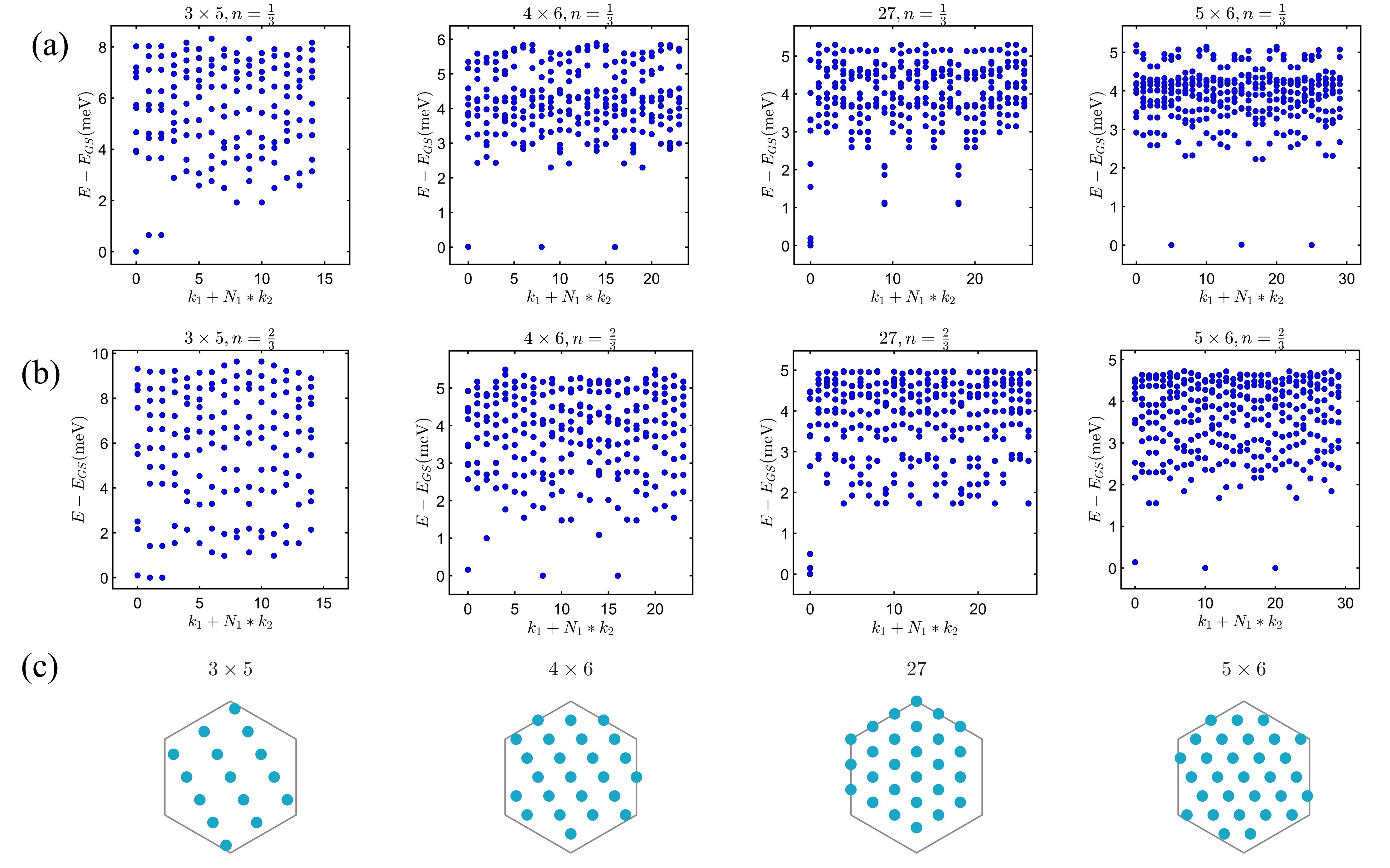}
    \caption{(a-b)Collection of many-body spectra for different clusters: $3\times 5$, $4\times 6$, $27$, $5\times 6$ obtained with $\tilde{\lambda}=0.6$, $\tilde{V}_z=0.27$, and $\tilde{U}_0=0.03$. $n=1/3$ in (a) and $n=2/3$ in (b). (c)Momentum space mesh diagrams of the finite-size clusters used in this work.}
    \label{many body collection}	
\end{figure}

We further investigate the effect of varying the Coulomb interaction strength on the FCI states. Figure~\ref{many body dielectric} shows the many-body spectra for $n=1/3$ on the $4\times6$ cluster for several values of the dielectric constant. Although the excitation gap decreases almost linearly with increasing dielectric constant, the threefold ground-state degeneracy remains robust. 

\begin{figure}
    \centering
    \includegraphics[width=0.99\columnwidth]{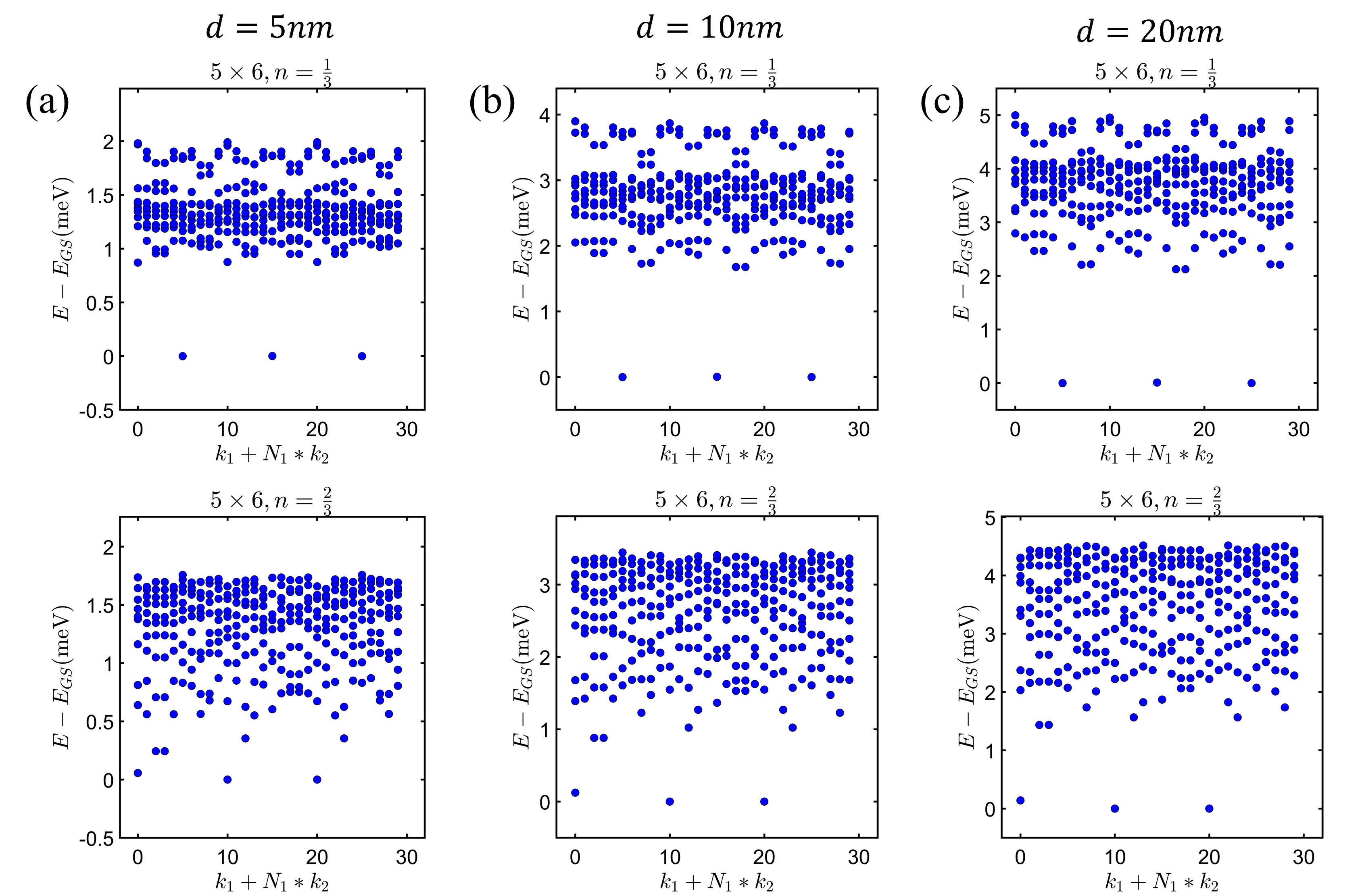}
    \caption{Collection of many-body spectra with interaction defined in equation (\ref{interaction new}) at different gate distance $d$, obtained with $\tilde{\lambda}=0.6$, $\tilde{V}_z=0.27$, and $\tilde{U}_0=0.03$.}
    \label{many body for gated}	
\end{figure}

Another important issue is that in real systems, the interaction may not exactly has the same form to the unscreened Coulomb potential we used in the main text. For existence, due to the existence of the gate, the electron interaction has the form 
\begin{equation}
V(\bm{q}) = \frac{e^2 \tanh(d|\bm{q}|)}{\epsilon |\bm{q}|}
\label{interaction new}
\end{equation}

where $d$ is the distance between two gates. In Fig~\ref{many body for gated}, we show the many body spectra when $d$ takes different values. We see that if $d$ is too small, it would make the gap smaller and FCI states less stable. And when $d$ is larger, it would not effect the stablity of the FCI states. 

\begin{table} 
\centering
\renewcommand{\arraystretch}{1.18}
\setlength{\tabcolsep}{0pt}
\caption{\label{tab:many_body_gap}
Many body gap for different cluster sizes.}
\begin{tabular}{|c|c|c|}
\hline
\multirow{2}{*}{\makebox[2.8cm][c]{\textbf{Cluster size}}}
& \multicolumn{2}{c|}{\makebox[4.4cm][c]{\textbf{Excitation gap (meV)}}} \\
\cline{2-3}
& \makebox[2.2cm][c]{$n = 1/3$}
& \makebox[2.2cm][c]{$n = 2/3$} \\
\hline
\makebox[1.5cm][c]{15 ($ 3\times 5$)}
& \makebox[2.2cm][c]{1.28}
& \makebox[2.2cm][c]{0.83} \\

\makebox[1.5cm][c]{24 ($4\times 6$)}
& \makebox[2.2cm][c]{2.25}
& \makebox[2.2cm][c]{0.89} \\

\makebox[1.5cm][c]{27}
& \makebox[2.2cm][c]{0.87}
& \makebox[2.2cm][c]{1.12} \\

\makebox[1.5cm][c]{30 ($5\times 6$)}
& \makebox[2.2cm][c]{2.27}
& \makebox[2.2cm][c]{1.49} \\
\hline
\end{tabular}
\end{table}

\section{Robustness of flat bands and FCI states against disorder and inhomogeneity(new)}

\begin{figure}
    \centering
    \includegraphics[width=0.8\columnwidth]{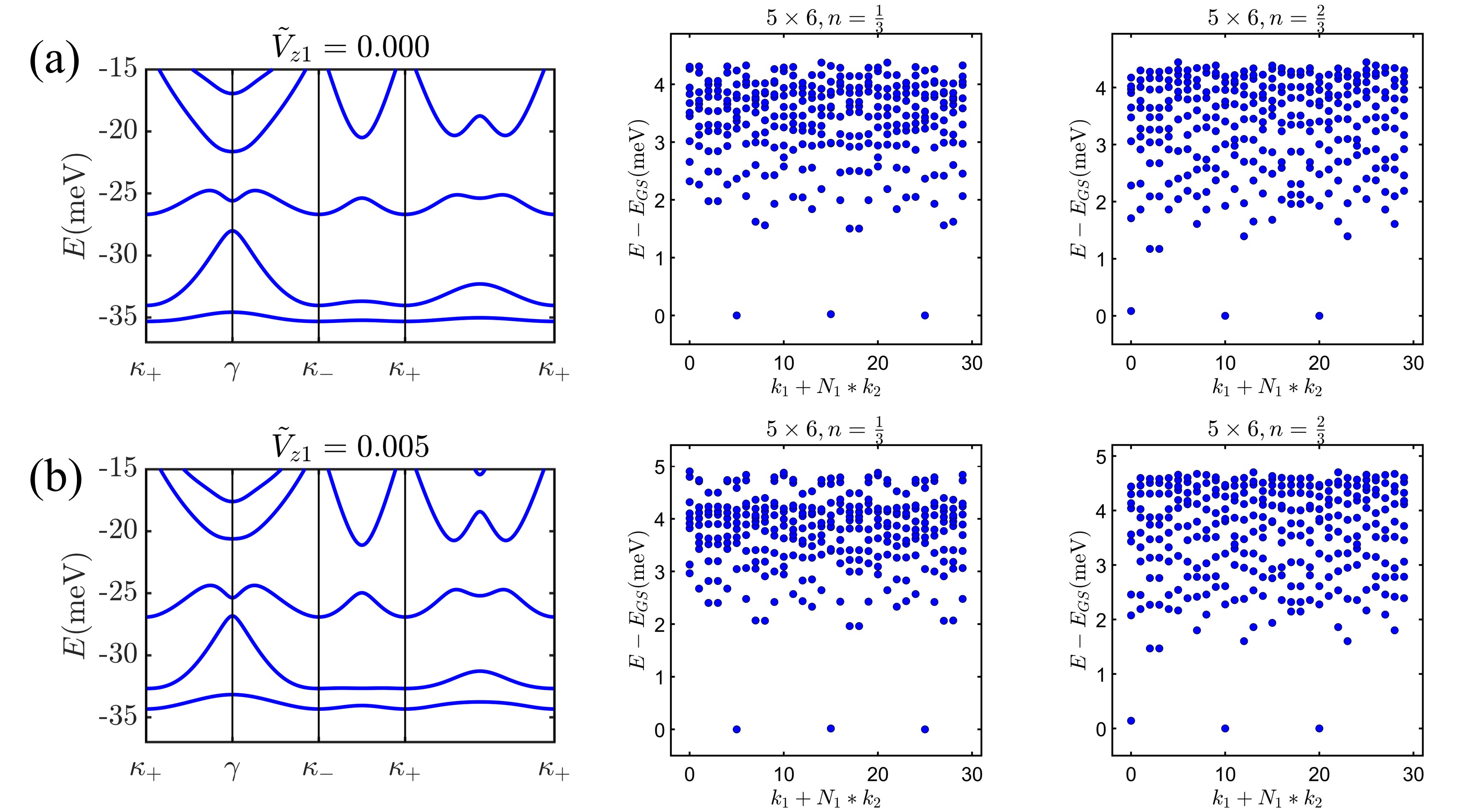}
    \caption{Band results and many body spectra in the presence of an inhomogeneous $V_z$, with $\tilde{V}_{z1}$=0(a), $\tilde{V}_{z1}$=0.005(b). Other parameters are $\tilde{\lambda}=0.6$, $\tilde{V}_{z0}=0.3$, $\tilde{U}_0=0.03$ for the many body spectra.}
    \label{inhomogenity of Vz}	
\end{figure}

In experimental realizations, disorder and spatial inhomogeneities are inevitable. Here we examine the stability of the flat bands and FCI states against two representative classes of perturbations. We first consider spatial fluctuations of the Zeeman field, $V_z$, which may arise from coupling to the moir\'e-periodic environment. We model the leading modulation by the lowest reciprocal lattice harmonics, 
\begin{equation} 
V_z(\bm{r})= V_{z0}+2V_{z1}\sum_{n=0}^{2}\cos(\bm{G}_n \cdot \bm{r}), 
\end{equation} 
where $\bm{G}_n=(4\pi/\sqrt{3}L)(\cos n\phi,\sin n\phi)$ with $\phi=2\pi/3$ are the primary reciprocal lattice vectors. Figure~\ref{inhomogenity of Vz} shows the corresponding single-particle band structures and many-body spectra for representative parameters. Both the flat band and the topological ground-state degeneracy remain robust against this modulation. 

\begin{figure}
    \centering    \includegraphics[width=0.8\columnwidth]{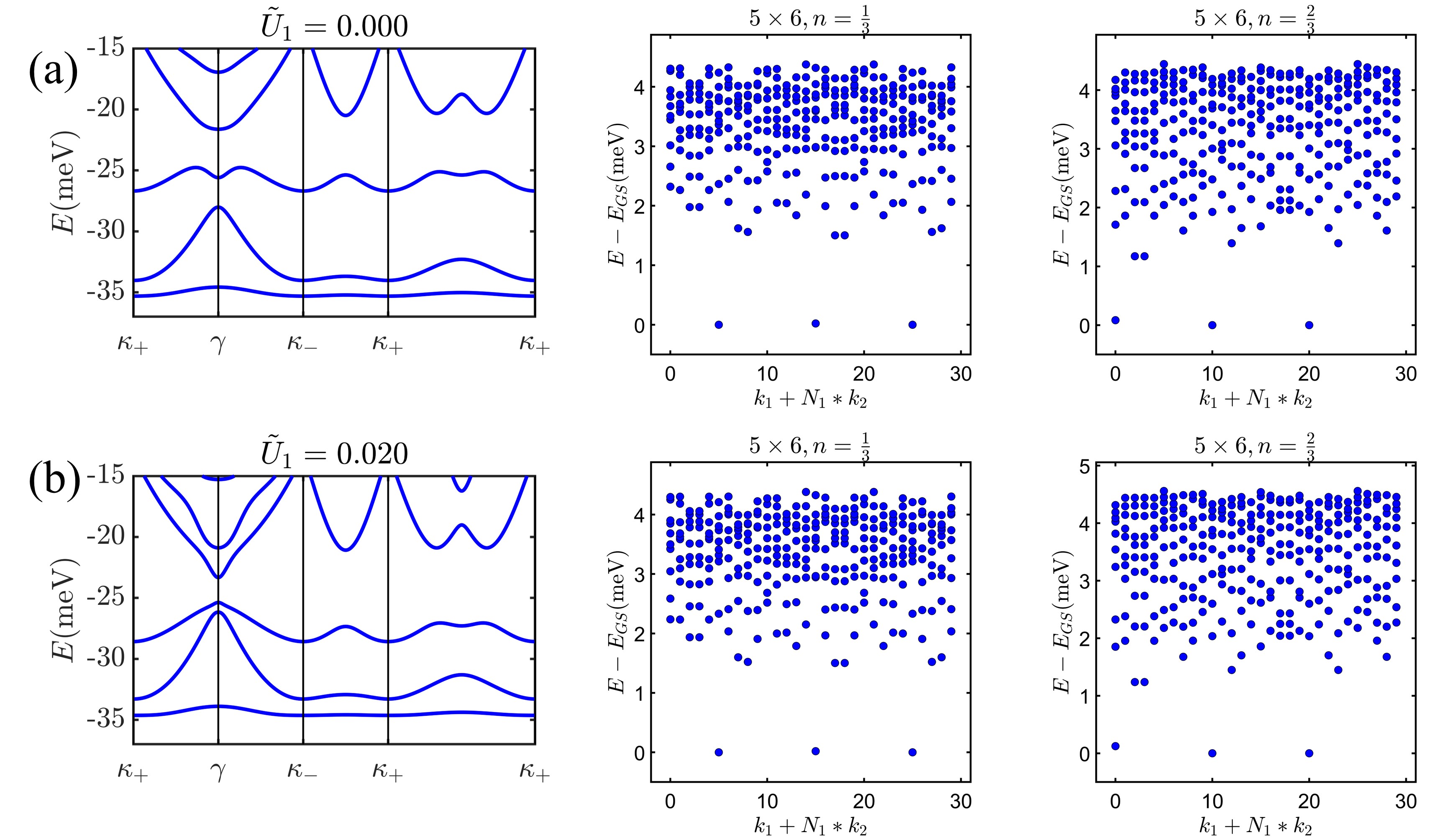}
    \caption{Band results and many body spectra with the second-order Fourier component of the superlattice potential $U(\bm{r})$, with $\tilde{U}_{1}$=0(a), $\tilde{U}_{1}$=0.02(b). Other parameters are $\tilde{\lambda}=0.6$, $\tilde{V}_{z}=0.3$, $\tilde{U}_0=0.03$ for the many body spectra.}
    \label{inhomogenity of U}	
\end{figure}

We next study inhomogeneity in the superlattice potential. As the first perturbation, we include second-harmonic components, 
\begin{equation} 
U(\bm{r})=2 U_0 \sum_{n=0}^2 \cos \left(\bm{G}_n \cdot \bm{r}\right) +2U_{1}\sum_{n=0}^2 \cos \left(\bm{G}_{2n} \cdot \bm{r}\right), \label{Vout} \end{equation} 
where $\bm{G}_{2n}= (4\pi/L)(\cos (n\phi+\pi/6), \sin (n\phi+\pi/6))$ with $\phi=2\pi/3$ are the second-shell reciprocal lattice vectors. As shown in Fig.~\ref{inhomogenity of U}, the flat-band structure and the many-body topological degeneracy persist, demonstrating the resilience of the FCI state against such lattice deformations. 

\begin{figure}
    \centering
    \includegraphics[width=0.8\columnwidth]{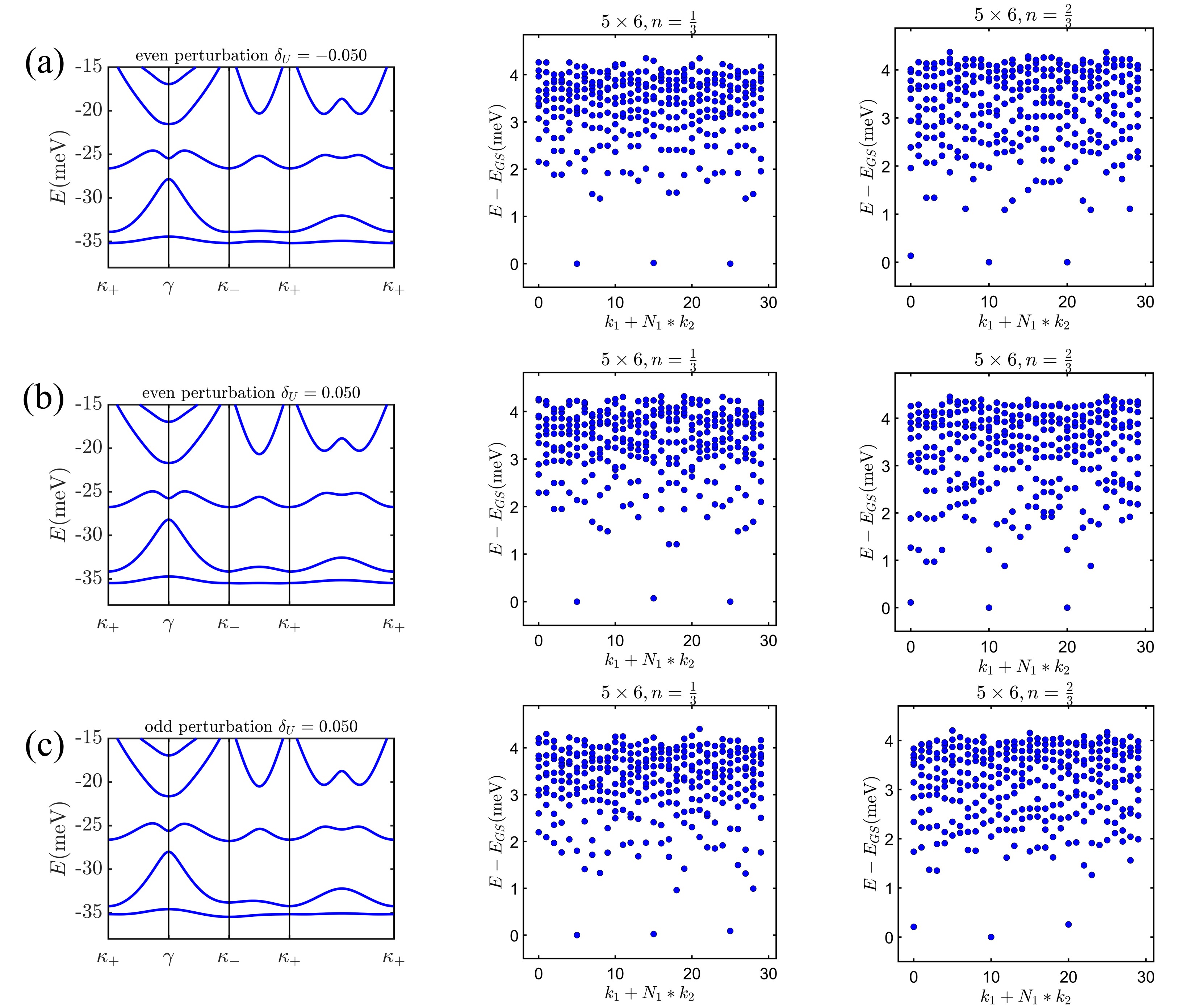}
    \caption{Band results and many body spectra with even(a,b) and odd(c) perturbation that breaks $C_3$ symmetry of the superlattice potential $U(\bm{r})$, with $\delta_U=0.05$. Other parameters are $\tilde{\lambda}=0.6$, $\tilde{V}_{z}=0.3$, $\tilde{U}_0=0.03$. }
    \label{inhomogenity of U2}	
\end{figure}

As the second perturbation, we introduce weak $C_3$-symmetry breaking in the superlattice potential. We consider two representative forms. The first one is the even form: 
\begin{equation} 
U(\bm{r})= 2U_0\delta_U\cos(\bm{G}_0 \cdot \bm{r}) +2 U_0 \sum_{n=0}^2 \cos \left(\bm{G}_n \cdot \bm{r}\right), 
\end{equation} 

while the second one is the odd form: \begin{equation} 
U(\bm{r})= 2U_0 \delta_U\sin(\bm{G}_0 \cdot \bm{r})+2 U_0 \sum_{n=0}^2 \cos \left(\bm{G}_n \cdot \bm{r}\right). 
\end{equation} 
The resulting band structures and many-body spectra are shown in Fig.~\ref{inhomogenity of U2}. For small perturbations, the flat band remains robust and the topological ground-state degeneracy is preserved.

\section{Candidate materials and engineering conditions}

In the main text, Table~I summarizes the representative candidate materials and the corresponding engineering conditions for realizing topological flat miniband by setting the dimensionless parameters $(\alpha_1,\alpha_2) = (0.4,0.12)$.  As illustrated in Fig.~4(c) of the main text and reproduced in Fig.~\ref{new ratio}, the optimal condition traces out an arc in the parameter space spanned by $\tilde{\lambda}$ and $\tilde{V}_z$. This implies that $\alpha_{1,2}$ can assume values anywhere along this arc, providing a flexible range of engineering conditions. To complement Table~I of the main text, we list the corresponding parameter ranges in Table \ref{Table1}.

Regarding the experimental realization of the dielectric superlattice, we discuss several possible routes. Epitaxial growth would be an ideal approach, as it can in principle provide a clean periodic modulation with high crystalline quality and minimal disorder. However, achieving a controllable superlattice with a short periodicity of $L\sim 5\text{--}10{\rm nm}$ by epitaxial growth remains technologically challenging.

Moiré superlattices provide a promising alternative route to access this length scale. For example, one may use doped moiré heterostructures, such as transition-metal-dichalcogenide (TMD) heterobilayers~\cite{wu2018hubbard}, where carriers are naturally localized near high-symmetry stacking regions, such as the (AA) sites, forming an emergent triangular lattice. In this setting, both the target periodicity and the strength of the effective potential can be tuned by the twist angle, lattice mismatch, interlayer displacement field, and carrier density. Another possible route is to exploit charge order formed at the interface between two-dimensional materials~\cite{tseng2022gate, wang2022quantum, lu2023synergistic, yang2023unconventional, tan2024designing}. Such interfacial charge ordering can generate a spatially periodic electrostatic landscape for the active electronic layer and thus serve as an intrinsic dielectric superlattice potential.

We also want to emphasize that, while achiving short periodicity is possible, as discussed in the last paragraph, because the FCI phase in our model remains stable over a broad parameter choice as shown in Table~\ref{Table1}, the shortest period is not a strict requirement; larger lattice constants can also be used, providing additional flexibility for experimental implementation.

We further comment on the achievable potential amplitude. In moiré- or charge-order-based realizations, the effective periodic potential is generated by spatial modulation of the local electrostatic environment, including local band alignment, screening, charge redistribution, and interlayer hybridization. Its amplitude is therefore not limited to the bare externally applied gate voltage, but is controlled by the self-consistent electrostatic potential felt by the low-energy carriers. Experimentally, artificial superlattices have already demonstrated sizable effective periodic modulations\cite{wang2024dispersion, wang2026correlated}. In particular, an artificial kagome superlattice in graphene has realized an effective onsite modulation up to $V_{\rm on-site}=168{\rm meV}$\cite{wang2024dispersion}, where $V_{\rm on-site}$ denotes the extremum of the total real-space superlattice potential. Translated into the Fourier-amplitude convention used in our model, this corresponds to $U_0\sim 18{\rm meV}$, or equivalently $\tilde U_0\sim 0.09$. This value is well above the potential amplitude required in our phase diagram to obtain a flat Chern band and stabilize the FCI phase. 

% For instance, in doped twisted bilayer transition metal dichalcogenides (TMDs), charge localization at specific high-symmetry stacking sites induces an effective triangular lattice potential. 

\begin{table}[t]
\caption{\label{table1}
Candidate materials and the corresponding engineering parameter range for realizing topological flat minibands. The SOC strength $\lambda$ and electron effective mass $m^*$ are reproduced from Table~I of the main text. The ranges of superlattice period $L$, Zeeman coupling $V_z$, and band gap $\Delta$ are estimated from Eq.~(6) in the main text by varying $(\alpha_{1},\alpha_{2})$ from $(0.35,0.08)$ to $(0.65,0.395)$ along the arc shown in Fig.~4(c) of the main text}
\begin{ruledtabular}
\begin{tabular}{l l l   |  l l l l}
    Material & $\lambda$ (eV$\cdot$\AA) & $m^*/m_e$ & $L$ (nm) & $U_0$ (meV)& $V_z$ (meV) & $\Delta$ (meV) \\
\hline
    h-TaN & 4.23 & 0.063 & 7.50--13.9 & 72.57--21.13  &193--277 & 24--3.6 \\
    h-NbN & 2.90 & 0.094 & 7.32--13.6 & 51.06--15.70 &136--194 & 17--2.5 \\
    AlBi  & 2.80 & 0.043 & 16.7--31.0 & 21.45--6.22 &57.5--82.3 & 7.2--1.1 \\
    PbSi  & 2.70 & 0.019 & 39.4--73.1 &8.72--2.54 &23.5--33.7 & 2.9--0.4 \\
    BiSb  & 2.30 & 0.037 & 23.2--43.1 & 12.91--3.74 &34.0--48.6 & 4.2--0.6 \\
    PbBi  & 1.60 & 0.119 & 10.5--19.5 & 19.60--5.68 &52.2--74.8 & 6.5--1.0 \\
    GaTe  & 1.00 & 0.229 & 8.75--16.3 & 14.67--4.23 &39.2--56.1 & 4.9--0.7 \\
    BiTeI & 1.97 & 0.157 & 6.46--12.0 & 39.25--11.38  &104--149.6 & 13--1.9 \\
    SbTeI & 1.39 & 0.134 & 10.7--19.9 &16.76--4.85 &44.4--63.6 & 5.5--0.8 \\
    TiS2Se & 1.08 & 0.522 & 3.55--6.59 & 39.10--11.34 &104--150 & 13--1.93 \\
    WSeTe & 0.92 & 0.936 & 2.32--4.31 &51.05--14.80 &136--194 & 17--2.5 \\
    SnSTe & 0.755 & 0.186 & 14.3--26.5 & 6.76--1.97 &18.1--26.0 & 2.3--0.3 \\
    ZrS2Se & 0.717 & 0.563 & 4.95--9.20 &18.64--5.40 & 49.6--71 & 6.2--0.9 \\
\end{tabular}
\end{ruledtabular}
\label{Table1}
\end{table}

\begin{table}[t]
\caption{\label{table1}
Candidate 2D Rashba materials for realizing topological flat bands. The SOC strength $\lambda$ and electron effective mass $m^*$ are taken from Ref.~\cite{bordoloi2024promises}, while the superlattice period $L$, potential depth $U_0$, Zeeman coupling $V_z$ (meV), and band gap $\Delta$ are estimated from Eq.~(\ref{eq:opcondition}) by varying $(\alpha_{1},\alpha_{2})$ from $(0.35,0.08)$ to $(0.65,0.395)$ along the arc shown in Fig.~4(c)
}
\begin{ruledtabular}
\begin{tabular}{c c c |  c c c c}
    Materials & $\lambda$ (eV$\cdot$\AA) & $m^*/m_e$ & $L$ (nm) & $U_0$ (meV)& $V_z$ (meV) & $\Delta$ (meV) \\
\hline
    h-TaN & 4.23 & 0.063 & 7.5 -- 13.9 & 72.6 -- 21.1  &193 -- 277 & 24 -- 3.6 \\
    h-NbN & 2.90 & 0.094 & 7.3 -- 13.6 & 51.1 -- 15.7 &136 -- 194 & 17 -- 2.5 \\
    AlBi  & 2.80 & 0.043 & 16.7 -- 31.0 & 21.6 -- 6.2 &57.5 -- 82.3 & 7.2 -- 1.1 \\
    PbSi  & 2.70 & 0.019 & 39.4 -- 73.1 &8.7 -- 2.5 &23.5 -- 33.7 & 2.9 -- 0.4 \\
    BiSb  & 2.30 & 0.037 & 23.2 -- 43.1 & 12.9 -- 3.7 &34.0 -- 48.6 & 4.2 -- 0.6 \\
    PbBi  & 1.60 & 0.119 & 10.5 -- 19.5 & 19.6 -- 5.7 &52.2 -- 74.8 & 6.5 -- 1.0 \\
    GaTe  & 1.00 & 0.229 & 8.8 -- 16.3 & 14.7 -- 4.2 &39.2 -- 56.1 & 4.9 -- 0.7 \\
    BiTeI & 1.97 & 0.157 & 6.5 -- 12.0 & 39.2 -- 11.4  &104 -- 150 & 13 -- 1.9 \\
    SbTeI & 1.39 & 0.134 & 10.7 -- 19.9 &16.8 -- 4.9 &44.4 -- 63.6 & 5.5 -- 0.8 \\
    TiS$_2$Se & 1.08 & 0.522 & 3.6 -- 6.6 & 39.1 -- 11.3 &104 -- 150 & 13 -- 1.93 \\
    WSeTe & 0.92 & 0.936 & 2.3 -- 4.3 &51.1 -- 14.8 &136 -- 194 & 17 -- 2.5 \\
    SnSTe & 0.76 & 0.186 & 14.3 -- 26.5 & 6.8 -- 2.0 &18.1 -- 26.0 & 2.3 -- 0.3 \\
    ZrS$_2$Se & 0.72 & 0.563 & 5.0 -- 9.2 &18.6 -- 5.4 & 49.6 -- 71 & 6.2 -- 0.9 \\
\end{tabular}
\end{ruledtabular}
\label{Table1}
\end{table}

\section{Square superlattice potential}
\begin{figure}[t]
    \centering
    \includegraphics[width=0.9\columnwidth]{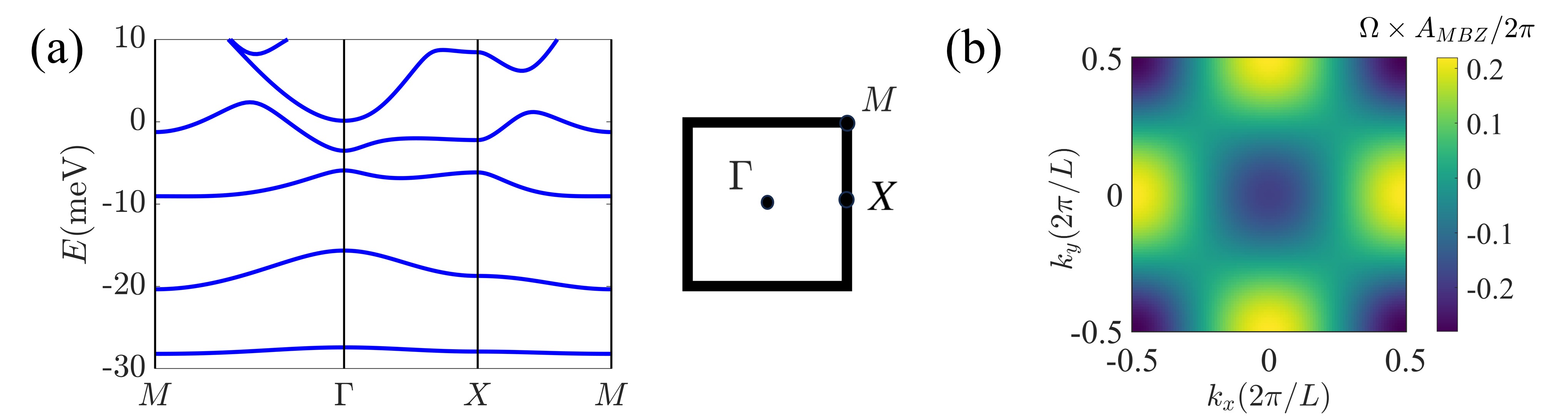}
    \caption{(a) Miniband structure for a square superlattice potential with period $L=15$ nm. (b) Berry curvature of the lowest-energy mininband. These results are obtained by choosing $m^*=0.2m_e$, $\tilde{\lambda}=0.4$, $\tilde{V}_z=0.12$, and $\tilde{U}_0=0.052$.}
    \label{square band}	
\end{figure}

In this section, we consider the case of a square superlattice potential, given by
\begin{equation}
U(\bm{r})=2 U_0 \sum_{n=0}^1 \cos \left(\bm{G}'_n \cdot \bm{r}\right),
\end{equation}
where the reciprocal lattice vectors are defined as $\bm{G}'_n=(2\pi/L)[\cos(n \pi/2),\sin(n\pi/2)]$. As shown in Fig~\ref{square band}(a), the square superlattice potential also produces an isolated flat miniband that is well separated from other bands. However, in contrast to the triangular superlattice case, this miniband is topologically trivial, carrying Chern number $\mathcal{C} = 0$, as shown in Fig~\ref{square band}(b).  

\bibliography{sup}